\newif\ifconfver
\confvertrue        

\newif\ifonecoltab

\newif\ifplainver  

\ifplainver
    \confverfalse                         
\fi

\ifconfver
     \documentclass[10pt,journal]{IEEEtran}
\else
    \ifplainver
        \documentclass[11pt]{article}
        \usepackage{fullpage} 
        \usepackage{amssymb}
    \else
        \documentclass[12pt,draftcls,onecolumn]{IEEEtran}
    \fi
\fi

\usepackage{calc,amsfonts,amsmath,bm,url,color,theorem,graphicx,cite,epstopdf,nicefrac,bbold}
\usepackage{psfrag,subcaption,float}
\usepackage[algoruled,linesnumbered]{algorithm2e}
\usepackage{multirow}
\usepackage[normalem]{ulem}
\usepackage{stmaryrd}
\usepackage{xcolor}
\usepackage{psfrag,framed,mdframed}
\usepackage{comment}
\usepackage{mdframed}
\usepackage{threeparttable}
\usepackage{amsmath,mathtools}
\usepackage{amssymb}
\usepackage{adjustbox}
\usepackage{placeins}
\usepackage{pifont} 

\usepackage{tikz}
\usetikzlibrary{positioning} 

\usepackage{tikz-cd}

\definecolor{orange}{RGB}{255,107,0}

\newcommand{\X}{\boldsymbol{X}}

\newcommand{\A}{\boldsymbol{A}}
\newcommand{\B}{\boldsymbol{B}}

\renewcommand{\S}{\boldsymbol{S}}

\newcommand{\x}{\boldsymbol{x}}

\newcommand{\z}{\boldsymbol{z}}
\renewcommand{\c}{\boldsymbol{c}}
\newcommand{\y}{\boldsymbol{y}}

\newcommand{\w}{\boldsymbol{w}}

\newcommand{\T}{{\!\top\!}}

\newcommand{\tX}{\underline{\bm X}}
\newcommand{\tY}{\underline{\bm Y}}

\newcommand{\g}{\bm{g}}

\newcommand{\pM}{^{({\rm M})}}
\newcommand{\pH}{^{({\rm H})}}

\DeclareMathOperator*{\minimize}{\textrm{minimize}}

\usepackage{xcolor}
\usepackage{psfrag,framed}
\usepackage{lipsum}
\usepackage{chapterbib}
\PassOptionsToPackage{normalem}{ulem}
\usepackage{ulem}
\definecolor{shadecolor}{RGB}{220,220,220}

\newtheorem{Fact}{Fact}
\newtheorem{Lemma}{Lemma}

\newtheorem{Theorem}{Theorem}
\newtheorem{Def}{Definition}

\theorembodyfont{\rmfamily}

\newtheorem{Assumption}{Assumption}
\newtheorem{Remark}{Remark}

\usepackage{extarrows}

\definecolor{orange}{RGB}{255,107,0}

\newcommand{\cmark}{\textcolor{green!60!black}{\ding{51}}}
\newcommand{\xmark}{\textcolor{red}{\ding{55}}}

\begin{document}

\newcommand{\papertitle}{
Unregistered Spectral Image Fusion: Unmixing, Adversarial Learning, and Recoverability
}

\newcommand{\paperabstract}{%
This paper addresses the fusion of a pair of \textit{spatially unregistered} hyperspectral image (HSI) and multispectral image (MSI) covering roughly overlapping regions. HSIs offer high spectral but low spatial resolution, while MSIs provide the opposite. The goal is to integrate their complementary information to enhance both HSI spatial resolution and MSI spectral resolution.
While \textit{hyperspectral-multispectral fusion} (HMF) has been widely studied, the unregistered setting remains challenging. Many existing methods focus solely on MSI super-resolution, leaving HSI unchanged. Supervised deep learning approaches were proposed for HSI super-resolution, but rely on accurate training data, which is often unavailable. Moreover, theoretical analyses largely address the co-registered case, leaving unregistered HMF poorly understood.
In this work, an unsupervised framework is proposed to simultaneously super-resolve both MSI and HSI. The method integrates coupled spectral unmixing for MSI super-resolution with latent-space adversarial learning for HSI super-resolution. Theoretical guarantees on the recoverability of the super-resolution MSI and HSI are established under reasonable generative models---providing, to our best knowledge, the first such insights for unregistered HMF. The approach is validated on semi-real and real HSI-MSI pairs across diverse conditions.

}


\ifplainver

    \date{\today}

    \title{\papertitle}

    \author{
    Jiahui Song, Sagar Shrestha, and Xiao Fu\\
    School of Electrical Engineering and Computer Science\\
    Oregon State University\\
    Email: (lyuqi, xiao.fu)@oregonstate.edu
    }

	\date{}

    \maketitle

\else
    \title{\papertitle}

    \ifconfver \else {\linespread{1.1} \rm \fi

\author{   Jiahui Song, Sagar Shrestha, and Xiao Fu\\
	
	\thanks{First submission on June 25, 2025; revised on Jan 31, 2026.

		The authors are with the School of Electrical Engineering and Computer Science, Oregon State University, Corvallis, OR 97331, United States. email: (songjiah, shressag, xiao.fu)@oregonstate.edu (\textit{Corresponding author: Xiao Fu})

        This work is supported in part by the National Science Foundation (NSF) under Project ECCS-2450987.

        This paper has supplementary downloadable material available at http://ieeexplore.ieee.org., provided by the author. The material includes proofs of the lemmas and theorems. This material is 402.2KB in size.

	}
}

\maketitle

\ifconfver \else
\begin{center} \vspace*{-2\baselineskip}

\end{center}
\fi

\begin{abstract}
	\paperabstract
\end{abstract}

\begin{IEEEkeywords}\vspace{-0.0cm}%
	Unregistered image fusion, hyperspectral imaging, tensor factorization, adversarial learning, recoverability
\end{IEEEkeywords}

    \ifconfver \else \IEEEpeerreviewmaketitle} \fi

 \fi

\ifconfver \else
    \ifplainver \else
        \newpage
\fi \fi
\section{Introduction}
Remote sensing leverages various imaging modalities, such as \textit{hyperspectral images} (HSIs) and \textit{multispectral images} (MSIs) \cite{yokoya2017hyperspectral}, to capture rich information in both spatial and spectral domains. However, due to hardware limitations, HSIs typically provide high spectral but low spatial resolution, whereas MSIs offer the opposite. To address this trade-off, \textit{hyperspectral and multispectral image fusion} (HMF) techniques aim to generate \textit{super-resolution spectral images} (SRIs) with high resolution in both domains by combining HSI and MSI pairs acquired over similar regions \cite{yokoya2017hyperspectral}.

HMF methods have been developed from diverse perspectives, ranging from early approaches such as component substitution and multiresolution analysis \cite{carper1990use,aiazzi2003mtf}, to more recent strategies based on coupled matrix and tensor decomposition \cite{kanatsoulis2018hyperspectral,ding2020hyperspectral,yokoya2012coupled,wei2015hyperspectral,simoes2014convex}, and most recently, deep learning-based methods \cite{zhang2020deep,nie2020unsupervised,zheng2021nonregsrnet,qu2021unsupervised}. In particular, matrix and tensor factorization approaches \cite{liu2019there,kanatsoulis2018hyperspectral,ding2020hyperspectral,prevost2020hyperspectral,li2018hyperspectral} have established analytical frameworks with performance guarantees. These methods model the observed HSI and MSI as degraded versions of an underlying super-resolution image (SRI), and cast HMF as an inverse problem to recover the SRI. Properties of matrix and tensor factorizations are then exploited to analyze the recoverability of the SRI. However, the vast majority of existing methods assume that the HSI and MSI are spatially co-registered—i.e., the images cover exactly the same spatial region and the objects share a common underlying continuous coordinate system. In practice, this assumption can be difficult to satisfy, especially when spatial misalignment in the acquired raw images is non-negligible \cite{qu2021unsupervised,zhu2023advances}.

\begin{figure}[t!]
    \centering
    \includegraphics[width=.99\linewidth]{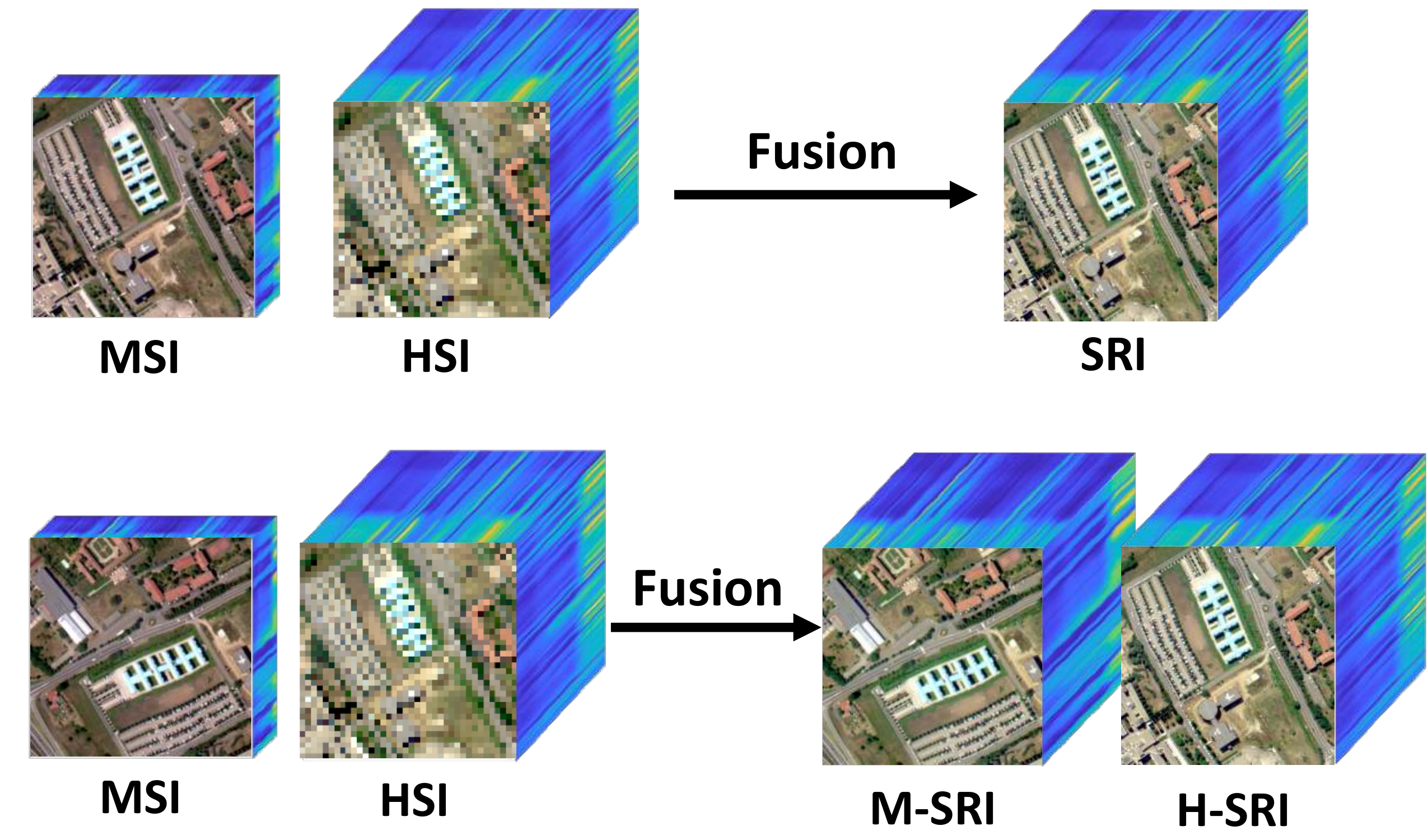}
    \caption{Top: Coregistered HMF. Bottom: Unregistered HMF. M-SRI and H-SRI are SRIs covering the spatial regions of the MSI and the HSI, respectively. In all the images, the third mode is the spectral domain.}
    \label{fig:HMF}
\end{figure}

Hyperspectral-multispectral fusion with spatially {\it unregistered} HSI and MSI pairs (hereafter referred to as \textit{unregistered HMF}) is a practically important and challenging problem (see Fig.~\ref{fig:HMF} for illustration). Many existing approaches formulate joint optimization problems to simultaneously estimate the spatial registration functions (e.g., coordinate transformations) and perform image fusion. For example, the works in \cite{zhou2019integrated,ying2021unaligned,fu2020simultaneous} used classical models for coordinate registration and SRI recovery simultaneously, while more recent approaches leverage neural networks to compensate for both spatial misalignment and imaging degradations \cite{hu2024unsupervised,nie2020unsupervised,zheng2021nonregsrnet,qu2021unsupervised}.

\noindent
\textbf{Challenges.} Despite growing interest in unregistered HMF, several critical challenges remain open.

\noindent
(i) \textit{Limited Problem Scope.} Many existing works adopt narrow definitions of “unregistration” and limited fusion objectives. For example, \cite{zhou2019integrated,ying2021unaligned,guo2023stereo,fu2020simultaneous,nie2020unsupervised,fang2024deep} address HSI and MSI pairs that are misaligned in coordinate systems but still cover exactly the same spatial region---i.e., unregistration is limited to rigid transformations such as rotation and shift. Other works \cite{chen2020unregistered,qu2021unsupervised,hu2024unsupervised} allow partial spatial non-overlap between the HSI and MSI, but only super-resolve the MSI's spectral information. That is, they estimate the super-resolution image (SRI) over the MSI region, while the corresponding high-resolution content over the HSI region remains unmodeled.

\noindent
(ii) \textit{Dependence on Training Data.} Several recent methods \cite{li2020mixed,liang2023blind} apply deep learning to enhance the spatial resolution of the HSI. These methods in principle can super-resolve the HSI without the assistance of a corresponding MSI.
However, these approaches require large amounts of training data, which may be unavailable in practice.

\noindent
(iii) \textit{Lack of Theoretical Understanding.}  Fundamental aspects of unregistered HMF remain largely unexplored. In particular, theoretical guarantees on recoverability of the underlying high-resolution images associated with the HSI and MSI have only been established for the co-registered setting \cite{kanatsoulis2018hyperspectral,ding2020hyperspectral,liu2019there,prevost2020hyperspectral,prevost2022hyperspectral}, and remain absent for the unregistered case.

\medskip

\noindent
\textbf{Contributions.}
This work develops an unsupervised framework for unregistered HMF with theoretical guarantees. The goal is to recover the high-resolution content over both the MSI-covered and HSI-covered regions. Our main contributions are:

\noindent
$\bullet$ \textit{An Unsupervised Framework for Unregistered HMF.}
We adopt the \textit{linear mixture model} (LMM) \cite{ma2013signal}, assuming shared endmembers across the MSI and HSI regions. A  coupled spectral unmixing criterion (via block-term tensor decomposition (BTD)) is used to recover the high-resolution image over the MSI region. To reconstruct the high-resolution image over the HSI region, we take a multimodal generative model learning perspective and introduce a collaborative adversarial learning loss that aligns the distribution of transformed HSI abundance patches with that of the MSI abundance patches, enhancing the spatial resolution of the HSI. The framework requires neither spatial co-registration nor training data.

\noindent
$\bullet$ \textit{Recoverability Guarantees.}
We provide the first theoretical recovery guarantees for unregistered HMF. We prove that: (i) The BTD-based coupled spectral unmixing stage can recover the high-resolution image over the MSI region under mild assumptions, such as spatial smoothness leading to low-rank abundance map. While coupled tensor models were previously applied to co-registered HMF \cite{kanatsoulis2018hyperspectral,ding2020hyperspectral,prevost2020hyperspectral}, our results show that it remains effective in the unregistered setting; and (ii) the proposed adversarial learning loss, under a reasonable multimodal patch generative model, enables recovery of the high-resolution image over the HSI region. To our knowledge, such guarantees have not been established in the unregistered setting.

\medskip

The proposed method is validated on both semi-real and real datasets, and achieves strong performance under a variety of conditions.

A preliminary version of this work was published at ICASSP~2026 \cite{song2026unregistered}, where only pixel misalignment scenarios (without angular mismatches) were considered.
The present journal version substantially extends that work by providing detailed proofs, new theoretical results accounting for angular mismatches, and a robust recoverability theorem.
In addition, it includes more extensive simulations and real-data experiments.

\smallskip
\noindent {\bf Notation.} 
We largely follow the established convention in signal processing.
In addition, $x,\bm x,\bm X$, and $\tX$ represent a scalar, vector, matrix, and tensor, respectively; 
both $[\X]_{i,j}$ and $\X(i,j)$ represent the $(i,j)$th element of $\X$; 
the Matlab notations
$\X(i,:)$ and $\X(:,j)$ represent the $i$th row and $j$th column of the matrix, respectively; similarly, $\tX(:,:,k)$ represents the $k$th mode-3 slab of the tensor $\tX$;
${\rm vec}(\bm X)$ is the vectorization operator that concatenate all the columns of $\bm X$; 
$\otimes$ and $\circ$ represent the Kronecker product and outer product (i.e., for $\X\in \mathbb{R}^{I\times J}$ and $\y\in \mathbb{R}^K$, the outer product $\X\circ \y \in \mathbb{R}^{I\times J \times K}$ is a third-order tensor such that $[\X\circ \bm y]_{i,j,k}=\X(i,j)\y(k)$), respectively;
[$K$] denotes the integer set $\{1,2,\cdots,K\}$ for a positive integer $K$; 
for a continuous random variable $\bm x \in \mathcal{X}$, $p_{\bm x}$ denotes its probability density function (PDF);
for a measurable function $\bm f: \mathcal{X} \to \mathcal{Y}$, the ``push-forward'' notation $[\bm f]_{\#}p_{\bm x}$ expresses the PDF of $\bm f(\bm x)$; ${\rm dom}(\bm x)$ denotes the domain of the random variable $\x$;
and for two functions $\bm f:\mathcal{X} \rightarrow \mathcal{Z}$ and $\bm g:\mathcal{Z} \rightarrow \mathcal{Y}$, their composition is denoted by $\bm f \bullet \bm g:\mathcal{X}\rightarrow \mathcal{Y}$ and defined as $(\bm f \bullet \bm g)(\bm x) = \bm f(\bm g (\bm x))$ for all $\bm x \in \mathcal{X}$.

\section{Background and Problem Statement}
Assume that there is a pair of HSI and MSI acquired over similar spatial regions, denoted by 
\begin{align*}
   &\text{HSI}:~\tY\pH\in \mathbb{R}^{I\pH \times J\pH \times K\pH }\\
   &\text{MSI}:~ \tY\pM\in\mathbb{R}^{I\pM\times J\pM\times K\pM},  
\end{align*}
respectively. 
Here, $I\pH \times J\pH$ and $K\pH$ denote the spatial and the spectral dimensions of the HSI, respectively; that is, $\tY\pH(i,j,:)\in\mathbb{R}^{K\pH}$ denotes the
HSI pixel measured over $K\pH$ bands at location $(i,j)$. Similar notations are used for $\tY\pM$ as well.
Note that when the HSI and MSI cover similar regions, 
we normally have
$$I\pH  J\pH \ll I\pM J\pM,\quad K\pH \gg K\pM.$$
This is because the HSI has a low spatial resolution but a high spectral resolution, while the MSI has the opposite. Hence, to describe similar spatial regions, the MSI uses many more pixels.
In addition, HSI sensors often operates on hundreds of wavelengths (i.e., $K\pH \geq 100$) yet MSI sensors have $K\pM$'s ranging from single-digit wavelengths to a couple of dozens.

\subsection{Coregistered HMF}
\begin{figure}[t!]
    \centering
    \includegraphics[width=.99\linewidth]{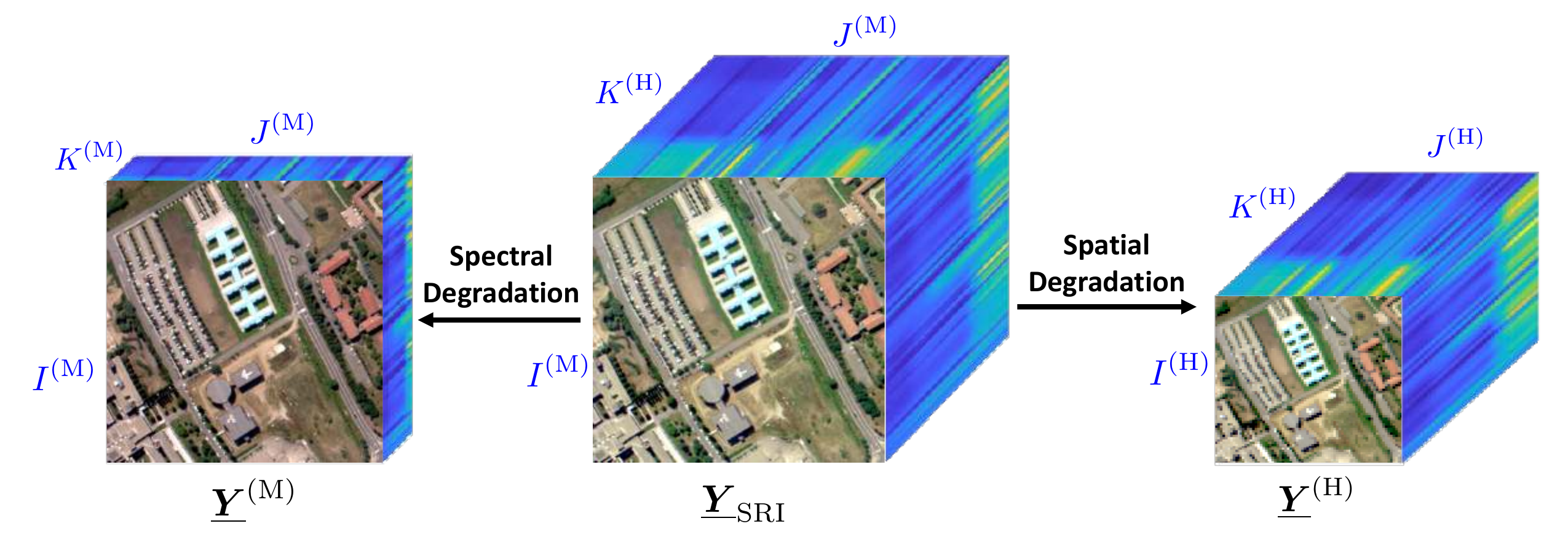}
    \caption{Data acquisition of co-registered HMF setting; see \cite{ding2020hyperspectral,kanatsoulis2018hyperspectral,wei2015hyperspectral}. }
    \label{fig:acquire_coregi}
\end{figure}

Fig.~\ref{fig:acquire_coregi} shows the data acquisition model of co-registered HMF from the fusion literature (see, e.g., \cite{kanatsoulis2018hyperspectral,ding2020hyperspectral,liu2019there,wei2015hyperspectral,simoes2014convex,yokoya2012coupled}). There, the $I\pH J\pH$ pixels of the HSI and the $I\pM J\pM$ pixels of the MSI correspond to {\it exactly} the same underlying spatial region and belong to the same continuous coordinate system. 
Both the HSI and the MSI are considered as degraded versions of the same SRI, denoted by $\tY_{\rm SRI}\in\mathbb{R}^{I\pM\times J\pM \times K\pH}
$.
A widely used degradation model for the MSI and the HSI is as follows \cite{kanatsoulis2018hyperspectral,ding2020hyperspectral,prevost2020hyperspectral,prevost2022hyperspectral}:
\begin{subequations}\label{eq:degredation}
\begin{align}
\tY\pM(i,j,:) = \bm P\pM \tY_{\rm SRI}(i,j,:),\\
\tY\pH(:,:,k) =\bm P_{\rm L} \tY_{\rm SRI}(:,:,k) \bm P_{\rm R}^\T,    
\end{align}    
\end{subequations}
respectively, where $(i,j)\in [I\pM]\times[J\pM]$ and $k\in [K\pH]$,
the matrix $\bm P\pM \in \mathbb{R}^{K\pM\times K\pH}$ is the spectral degradation operator applied onto every pixel of the SRI, i.e., $\tY_{\rm SRI}(i,j,:)$ for all $(i,j)$, and $\bm P_{\rm L} \in \mathbb{R}^{I\pH\times I\pM} $ and $\bm P_{\rm R} \in \mathbb{R}^{J\pH\times J\pM} $ the 
spatial degradation operators applied onto every band of the SRI, i.e., $\tY(:,:,k)$ for all $k$.
Note that many HMF methods in this domain assume that the degradation operators are known or estimated \cite{kanatsoulis2018hyperspectral,ding2020hyperspectral,liu2019there,wei2015hyperspectral,simoes2014convex,yokoya2012coupled}. Some recent works, e.g., \cite{kanatsoulis2018hyperspectral,ding2020hyperspectral}, relaxed the knowledge of the spatial degradation operators and argue that the $\bm P\pM$ is relatively easy to acquire, e.g., by inspecting the sensor specifications.

Under Eq.~\eqref{eq:degredation}, the co-registered HMF problem boils down to recovering $\tY_{\rm SRI}$ from fusing $\tY\pM$ and $\tY\pH$, which is an ill-posed inverse problem. 
The recoverability of $\tY_{\rm SRI}$ is nontrivial to establish in general. Nonetheless, in recent years, some works showed that $\tY_{\rm SRI}$ is recoverable via modeling the spectral images as low-rank tensors \cite{kanatsoulis2018hyperspectral,ding2020hyperspectral,prevost2020hyperspectral}, or low-rank matrices \cite{liu2019there,li2018hyperspectral}, respectively.

\begin{figure}[t!]
\centering
\includegraphics[width=\linewidth]{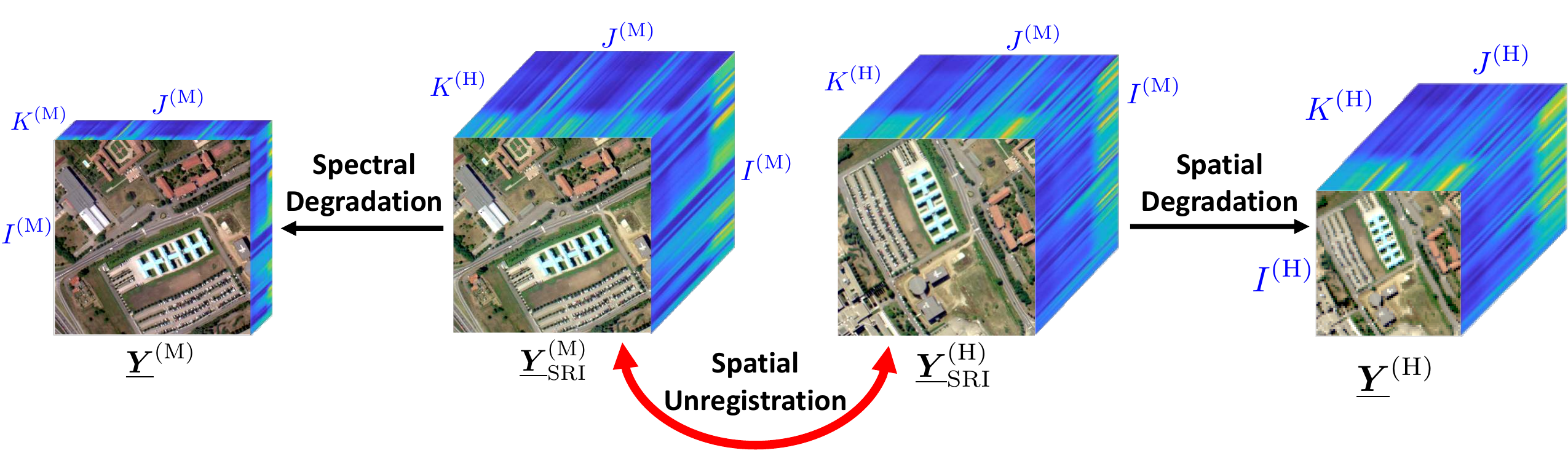}
\caption{Data acquisition model of unregistered HMF setting.}\label{fig:acquire_unregi}
\end{figure}

\subsection{Unregistered HMF}

For unregistered $\tY\pH$ and $\tY\pM$, we propose a degradation model as follows (see Fig.~\ref{fig:acquire_unregi}):
\begin{align}\label{eq:deg}
    & \tY_{\rm SRI}\pM \xRightarrow{\text{spectral degrad.}} \tY\pM,\quad \tY_{\rm SRI}\pH \xRightarrow{\text{spatial degrad.}} \tY\pH,
\end{align}   
where $\tY_{\rm SRI}\pM$ and $\tY_{\rm SRI}\pH$ are two different super-resolution spectral images with roughly overlapped spatial regions. 
Both of $\tY_{\rm SRI}\pM$ and $\tY_{\rm SRI}\pH$ have the MSI's spatial resolution and the HSI's spectral resolution.
However, these two images do not share the same underlying spatial coordinate system due to unregistration.
Spectral and spatial degradations are applied to $\tY_{\rm SRI}\pM$ and $\tY_{\rm SRI}\pH$, respectively, to produce the observed $\tY\pM$ and $\tY\pH$. 
Under \eqref{eq:deg},
unregistered HMF aims at recovering $\tY_{\rm SRI}\pM$ and $\tY_{\rm SRI}\pH$ by integrating the information in $\tY\pM$ and $\tY\pH$. 
In other words, the goal is to super-resolve 
the and spectral resolution of the MSI and
the spatial resolution  of the HSI, respectively.

A natural approach to handling unregistered HMF is to first apply a co-registration algorithm to the HSI-MSI pair and then use existing co-registered HMF methods (see, e.g., \cite{zhou2019integrated,ying2021unaligned}). Traditional co-registration techniques typically seek transformations of the pixel coordinates to correct mismatches between two images \cite{ying2021unaligned,fu2020simultaneous}, though they often perform poorly when faced with relatively large spatial misalignments (e.g., mismatches more than several pixels \cite{qu2021unsupervised}).
Learning-based one-stage approaches have also been proposed (see, e.g., \cite{guo2023stereo,zheng2021nonregsrnet,qu2021unsupervised,ying2021unaligned}), employing neural networks to capture key aspects of unregistered HMF, such as degradation and coordinate transformation functions. With the rapid advancement of deep learning, these methods sometimes show promising performance.

\subsection{Challenges}
Although some progress has been made, several critical challenges in unregistered HMF remain unresolved.

On the methodology side, there is no established method for simultaneously recovering both $\tY_{\rm SRI}\pM$ and $\tY_{\rm SRI}\pH$. Most existing works aim solely to recover $\tY\pM_{\rm SRI}$, without addressing the recovery of $\tY\pH_{\rm SRI}$---see, e.g., \cite{chen2020unregistered,qu2021unsupervised,hu2024unsupervised}. Some hyperspectral super-resolution methods (which are not fusion-based), such as \cite{liang2023blind,li2020mixed}
, attempt to recover $\tY_{\rm SRI}\pH$ from $\tY\pH$ alone using training data synthesized through assumed spatial degradation models. However, these approaches rely heavily on the accuracy of the assumed degradation process to generate training data, which may not hold in practice. As a result, how to jointly recover $\tY\pM_{\rm SRI}$ and $\tY\pH_{\rm SRI}$ from spatially misaligned and partially overlapping $\tY\pM$ and $\tY\pH$ remains an open problem.

On the theory side, unlike the co-registered HMF setting---where recoverability has been well studied \cite{kanatsoulis2018hyperspectral,ding2020hyperspectral,liu2019there,li2018hyperspectral,prevost2020hyperspectral}---existing work on unregistered HMF has largely focused on methodological innovations and empirical results. To the best of our knowledge, no theoretical framework has been developed to analyze the recoverability of either $\tY\pM_{\rm SRI}$ or $\tY\pH_{\rm SRI}$ in the unregistered setting. Even the aforementioned methods that focus on recovering $\tY\pM_{\rm SRI}$ (e.g., \cite{chen2020unregistered,qu2021unsupervised,hu2024unsupervised}) do not provide theoretical guarantees.

\subsection{Problem Statement of This Work}
\begin{figure}[t!]
    \centering
    \includegraphics[width=1\linewidth]{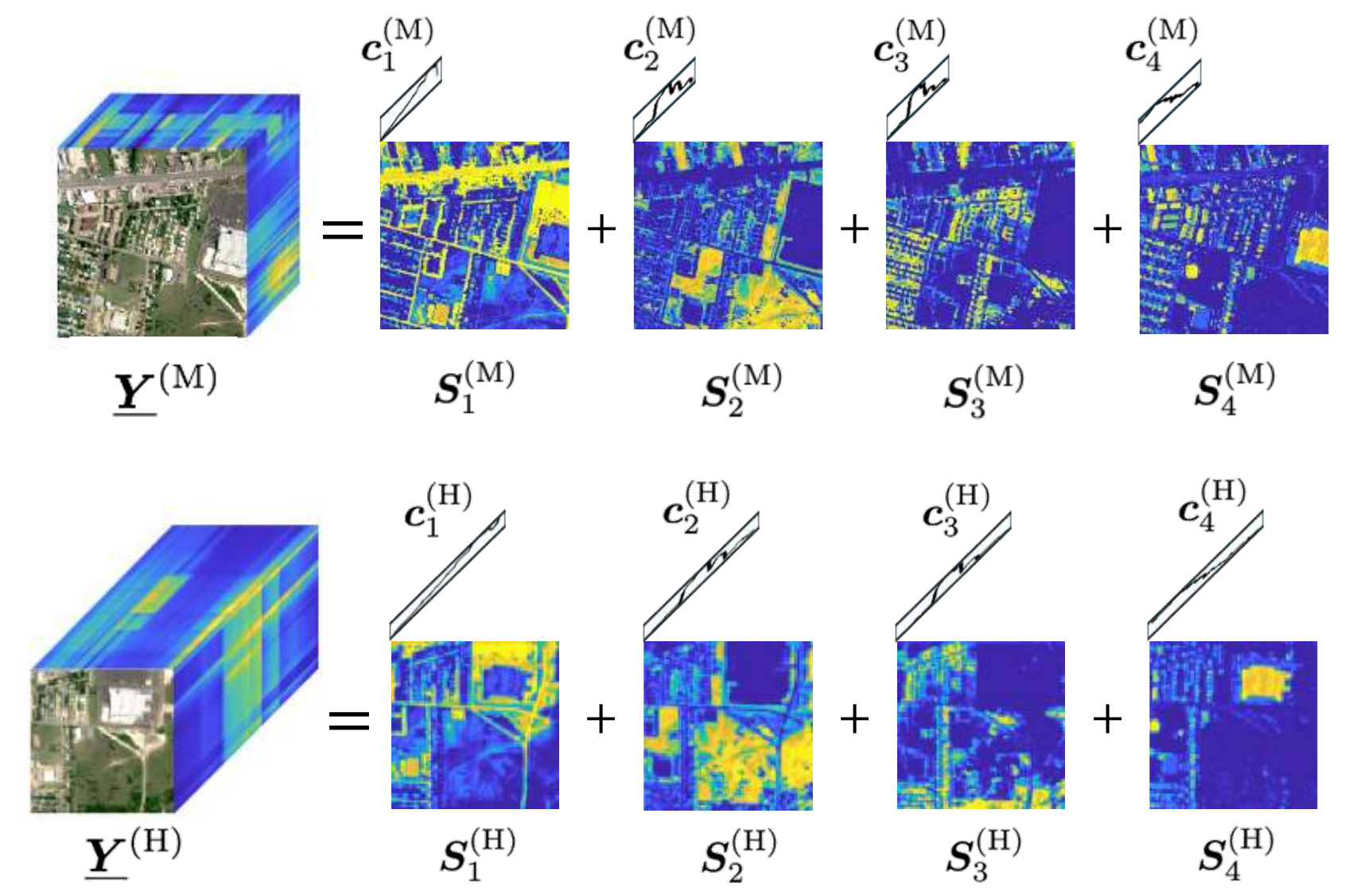}
    \caption{LMM representations of the MSI (top) and the HSI (bottom) in a case where $R=4$.}
    \label{fig:lmm}
\end{figure}

We start by employing the following classical LMM of spectral images \cite{fu2019nonnegative,ma2013signal}.
Assume that there are $R$ and $R'$ materials (e.g., water, soil, and vegetation) captured in the HSI and MSI, the LMM expresses the spectral images as follows:
\begin{subequations}\label{eq:lmm}
    \begin{align}
      \tY\pH &=\sum_{r=1}^{R} {\bm S}_r\pH \circ  {\c}_r\pH.\label{eq:lmmH}\\
    \tY\pM &= \sum_{r=1}^{R'} \bm S_r\pM \circ {\c}_r\pM  \label{eq:lmmM},
\end{align}
\end{subequations}
where $\c_r\pH\in\mathbb{R}^{K\pH}$ and $\c\pM\in \mathbb{R}^{K\pM}$ are the hyperspectral and multispectral signatures of material $r$ (also called endmember $r$), respectively, and $\S_r\pH\in\mathbb{R}^{I\pH\times J\pH}$ and $\S_r\pM\in\mathbb{R}^{I\pM\times J\pM}$ are the abundance maps of endmember $r$ over the spatial regions of the HSI and MSI, respectively.
Taking the HSI as an example, $[{\bm S}_r^{({\rm H})}]_{i,j}$ represents the proportion (abundance) of endmember $r$ contained in the pixel $(i,j)$. By such physical meaning, all the factors $\{\S\pH,\S_r\pM\}$ and $\{\c_r\pH,\c_r\pM\}$ are nonnegative, and the abundance maps satisfy \cite{ma2013signal,fu2014self,bioucas2009variable}:
\begin{align}\label{eq:physicalmeaning}
   \sum_{r=1}^{R} [{\bm S}_r^{({\rm H})}]_{i,j}=1,\quad \sum_{r=1}^{R'} [{\bm S}_r\pM]_{i,j}=1.
\end{align}
The LMM representations of the $\tY\pM$ and $\tY\pH$ are shown in Fig.~\ref{fig:lmm}.
Under the LMM, we develop our methods based on the following assumptions:
\begin{Assumption}\label{assumption:R}
    The HSI and MSI cover similar regions so that $\{ \c_r\pH \}_{r=1}^R$ and $\{ \c_r\pM \}_{r=1}^{R'}$ represent the same set of materials with $R=R'$.
\end{Assumption}

\begin{Assumption}\label{assumption:P}
    The spectral degradation operator $\bm P\pM \in \mathbb{R}^{K\pM \times K\pH}$ [cf. Eq.~\eqref{eq:degredation}] is known or previously estimated.
\end{Assumption}
Note that Assumption~\ref{assumption:R} is arguably mild: when the the HSI and the MSI cover roughly overlapped regions, it is reasonable to believe that both images contain the same materials.
Assumption~\ref{assumption:P} is often supported by the availability of sensor specifications; see, e.g., discussions in \cite{simoes2014convex,kanatsoulis2018hyperspectral}. This assumption leads to an important relationship under \eqref{eq:lmm} and Assumption~\ref{assumption:R} \cite{kanatsoulis2018hyperspectral,ding2020hyperspectral}:
\begin{align}\label{eq:c_relation}
    \c_r\pM=\bm P\pM \c_r\pH,~\forall r\in[R].
\end{align}
In practice, when $\bm P\pM$ is unknown, we will propose a simple heuristic to estimate it; see Sec.~\ref{sec:PMestimate}.

Under these assumptions, the LMM representations of the SRIs are
\begin{align}\label{eq:lmmrepresentation}
     \tY\pH_{\rm SRI} =\sum_{r=1}^R \widetilde{\S}_r\pH \circ \c\pH_r,\quad    \tY\pM_{\rm SRI} = \sum_{r=1}^R{\S}_r\pM \circ \c\pH_r,
\end{align}
where $\widetilde{\S}_r\pH$ is the super-resolved version of ${\S}_r\pH$.
Therefore, unregistered HMF of $\tY\pM$ and $\tY\pH$ in \eqref{eq:lmm} can be considered as converting $\{\c_r\pM\}$ and $\{ \S_r\pH \}$ 
to their counterparts $\{ \c_r\pH\}$ and $\{  \widetilde{\S}_r\pH\}$---so that one can ``recover'' the two SRIs in \eqref{eq:lmmrepresentation}. We emphasize that the images lack spatial alignment. Consequently, the unregistered HMF problem is a much harder task compared to those in \cite{kanatsoulis2018hyperspectral,wei2015hyperspectral,ding2020hyperspectral,yokoya2012coupled,simoes2014convex}. 
In this work, we take a divide-and-conquer approach to tackle the unregistered HMF problem. That is, we solve two ``sub-tasks'', namely, {\it multispectral super-resolution} (MSR) and {\it hyperspectral super-resolution} (HSR), in a sequential manner.

\section{MSR via Coupled Spectral Unmixing}
We first deal with the problem of sharpening the spectral resolution of $\tY\pM$ to recover $\tY_{\rm SRI}\pM$---i.e., the MSR task.
Under \eqref{eq:lmm}, MSR boils down to {\it unmixing} $\{\S_r\pM\}_{r=1}^R$ from the MSI and $\{\c_r\pH\}_{r=1}^R$ from the HSI, respectively. Then, the high-resolution abundance maps can be ``re-assembled'' with the high-resolution endmembers to recover $\tY\pM_{\rm SRI}$ via $\widehat{\tY}\pM_{\rm SRI} = \sum_{r=1}^R \widehat{\S}\pM_r\circ \widehat{\c}\pH_r$,
where $\widehat{\S}\pM_r$ and $\widehat{\c}\pH_r$ are the estimated MSI abundance and HSI spectrum of material $r$, respectively.

\subsection{Problem Formulation for MSR}
Many spectral unmixing approaches can potentially serve our purpose (see, e.g., \cite{ding2023fast,fu2014self,bioucas2009variable}).
In this work, we propose to use a tensor-based unmixing framework, which works under the following relatively mild assumption:
\begin{Assumption}\label{assumption:tensor}
    The abundance maps of the HSI and MSI are low-rank matrices; i.e., 
    \begin{align}
        {\rm rank}(\S_r\pM) \leq L_r\pM,~{\rm rank}(\S_r\pH)\leq L_r\pH,~\forall r\in[R],
    \end{align}
    where $L_r\pH$ and $L_r\pM$ denote the rank upper bounds of the HSI and MSI abundance maps associated with material $r$, respectively.
\end{Assumption}
Assumption~\ref{assumption:tensor} is often used in spectral image analysis (see, e.g., \cite{qian2016matrix,ding2020hyperspectral,ding2023fast,xiong2018hyperspectral,prevost2022hyperspectral}), which stems from the fact that the abundance maps exhibit spatial smoothness in many cases.
For MSR, we formulate the following criterion:
\begin{subequations} \label{eq:MSR}
\begin{align} 
    &\minimize_{\{\bm S_r\pH, \c_r\pH, \bm S_r\pM\}_{r=1}^{R}} \left\|\underline{\bm Y}\pH - \sum_{r=1}^{R} \bm S_r\pH \circ \bm c_r\pH \right\|_{\rm F}^2 \label{eq:obj_MSR}\\
    & \qquad\qquad\qquad\quad +  \left\| \underline{\bm Y}\pM - \sum_{r=1}^R \bm S_r\pM \circ (\bm P\pM \bm c_r\pH) \right\|_{\rm F}^2,  \nonumber\\
    &{\rm subject~to:}~{\rm rank}(\S_r\pM) \leq L_r\pM,~\forall r\in[R],\\
    &~~~~~~~~~~~~~~~~{\rm rank}(\S_r\pH)\leq L_r\pH,~\forall r\in[R],\\
    &~~~~~~~~~~~~~~~~\sum_{r=1}^R \S_r\pM =\bm 1\bm 1^\T,~~\sum_{r=1}^R \S_r\pH =\bm 1\bm 1^\T,\\
    &~~~~~~~~~~~~~~~~\S_r\pM,\S_r\pH,\c_r\pH\geq \bm 0,~\forall r\in [R].
\end{align} 
\end{subequations}
Note that the last two constraints are added according to the physical meaning of the factors; see \eqref{eq:physicalmeaning}.

\subsection{MSR Recoverability}
Denote $\{\widehat{\bm S}_r\pH, \widehat{\c}_r\pH, \widehat{\bm S}_r\pM\}_{r=1}^{R}$ as any optimal solution of \eqref{eq:MSR}.
Also represent the low-rank ground-truth abundances using factorization-based forms, i.e., $$\S_r\pH=\A_r\pH(\B_r\pH)^\T,~\S_r\pM=\A_r\pM(\B_r\pM)^\T$$ where $\A_r\pH\in\mathbb{R}^{I\pH\times L_r\pH}$, $\B_r\pH\in\mathbb{R}^{J\pH\times L_r\pH}$, $\A_r\pM\in\mathbb{R}^{I\pM\times L_r\pM}$, and $\B_r\pM\in\mathbb{R}^{J\pM\times L_r\pM}$. 
For ease of exposition, let $L_r\pH = L\pH$ and $L_r\pM = L\pM$ for all $r$.
We show the following:
\begin{Theorem}[MSR Recoverability]\label{theorem:MSR}
    Suppose that Assumptions~\ref{assumption:R}-\ref{assumption:tensor} hold, that $(\A_r\pH,\B_r\pH)$, $(\A_r\pM,\B_r\pM)$, and $\bm c_r\pH$ for all $r$ are drawn from any joint absolutely continuous distribution, that $\bm P\pM$ has full row rank, and that $K\pM \geq 2$. 
    Suppose that the following inequalities hold:
    \begin{align*}
         &I\pH J\pH \geq (L\pH)^2 R,~I\pM J\pM \geq (L\pM)^2 R,\\
         &\min\big( \lfloor\nicefrac{I\pH}{L\pH}\rfloor,R\big) + \min\big(\lfloor\nicefrac{J\pH}{L\pH}\rfloor,R\big) + \min\big(K\pH,R\big) \nonumber \\
         &\qquad \geq 2R +2,\\
         &\min\big( \lfloor\nicefrac{I\pM}{L\pM}\rfloor,R\big) + \min\big(\lfloor\nicefrac{J\pM}{L\pM}\rfloor,R\big) + \min\big(K\pM,R\big) \\
         & \qquad \geq 2R +2.
    \end{align*}
Then, we have
    \begin{align}\label{eq:theoremmsr_result}
      \widehat{\S}_r\pH = \S_{\bm \pi(r)}\pH,~ \widehat{\S}_r\pM= \S_{\bm \pi(r)}\pM,~ \widehat{\c}_r\pH = \c_{\bm \pi(r)}\pH,
    \end{align}
    where $\bm \pi$ is a permutation of $\{1,\ldots,R\}$. Consequently, 
    letting $\widehat{\tY}\pM_{\rm SRI} =\sum_{r=1}^R \widehat{\S}_r\pM \circ \widehat{\c}_r\pH$,
    we have $\widehat{\tY}\pM_{\rm SRI} =\tY\pM_{\rm SRI},$
    almost surely.
\end{Theorem}
The detailed proof is provided in the supplementary material.
A proof sketch is as follows.
Under the signal model in \eqref{eq:lmm} and the stated assumptions, both the MSI and HSI tensors admit block-term tensor decompositions with multilinear LL1 components (LL1 decompositions) \cite{de2008decompositions,ding2020hyperspectral}. Moreover, the imposed conditions guarantee that these LL1 decompositions are unique up to scaling and permutation ambiguities. The scaling ambiguity is eliminated by the sum-to-one constraint on the abundance maps, while the permutation ambiguity is matched through $\bm P\pM$; see the detailed proof and similar arguments in \cite{ding2020hyperspectral,zhang2020spectrum}) in the supplementary material. Consequently, $\widehat{\tY}_{\rm SRI}\pM$ in \eqref{eq:lmmrepresentation} can be recovered by re-assembling the estimated HSI endmembers with the MSI abundance maps.

The above result extends \cite[Theorem 2]{ding2020hyperspectral}, which analyzed SRI recoverability of coupled tensor decomposition under when the HSI and MSI are co-registered. In contrast, our setting allows for unregistered spatial domains, which permits the ranks of $\S_r\pM$ and $\S_r\pH$ to differ---a scenario that does not arise in the co-registered case considered in \cite{ding2020hyperspectral}; see the detailed proof of the theorem.
The conditions in Theorem~\ref{theorem:MSR} favor cases where the number of materials and the rank of the abundance maps are moderate. The low rank condition in Assumption~\ref{assumption:tensor} is considered a reasonable approximation for real-world spectral images, especially when the materials' abundances change slowly and smoothly in space; see discussions and illustrations in \cite{ding2020hyperspectral,qian2016matrix,prevost2022hyperspectral}.

\begin{Remark}
 Although we proposed a coupled tensor factorization approach to attain \eqref{eq:theoremmsr_result}, other coupled spectral unmixing techniques---such as coupled matrix factorization methods (e.g., \cite{li2018hyperspectral,liu2019there,yokoya2012coupled}), with suitable modifications to handle the unregistered case---could also be considered. These matrix factorization-based approaches may relax the low-rank assumption on the abundance maps but typically require alternative conditions, such as the sufficiently scattered condition \cite{fu2019nonnegative} or sparsity-related assumptions \cite{li2018hyperspectral,liu2019there}. A detailed discussion of alternatives is beyond the scope of this work.
\end{Remark}

\section{HSR via Abundance Adversarial Learning}
The more challenging task is to enhance the spatial resolution of $\tY\pH$, i.e., the HSR task.
Note that in the spectral unmixing stage, we identify $\{\S_r\pH,\c_r\pH\}$ such that $\tY\pH=\sum_{r=1}^R \S_r\pH \circ \c_r\pH$. Under the LMM, the spatial information of the HSI is completely captured in $\S_r\pH$ for $r=1,\ldots,R$. Therefore, the task amounts to enhancing the resolution of the abundance maps $\S_r\pH$ for all $r$.

\subsection{Patch-based Distribution Matching}
In this work, we aim to enhance the spatial resolution of patches from $\S_r\pH$ so that the resulting resolution matches that of the corresponding MSI patches. We focus on patches because super-resolution is inherently a local spatial operation. To illustrate our idea, let $\S_r\pH[{\cal H}^{\bm w}]$ and $\S_r\pM[{\cal M}^{\bm w}]$ denote two patches extracted from the HSI and MSI, respectively, where ${\cal H}^{\bm w} \in \mathbb{R}^{B\pH \times B\pH}$ and ${\cal M}^{\bm w} \in \mathbb{R}^{B\pM \times B\pM}$ represent the pixel index sets of the patches centered at the coordinate $\bm w \in \mathbb{R}^{2}$, which is defined as follows:
\begin{Def}\label{def:w}
The coordinate $\bm w \in \mathbb{R}^{2}$ is a {\it global} coordinate in the ground-truth continuous spatial domain, where the $w_1$- and $w_2$-axes are aligned with the north and east directions, respectively (cf. latitude and longitude in the geographic coordinate system).    
\end{Def}
Consider the case where these two patches correspond to the same underlying spatial region. In this case, $\S_r\pM[{\cal M}^{\bm w}]$ can be interpreted as a high-resolution counterpart of $\S_r\pH[{\cal H}^{\bm w}]$.
We postulate the existence of an unknown nonlinear function $\bm f^\star(\cdot): \mathbb{R}^{B\pH \times B\pH} \rightarrow \mathbb{R}^{B\pM \times B\pM}$ that maps the low-resolution HSI patch to its high-resolution MSI counterpart:
\begin{align}\label{eq:flink}
    \S_r\pM[{\cal M}^{\bm w}] = \bm f^\star\left( \S_r\pH[{\cal H}^{\bm w}] \right), \quad \forall r \in [R].
\end{align}
Our objective then amounts to recovering $\bm f^\star$, whose existence will be discussed under a multimodal generative model shortly (see Lemma~\ref{lemma:bijective_function_f}).

In unregistered HMF, a core challenge lies in the fact that the correspondence between ${\cal M}^{\bm w}$ and ${\cal H}^{\bm w}$ is \textit{unknown}. 
A major reason is that the association of patches and the underlying coordinate $\bm w$ is not available in the unregistered HMF case.
As a result, regression-based approaches for learning $\bm f^\star$ in \eqref{eq:flink}---such as formulations of the form
\begin{align}\label{eq:patch_match}
\bm f^\star \approx \arg\min_{\bm f}\sum_{ \bm w }\|\bm S_r\pM[{\cal M}^{\bm w}] - \bm f(\bm S_r\pH[{\cal H}^{\bm w}]) \|^2,
\end{align}
are not applicable. 
To overcome this difficulty, an alternative is to find a function $\bm f$ such that the \textit{distributions} of $\bm S_r\pM[{\cal M}^{\bm w}]$ and $\bm f(\bm S_r\pH[{\cal H}^{\bm w'}])$ are matched, rather than relying on paired patch matching as in \eqref{eq:patch_match}.
Note that if one cannot distinguish the distributions $\bm S_r\pM[{\cal M}^{\bm w}]$ and $\bm f(\bm S_r\pH[{\cal H}^{\bm w'}])$, it means that they visually belong to the same feature space---the resolution of $\bm f(\bm S_r\pH[{\cal H}^{\bm w}])$ matches that of $\bm S_r\pM[{\cal M}^{\bm w'}]$, and therefore (hopefully) $\bm f=\bm f^\star$.
In other words, assume that
\begin{align}
\S_r\pH[{\cal H}^{\bm w}]\sim p_r\pH,~\S_r\pM[{\cal M}^{\bm w'}]\sim p_r\pM,
\end{align}
where $p_r\pH$ and $p_r\pM$ are joint distributions of the pixels of the HSI and MSI abundance patches of material $r$, respectively.
Our learning objective boils down to finding $\bm f$ such that
\begin{align}\label{eq:abundanceDM}
    [\bm f]_{\#} p_r\pH = p_r\pM, ~\forall r\in[R].
\end{align}
Distribution matching objective such as Eq.~\eqref{eq:abundanceDM} can be realized using tools like the \textit{generative adversarial networks} (GANs)\cite{goodfellow2014generative}. Let $\widehat{\bm f}$ be any solution of \eqref{eq:abundanceDM}. Then,
one can estimate the super-resolved version of $\S_r\pH[{\cal H}^{\bm w}]$ using
\begin{align}\label{eq:patch_recover}
     \widehat{\S}_r[{\cal M}^{\bm w}] &= \widehat{\bm f}(\S_r\pH[{\cal H}^{\bm w}]),~\forall \bm w \\
     \widehat{\tY}\pH_{\rm SRI}[{\cal M}^{\bm w}] & = \sum_{r=1}^R \widehat{\bm f}(\S_r\pH[{\cal H}^{\bm w}]) \circ \c_r\pH,~\forall {\cal H}.
\end{align} 
Note that if $\widehat{\bm f}=\bm f^\star$, then we provably recover the super-resolved HSI patch by patch.

\begin{Remark}
To implement the above idea, there are a couple of critical challenges to overcome.
First, how should one define (and subsequently sample) the patches ${\cal M}^{\bm w}$ and ${\cal H}^{\bm w'}$ to ensure the existence of a valid super-resolution signal $\bm f^\star$ under reasonable generative models?
Second---and more critically---under what conditions does the method guarantee identifiability of $\bm f^\star$, and hence the recoverability of the high-resolution $\tY_{\rm SRI}\pH$?
The answers to both questions are far from obvious. In the following subsection, we develop implementation strategies based on a reasonable generative model, which offer insights into these fundamental issues.
\end{Remark}

\subsection{Angle Randomization and Recoverability}
To realize the idea in \eqref{eq:abundanceDM}, we hope to sample $\{ {\cal M}_\ell \}_{\ell=1}^L$ and $\{ {\cal H}_t \}_{t=1}^T$ such that
there is an underlying one-to-one mapping between the HSI and MSI patches. This is necessary to ensure that there exists an ${\bm f}^\star$ that maps any HSI patch uniquely to its target MSI representation.

To find such patch samples, some nuanced caveats exist: Notably, the \textit{orientations} of coordinate systems of the MSI and HSI are in general unknown due to unregistration. If one simply samples square patches from both images, these patches may never correspond to the same underlying regions.
A remedy is as follows: Assume that the MSI and HSI patches are associated with $D\times D$m$^2$ areas in space.
One can randomize the rotation of the sampled patches, so that every possible rotation of a spatial $D\times D$m$^2$ window is covered in $\{ {\cal M}_\ell \}$ and $\{ {\cal H}_t \}$.
However, as any $N\times$90 degree rotation (where $N\in \mathbb{Z}_+$) of the $D\times D$m$^2$ square region covers the same spatial area, such replicates still lacks one-to-one correspondence.

\begin{figure}[t!]
    \centering
    \includegraphics[width=1\linewidth]{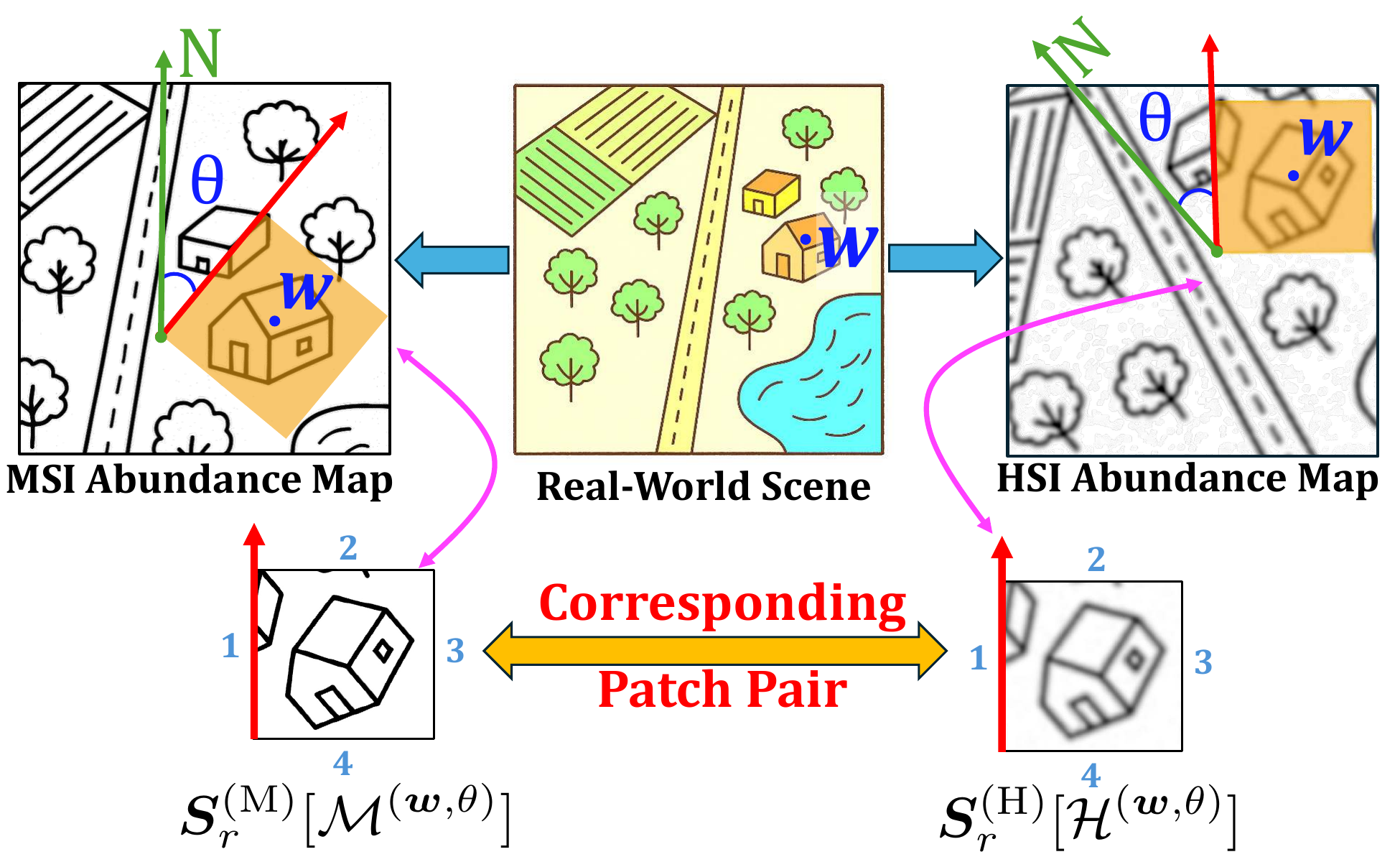}
    \caption{Patch boundaries and patch rotation angle. $\bm w$ is a continuous global coordinate (e.g., altitude and longitude from the geographic coordinate system). }
    \label{fig:patch_rotation_angle}
\end{figure}

To circumvent the above subtle but critical challenges, we define the following notions:
\begin{Def}[Patch Boundary Indexing]\label{def:patchindex} For any square patch, the boundaries are indexed by integers $\{1,2,3,4\}$. The boundaries are defined to be directional---i.e., the boundaries form a clockwise directional loop (see Fig.~\ref{fig:patch_rotation_angle}). 
\end{Def}

\begin{Def}[Patch Rotation Angle]\label{def:patchangle}
When the first boundary of a patch points to the (unknown) ground-truth north, we define the rotation angles of the patches as $\angle {\cal M}^{\bm w} =\angle {\cal H}^{\w'}=0$. In addition, $\angle {\cal M}^{\w} =\angle {\cal H}^{\w'}=\theta$ when the first boundary rotates clockwise by $\theta$ degrees away from north. A patch centered at $\w$ and rotated by $\theta$ is denoted as ${\cal M}^{(\bm w,\theta)}$ (or ${\cal H}^{(\bm w,\theta)}$); also see Fig.~\ref{fig:patch_rotation_angle}.
\end{Def}

The above two definitions will assist us to establish one-to-one correspondence of the sampled patches, under careful angle randomization and reasonable model assumptions. To move forward, we assume the following:

\begin{Assumption}[Patch Generative Model]\label{assumption:rotate_generative_model} Consider any coordinate $\bm w\in \mathbb{R}^2$ in the underlying continuous space and a random angle $\theta$.
A pair of squared MSI and HSI patches centered at $\bm w$ and rotated with $\theta$
is generated by 
\begin{subequations}\label{eq:genmodel}
\begin{align}
    &\bm w\sim p(\bm w),~\theta\sim {\rm unif}[0,360^\circ]\\
& \bm z_r\sim p({\bm z_r|\bm w,\theta}) \in\mathbb{R}^d,~\forall r\in[R],\\
    &  {\S}_r\pM[{\cal M}^{(\bm w,\theta)}]=\bm g\pM\left(\z_r\right) \in \mathbb{R}^{B\pM\times B\pM},\\
 & {\S}_r\pH[{\cal H}^{(\bm w,\theta)}]  =\bm g\pH\left(\z_r \right)\in \mathbb{R}^{B\pH\times B\pH},
\end{align}    
\end{subequations}
where $d\leq (B\pH)^2$, and $\bm g\pH$ and $\bm g\pM$ are both continuous and bijective functions.
\end{Assumption}
Under this model, $\z_r\in {\cal Z}$ denotes a $d$-dimensional latent representation of material $r$'s abundance patch centered at $\bm w$ with rotation $\theta$.
Note that all the patches $\bm{S}\pM_r[\mathcal{M}^{(\bm w,\theta)}]$ and $ \bm{S}\pH_r[\mathcal{H}^{(\bm w,\theta)}]$ are unique under a pair of $(\bm w, \theta)$, given the Definitions~\ref{def:patchindex}-\ref{def:patchangle}, which makes the bijective assumptions on $\bm g\pM$ and $\bm g\pH$ possible.
Assumption~\ref{assumption:rotate_generative_model} postulates that the patches are a collection of samples generated using $\bm  w$ in the continuous 2D space and the angle $\theta$. In this model, we treat the domain of $p(\bm w)$ as $\mathbb{R}^2$ and the MSI and HSI abundance patches as realizations of the generative process in \eqref{eq:genmodel}.
As these realizations cover the union of the spatial regions of the two spectral images, $p(\bm w)$ can be considered as a distribution that has high density over the union and small density everywhere else. Under this assumption, it is not hard to see that:
\begin{Lemma}\label{lemma:bijective_function_f}
Under Assumption~\ref{assumption:rotate_generative_model}, there exists a continuous bijective function $\bm f^\star$: ${\rm dom}({\S}_r\pH[{\cal H}^{(\bm w,\theta)}])\rightarrow {\rm dom}({\S}_r\pM[{\cal M}^{(\bm w,\theta)}])$ such that:
 \begin{align}\label{eq:abundancemapping}
 &\bm f^\star({\S}_r\pH[{\cal H}^{(\bm w,\theta)}]) = \S_r\pM[{\cal M}^{(\bm w,\theta)}],~\forall r\in[R], 
\end{align}
for all $\w$ and $\theta$.
\end{Lemma}

\begin{figure}[!t]
    \centering
    \includegraphics[width=1\linewidth]{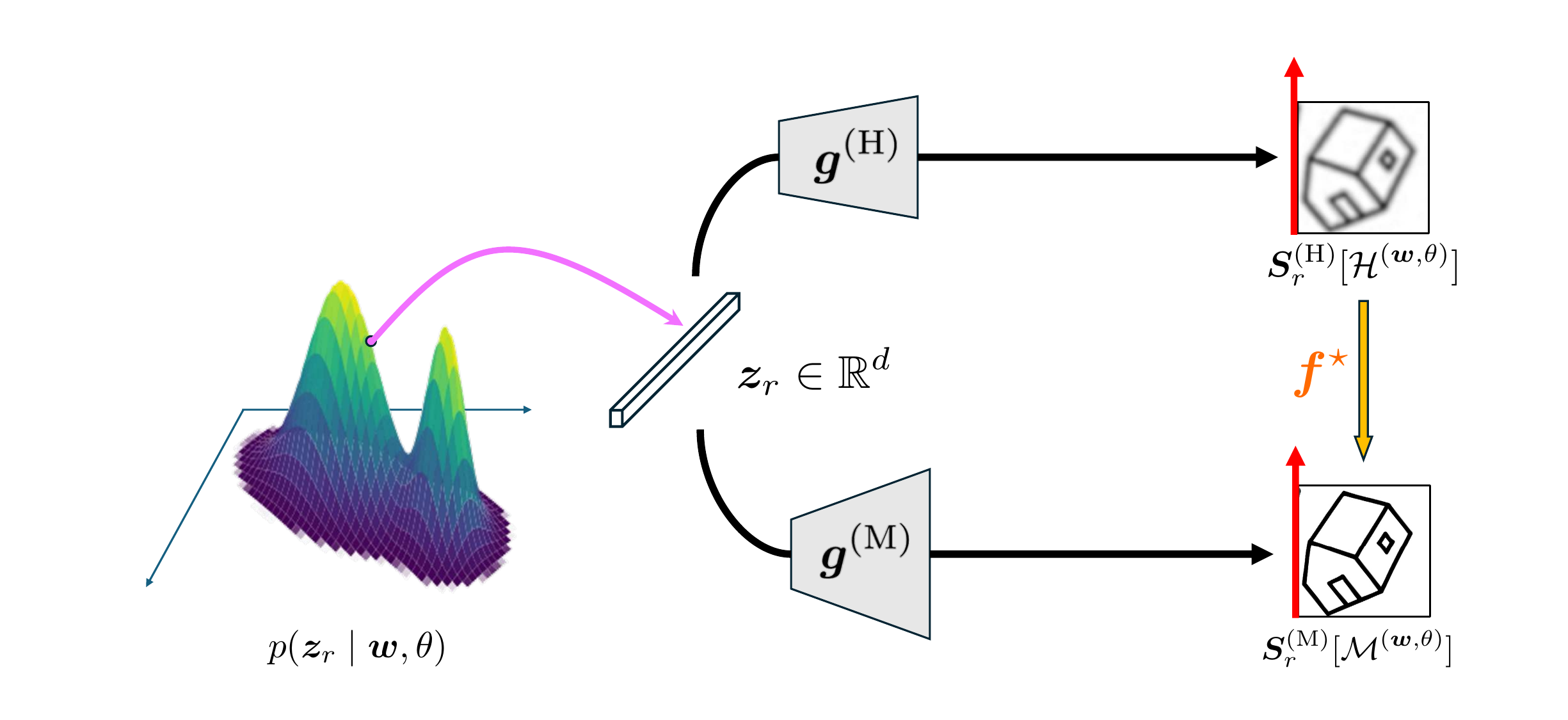}
    \caption{Illustration of the multi-modal patch generative model in Assumption~\ref{assumption:rotate_generative_model}.}
    \label{fig:mmillus}
\end{figure}

\begin{Remark}
Assumption~\ref{assumption:rotate_generative_model} and Lemma~\ref{lemma:bijective_function_f} offer a generative perspective for modeling the MSI and HSI patches; see Fig.~\ref{fig:mmillus}. Under this view, the pair $\bm{S}\pM_r[\mathcal{M}^{(\bm w,\theta)}]$ and $\bm{S}\pH_r[\mathcal{H}^{(\bm w,\theta)}]$ are two modality-specific representations of the same low-dimensional spatial content, encoded by $\bm z_r$. The generative functions $\bm g\pH$ and $\bm g\pM$ account for modality-dependent resolutions. The bijectivity of $\bm g\pH$ and $\bm g\pM$ ensures that the mapping from $\bm z_r$ to the MSI and HSI modalities preserves spatial content, and also guarantees the existence of a bijective $\bm f^\star$. Multimodal generative models of this form are common in vision and language processing; see, e.g., \cite{xie2023multi,kong2022partial,lyu2022understanding,eastwood2023self,von2024identifiable}.
The model in \eqref{eq:genmodel} renders the spatial ``degradation'' from $\tY_{\rm SRI}\pH$ to $\tY\pH$ fundamentally different from classical kernel-based degradation models that are widely used in multi-resolution image fusion. In those models, the low-spatial-resolution image is viewed as a blurred and downsampled version of its high-resolution counterpart; see, e.g., \cite{kanatsoulis2018hyperspectral,simoes2014convex,ding2020hyperspectral,wei2015hyperspectral,ding2025rethinking,li2018hyperspectral}. This classical viewpoint implies inevitable local information loss. In contrast, the generative perspective in \eqref{eq:genmodel} provides a ``content-preserving" relationship between modalities and enables provable, patch-wise super-resolution. On the other hand, we also acknowledge that the generative perspective in \eqref{eq:genmodel} may still be an idealized abstraction of practical HSI-MSI imaging systems, since local information loss induced by spatial resolution reduction may hinder exact bijectivity of the generative functions (especially $\bm g\pH$). Nevertheless, this model provides a useful theoretical lens for studying the recoverability of the HSR sub-task.

\end{Remark}

To learn such $\bm f^\star$, we use the idea in \eqref{eq:abundanceDM} to find $\bm f$ such that
the distributions of $\bm f({\S}_r\pH[{\cal H}^{(\bm w,\theta)}])$ and $ \S_r\pM[{\cal M}^{(\bm w',\theta')}]$ are identical.
To this end,
we use the following adversarial learning criterion on randomly sampled $\{ {\cal M}_\ell \}_{\ell=1}^L$ and $\{ {\cal H}_t\}_{t=1}^T$,
where now we have ${\cal M}_\ell$ and ${\cal H}_t$ represent patches with randomly sampled $(\w_\ell,\theta_\ell)$ and $(\w_t',\theta_t')$:
\begin{mdframed}
\begin{align}\label{eq:diversified_gan_theta}
    \min_{\bm f}\max_{ \{ \bm d_r\}_{r=1}^R }&~\sum_{r=1}^R \left[ \nicefrac{1}{L}\sum_{\ell=1}^L\log\left(\bm d_r \left( \S\pM_r[{\cal M}_\ell] \right) \right) \right. \\
    & \left. + \nicefrac{1}{T}\sum_{t=1}^T\log\left(1-\bm d_r\left(\bm f\left(\S_r\pH[{\cal H}_t]\right) \right)\right) \right], \nonumber
\end{align}    
\end{mdframed}
where $\bm d_r: \mathbb{R}^{B\pM \times B\pM} \rightarrow [0,1]$ for all $r\in[R]$ are the ``discriminators'', which are classifiers trained to distinguish  $\bm f(\S_r\pH[{\cal H}_t])$ from $\S\pM_r[{\cal M}_\ell]$. 
The discreminators work {\it adversarially} against the ``translator'' (or ``generator'') $\bm f$ that strives to make the transformed MSI abundance patches $\bm f\left(\S_r\pH[{\cal H}_t]\right)$ similar to the HSI patches $\S\pM_r[{\cal M}_\ell]$.
In implementation, both $\bm f$ and $\bm d_r$'s are represented using deep neural networks. It is well-known such GAN losses lead to distribution matching \cite{goodfellow2014generative,shrestha2024towards}:

\begin{Fact}
    Assume that $\bm f:\mathbb{R}^{B\pH\times B\pH}\rightarrow \mathbb{R}^{B\pM\times B\pM}$ and $\bm d_r:\mathbb{R}^{B\pM\times B\pM}\rightarrow [0,1]$ are represented by universal function approximators and that $T,L\rightarrow \infty$. Then, at the limit, 
    \begin{align}
    [\widehat{\bm f}]_{\#} p_r\pH = p_r\pM, ~\forall r\in[R]
\end{align}
is attained, where $\widehat{\bm f}$ is any optimal solution of \eqref{eq:diversified_gan_theta}.
\end{Fact}
The proof is by invoking \cite[Theorem~1]{goodfellow2014generative}; also see \cite{shrestha2024towards}.

We should remark that Eq.~\eqref{eq:diversified_gan_theta} is not an ordinary GAN loss, as we use a unified $\bm f$ to match $R$ pairs of distributions. This turns out to be the key for establishing theoretical guarantees. To see this, let us assume:

\begin{Assumption}[Sufficiently Diverse Abundances (SDA)]\label{assumption:SDA}
    For the set of conditional distributions $\{ p(\bm z_r|\bar{\theta}) \}_{r=1}^R$ (where $R\geq 2$) and any open, connected, and disjoint sets ${\cal A},{\cal B}\subseteq {\cal Z}$, there always exists an $({\cal A},{\cal B})$-dependent index $r'\in\{1,\ldots,R\}$ such that
   \begin{align}\label{eq:sda_key}
        \int_{\cal A} p(\z_{r'}|\bar{\theta})d \z \neq \int_{\cal B} p(\z_{r'}|\bar{\theta})d \z,~\forall \bar{\theta}.
    \end{align}
\end{Assumption}
Simply speaking, as $\bm g\pH$ and $\bm g\pM$ are bijective, \eqref{eq:sda_key} effectively requires that $\S_r\pH[{\cal H}^{(\bm w,\bar{\theta})}]$ ($\S_r\pM[{\cal M}^{(\bm w',\bar{\theta})}]$) for $r=1,\ldots,R$ have different distributions---that is, 
when there is no rotation mismatch, the abundance maps of different materials are sufficiently different.
Under this  assumption, we show:

\begin{Theorem}[HSR Recoverability]\label{theorem:HSR}
Assume that Eq.~\eqref{eq:theoremmsr_result} is attained, and that Assumptions~\ref{assumption:rotate_generative_model}-\ref{assumption:SDA} hold. Suppose that $p(\bm z_r \mid \theta)$ is continuous with respect to $\theta$.
In addition, assume that $\bm f,\bm d_r$ are universal function approximators.
Let $\widehat{\bm f}$ denote any optimal solution of \eqref{eq:diversified_gan_theta} when  $L,T\rightarrow \infty$. 
    Then,
    we have $\widehat{\bm f}=\bm f^\star$ and
    \begin{align}\label{eq:HSR_recovery}
        \tY\pH_{\rm SRI}[{\cal M}^{(\w,\theta)}] = \sum_{r=1}^R \widehat{\bm f}\left(\bm S_r\pH[{\cal H}^{(\w,\theta)}]\right)\circ \c_r\pH,
    \end{align}
    for any $\w,\theta$.
\end{Theorem}
Theorem~\ref{theorem:HSR} asserts that, as long as the abundance maps of the $R$ materials are sufficiently diverse, using \eqref{eq:diversified_gan_theta} provably recovers $\bm f^\star$. Interestingly, both the MSR and HSR stages leverage the classical LMM, a model deeply rooted in hyperspectral imaging, to enable recoverability and guide the learning process. The key idea of the proof is to link the patch generative model to distribution transfer analysis in \cite{shrestha2024towards}; see details in the supplementary material.

Theorem~\ref{theorem:HSR} relies on the SDA assumption (i.e., Assumption~\ref{assumption:SDA}). The assumption may be violated in some cases, e.g., when the densities are similar over very small-volume $\mathcal{A}$ and $\mathcal{B}$. Nonetheless, we show that recoverability holds up to bounded errors even if the SDA only holds approximately. To this end, we make the following assumption:

\begin{Assumption}[Relaxed $\eta$-SDA Condition]\label{assumption:relaxed_SDA}
Define $\mathcal{V}$ as the collection of pairs of disjoint non-empty, open and connected sets $(\mathcal{A},\mathcal{B})$ such that there exists at least one $\bar{\theta}\in [0, 2\pi]$ to make
\begin{equation}\label{eq:violateSDA}
        \int_{\mathcal{A}} p (\bm z_r \mid \bar{\theta}) d \bm z = \int_{\mathcal{B}} p (\bm z_r \mid \bar{\theta}) d \bm z,~~\forall r \in [R].
\end{equation}
 Let ${\rm dia}(\mathcal{A}) = \sup_{\bm w,\bm z \in \mathcal{A}}\|\bm w - \bm z\|_2$. 
 Then, the relaxed $\eta$-SDA condition holds if 
\begin{align}
\max_{({\cal A},{\cal B})\in \mathcal{V}} \max\{{\rm dia(\mathcal{A})}, {\rm dia(\mathcal{B})}\} \leq \eta
\end{align}
where $\eta \geq 0$.
\end{Assumption}
Note that \eqref{eq:violateSDA} means that the SDA in Assumption~\ref{assumption:SDA} is violated over ${\cal A}$ and ${\cal B}$, as there is not a single density $p(\z_r|\bar{\theta})$ that can satisfy the inequality \eqref{eq:sda_key}. 
The parameter $\eta$ denotes the maximal diameter of the SDA-violating regions. Importantly, $\eta$ is an intrinsic property of the data rather than a tunable parameter of the proposed method.

Under this relaxed assumption, we show the following:
\begin{Theorem}[Robust HSR Recoverability]\label{theorem:robust_HSR}
Let all assumptions in Theorem~\ref{theorem:HSR} hold with Assumption~\ref{assumption:SDA} replaced by Assumption~\ref{assumption:relaxed_SDA}.
In addition, assume that $\bm g\pM$ and $\bm g\pH$ satisfy bi-Lipchitz conditions, particularly,
\begin{subequations}
\begin{align}
        \| \bm g\pM (\bm z_1) - \bm g\pM (\bm z_2)\|_{\rm F} 
        &\leq \alpha\pM \|\bm z_1 - \bm z_2\|_2,
        \label{eq:lipschitz_g_M}\\
        \| \bm g\pH (\bm z_1) - \bm g\pH (\bm z_2)\|_{\rm F} 
        &\geq \alpha\pH \|\bm z_1 - \bm z_2\|_2,
        \label{eq:lipschitz_g_H}
\end{align}
\end{subequations}
for all $\bm z_1,\bm z_2 \in \mathcal{Z}$. 
Then, we have
 \begin{equation}
        \|\widehat{\bm f}(\bm h) - \bm f^{\star}(\bm h)\|_{\rm F} \leq \frac{2(\alpha\pM)^2}{\alpha\pH}\eta,
    \end{equation}
for $\forall \bm h \in {\rm dom}({\S}_r\pH[{\cal H}^{(\bm w,\theta)}])$.
Consequently, 
    \begin{align}
        &\left\| \tY\pH_{\rm SRI}[{\cal M}^{(\w,\theta)}] - \widehat{\tY}\pH_{\rm SRI}[{\cal M}^{(\w,\theta)}] \right\|_{\rm F}  \\ &\qquad \qquad  \qquad  \qquad \leq \frac{2(\alpha\pM)^2}{\alpha\pH}\eta \sum_{r=1}^R \|\bm c_r\pH\|_2, \nonumber
    \end{align}
holds for any $\bm w$ and $\theta$.
\end{Theorem}

Theorem~\ref{theorem:robust_HSR} shows that both the estimation error of the super-resolution function and the reconstruction error of the super-resolved patches scale linearly with the degree of violation of the SDA assumption---as long as the functions $\g\pH$ and $\g\pM$ preserve distances without collapsing or tearing. 
The proof is via integrating the robustness analysis of diversified distribution matching \cite{shrestha2024towards} with the patch generative model in \eqref{eq:genmodel}; see the supplementary material.

\medskip

In the sequel, we refer to our overall method (consisting of the MSR and HSR stages) as \textit{Factorized Representation for Enhanced Super-resolution using latent Component-adversarial Optimization} (\texttt{FRESCO}).

\section{Implementation of FRESCO}
In this section, we discuss the details of \texttt{FRESCO} implementation, i.e., practical reformulations and realizations of  \eqref{eq:MSR} and \eqref{eq:diversified_gan_theta}.

\subsection{Practical Implementation of Problem \eqref{eq:MSR}}
To handle Problem~\eqref{eq:MSR}, we use ideas from the co-registered HMF work \cite{ding2020hyperspectral} that also used a block-term tensor based formulation.
Specifically, we relax the hard constraints as regularization terms. The low-rank term is approximated by $$ {\cal L}^{\rm LR} =  \sum_{r=1}^R ({\rm tr}(\S_r\pM(\S_r\pM)^\T+\tau\bm I)^{\frac{p}{2}} +{\rm tr}(\S_r\pH(\S_r\pH)^\T+\tau\bm I)^{\frac{p}{2}}),$$ where $0<p\leq 1$ which is a nonconvex Shattern-$p$ function that is often used for low-rank promoting; also see \cite{ding2020hyperspectral}. The sum-to-one condition $\sum_{r=1}^R \bm S_r\pM =\bm 1\bm 1^\T$ is promoted by the regularization $$ {\cal L}^{\rm sto} = \left\| \sum_{r=1}^R \bm S_r\pM -\bm 1\bm 1^\T \right\|_{\rm F}^2 +  \left\| \sum_{r=1}^R \bm S_r\pH -\bm 1\bm 1^\T \right\|_{\rm F}^2.$$
The same regularization terms are used for $\S_r\pH$.
Like in \cite{ding2020hyperspectral}, we also employ a nonconvex $\ell_q$ function-based total variation regularization on each $\S_r\pM$ \cite{ding2020hyperspectral}:
$${\cal L}^{\rm TV} = \sum_{i=1}^{I\pM J\pM}  
        \left( [\widetilde{\bm s}_r]_i^2 + \varepsilon \right)^{\frac{q}{2}}
      + \left( [\check{\bm s}_r]_i^2 +\varepsilon\right)^{\frac{q}{2}},$$   where $\widetilde{\bm s}_r=(\bm H \otimes \bm I) \bm s_r$, $\check{\bm s}_r= (\bm I \otimes \bm H) \bm s_r$, $0<q\leq 1$, $\varepsilon>0$, $\bm s_r = \mathrm{{\rm vec}}(\bm S_r\pM)$, $\bm I \in \mathbb{R}^{J\pM \times J\pM}$, and
$\bm H$ is a circulant matrix whose first row is $[1,-1,0,\ldots,0]$. 
The overall objective is as follows:
\begin{align}
    {\cal L}^{\rm fitting} +\lambda^{\rm LR} {\cal L}^{\rm LR} + \lambda^{\rm sto} {\cal L}^{\rm sto} +\lambda^{\rm TV} {\cal L}^{\rm TV}
\end{align}
where we have ${\cal L}^{\rm fitting} = \eqref{eq:obj_MSR}$.
 
By our reformulation, Problem~\eqref{eq:MSR} is recast into a differentiable loss with only nonnegativity constraints. Hence, many off-the-shelf nonconvex optimization algorithms can be used to handle the problem. The formulation can be handled by the alternating projected gradient method proposed in \cite{ding2020hyperspectral}---although the algorithm was proposed in the context of co-registered HMF, the algorithmic framework can also be applied to tackle our coupled factorization problem.

\subsection{Practical Realization of Problem~\eqref{eq:diversified_gan_theta}}
As mentioned, the realization of \eqref{eq:diversified_gan_theta} is based on sampling patches ${\cal M}_\ell$ and ${\cal H}_t$ with randomized centers and rotation angles. To make the learned $\bm f$ bijective, we use a regularization as follows:

\begin{align*}
  {\cal L}^{\rm inv}(\bm g,\bm f)&= \sum_{r=1}^R \left( \nicefrac{1}{T}\sum_{t=1}^T\left\|\bm g\left(\bm f\left(\S_r\pH[{\cal H}_t]\right) \right) - \S_r\pH[{\cal H}_t]  \right\|_{\rm F}^2 \right.\\
  &\left. + \nicefrac{1}{L}\sum_{\ell=1}^L\left\|\bm f\left(\bm g\left(\S_r\pM[{\cal M}_\ell]\right) \right) - \S_r\pM[{\cal M}_\ell]  \right\|_{\rm F}^2\right).
\end{align*}
We observe that training GAN-based translation presents a challenging min-max optimization problem and divergence frequently happens. Particularly, when the updates explore disparate scales of $\bm f(\cdot)$ in the beginning phase, the algorithm could fail to converge.
To fend against this effect, we propose a scaling regularization term, preventing $\bm f$ and $\bm g$ from changing the global intensity of the patches:
\begin{align*}
  {\cal L}^{\rm scale}&(\bm g,\bm f) =\\
  &\sum_{r=1}^R\left(\nicefrac{1}{T}\sum_{t=1}^T \left( \mu\left(\bm f\left(\S_r\pH[{\cal H}_t]\right)\right) - \mu \left(\S_r\pH[{\cal H}_t]\right) \right)^2\right.\\
  &\left. + \nicefrac{1}{L}\sum_{\ell=1}^L \left( \mu \left(\bm g\left(\S_r\pM[{\cal M}_\ell]\right)\right) - \mu \left(\S_r\pM[{\cal M}_\ell]\right) \right)^2\right),
\end{align*}
where $\mu(\bm X)$ denotes the mean of $\bm X$. 
The scaling regularization is introduced to stabilize the GAN-based optimization. Empirically, during the early stages of training, the discriminator may not yet be sufficiently trained to penalize incorrect translations, causing the generator to occasionally drift toward solutions with undesirable global brightness changes (e.g., mapping dark regions to bright ones). Once such a drift occurs, the adversarial optimization can become trapped in a degenerate local equilibrium. The regularization discourages these intensity-shifting transformations and leads to more reliable optimization behavior across runs. As will be shown in the ablation study (see Sec.~\ref{sec:ablation}), this regularization term assists stabilizing the numerical behavior.

Denote the loss in \eqref{eq:diversified_gan_theta} as ${\cal L}^{\rm DM}(\bm f, \{ \bm d_r\}_{r=1}^R)$. Then, our overall loss for HSR is as follows
\begin{align}\label{eq:HSRstage}
\min_{\bm f,\bm g} \max_{\{\bm d_r\}_{r=1}^R} {\cal L}^{\rm DM} &(\bm f, \{ \bm d_r\}_{r=1}^R)\\ 
&+ \lambda^{\rm inv}  {\cal L}^{\rm inv}(\bm g,\bm f) + \lambda^{\rm scale}{\cal L}^{\rm scale}(\bm g,\bm f), \nonumber
\end{align}
where $\lambda^{\rm inv},\lambda^{\rm scale}\geq 0$ are regularization parameters.
Here, both $\bm f$ and $\bm g$ are represented by neural networks.
The function $\bm f$ is represented by an upscaling layer followed by a U-Net \cite{ronneberger2015u}, and the $\bm g$ function is a downsampling neural network; see the detailed structure in the supplementary material.

\subsection{Heuristic Remedy when $\bm P\pM$ is Unavailable}\label{sec:PMestimate}
Our development used the assumption that $\bm{P}\pM$ is known.
When this information is not available, the formulation in \eqref{eq:MSR} might not be able to align $\S_r\pM$ and $\S_r\pH$ that associate with the same material.

As a remedy, we propose a heuristic to estimate $\bm{P}\pM$ as follows
\begin{align}\label{eq:find_PM_unregistered_HMF}
  &\minimize_{\bm P\pM} \sum_{k=1}^{K\pM} \left( \left( \mu \big(\underline{\widetilde{\bm Y}}\pM(:,:,k)\big) -  \mu \big(\underline{\bm Y}\pM(:,:,k)\big) \right)^2  \right. \nonumber\\
 &\qquad + \left. \left({\rm var} \big( \underline{\widetilde{\bm Y}}\pM(:,:,k)\ \big) - {\rm var}  \big( \underline{\bm Y}\pM(:,:,k) \big) \right)^2 \right), \nonumber\\
  &{\rm subject~to} \quad \bm P\pM \geq \bm 0,\ \  \bm P\pM(\bm \varOmega) = \bm 0,
\end{align}
where $\widetilde{\tY}\pM(i,j,:)=\bm P\pM\tY\pH(i,j,:)$ for all $(i,j)\in[I\pH]\times [J\pH]$,
and ${\rm var}(\bm X)$ denotes the variance of all elements of the input matrix $\bm X$. 
In the above, the nonnegativity of $\bm P\pM$ holds as spectral degradation is often understood as a weighted sum process \cite{kanatsoulis2018hyperspectral,simoes2014convex}.
The index set $\bm \varOmega$ represents the indices $(i,j)$ where $\bm P\pM(i,j) = 0$, indicating that the $j$th band of $\tY\pH$ does not contribute to the $i$th band of $\tY\pM$ (as the MSI bands are considered local integrations of neighboring HSI bands \cite{kanatsoulis2018hyperspectral,simoes2014convex,ding2020hyperspectral}).

The idea of \eqref{eq:find_PM_unregistered_HMF} is to treat $\tY\pM(i,j,k)$ for all $k$ as i.i.d. Gaussian variables.
Then, the ground-truth $\bm P\pM$ should make the distributions of $\widetilde{\tY}(i,j,k)$ and $\underline{\bm Y}\pM(i,j,k)$ for any $k$ identical---which is enforced by matching their means and variances.
This idea avoids using any spatial alignment information to estimate $\bm P\pM$, which is needed in existing methods, e.g., that in \cite{simoes2014convex}.

\medskip

We note that in our framework $\bm P\pM$ is used to resolve the permutation ambiguity inherent in hyperspectral unmixing and to enforce a material-wise correspondence between the MSI and HSI abundance maps. However, recoverability in the MSR and HSR stages does not strictly depend on access to $\bm P\pM$, but only on whether such a correspondence can be established. In some cases, the abundance maps can be aligned without explicit knowledge of $\bm P\pM$, e.g., via visual inspection. In those cases, $\bm P\pM$ is not required by the \texttt{FRESCO} pipeline.

\subsection{Inference Strategy}
The model $\widehat{\bm f}$ is trained to super-resolve small patches. To convert the entire low-resolution abundance maps $\bm S\pH_r$ for all $r \in [R]$ into their high-resolution versions, we adopt a sliding-window strategy: overlapping patches are sequentially extracted from each $\bm S\pH_r$ with a stride of $1$, where stride refers to the ``step size'' by which the window moves across the spatial domain of the image. Specifically, the sliding window moves across the image one pixel at a time---first moving left to right along each row, and then moving down to the next row---similar to how convolution operations scan an image.
Then, each patch is processed independently by $\widehat{\bm f}$. The resulting high-resolution patches are then placed back into their corresponding locations within the high-resolution abundance maps. In regions where outputs overlap, pixel values are averaged to mitigate boundary artifacts and ensure smooth transitions. After applying this patch-based super-resolution process to all $\bm S\pH_r$ for $r \in [R]$, the high-resolution $\tY_{\rm SRI}\pH$ can be reconstructed via \eqref{eq:HSR_recovery}.

\section{Numerical Results}
In this section, we test the proposed method on semi-real and real-world datasets.

\subsection{Semi-Real Data Experiments}

\subsubsection{Data}
 We construct $\tY_{\rm SRI}\pH$ and $\tY_{\rm SRI}\pM$ using real hyperspectral datasets:  
i) \textbf{Pavia University} ($610\times340\times103$): 4 noisy bands removed; $R=4$ following~\cite{ding2020hyperspectral}.
ii) \textbf{Terrain} ($500\times307\times75$): noisy bands removed; $R=5$ following~\cite{ding2023fast}.
iii) \textbf{Indian Pines} ($837\times1600\times106$) \cite{baumgardner2015220}: water-absorption bands removed; we set $R=4$.

To simulate unregistered MSI and HSI, we extract two non-overlapping spatial regions from a dataset, apply distinct shifts and rotations (specified per experiment), and treat them as $\tY_{\rm SRI}\pM$ and $\tY_{\rm SRI}\pH$. To avoid undefined areas during rotation, we rotate the circumcircle enclosing the rectangular region and re-extract the original region post-rotation.
We set $s=\nicefrac{I\pM}{I\pH} = \nicefrac{J\pM}{J\pH} = 4$ for spatial degradation, and fix $K\pM=15$ for all experiments. The degradation matrix $\bm P\pM$ is constructed as in~\cite{ding2020hyperspectral,kanatsoulis2018hyperspectral}.

\subsubsection{Baselines}
As no existing method jointly recovers $\underline{\bm Y}_{\rm SRI}^{(\mathrm{M})}$ and $\underline{\bm Y}_{\rm SRI}^{(\mathrm{H})}$, we benchmark MSR and HSR separately.

\noindent\textbf{MSR Baselines:}  
We evaluate \texttt{u2MDN}~\cite{qu2021unsupervised}, \texttt{IARF}~\cite{zhou2019integrated}, \texttt{UHIF-RIM}~\cite{ying2021unaligned}, and \texttt{NBFusion} (co-registration via \texttt{NED}~\cite{chen2017normalized} followed by fusion using \texttt{BSTEREO}~\cite{kanatsoulis2018hyperspectral}). \texttt{u2MDN} uses the default neural network architecture in~\cite{qu2021unsupervised}.

\noindent\textbf{HSR Baselines:}  
We include \texttt{Lanczos} interpolation, and supervised methods \texttt{MC-Net}~\cite{li2020mixed} and \texttt{TSBSR}~\cite{liang2023blind}. The models are trained on CAVE~\cite{yasuma2010generalized} and AID~\cite{xia2017aid}, respectively, using blurred (Gaussian $5 \times 5$, $\sigma=1.7$) and $4\times$ downsampled image pairs.

\subsubsection{Performance Metrics}
We use standard metrics: PSNR, SSIM, and ERGAS~\cite{li2020mixed,liang2023blind,guo2023stereo,ying2021unaligned}. 
To assess perceptual similarity, we employ LPIPS~\cite{zhang2018unreasonable} and FID~\cite{heusel2017gans}, both computed on RGB bands. 
FID is estimated from 1,000 randomly sampled $80\times80$ RGB patches per image. 
Lower LPIPS and FID indicate better perceptual quality.

\subsubsection{Setup of The Proposed Method}
For MSR, we grid-search $\lambda^{\rm LR}$, $\lambda^{\rm TV}$, and $\lambda^{\rm sto}$ in $[10^{-4}, 10^{-2}]$ (divided into 5 uniform grids) and select the setting with the highest PSNR on the reconstructed MSI and HSI. That is,
$\bm{\lambda}^\star
= \arg\max_{\bm{\lambda} \in [10^{-4},\,10^{-2}]^3}
(
\mathrm{PSNR}(\widehat{\tY}\pM(\bm{\lambda}), \tY\pM)
+\mathrm{PSNR}(\widehat{\tY}\pH(\bm{\lambda}), \tY\pH)
),$
where $\bm{\lambda} \triangleq (\lambda^{\rm LR}, \lambda^{\rm TV}, \lambda^{\rm sto})$ and $\widehat{\tY}\pM(\bm{\lambda})$ and $\widehat{\tY}\pH(\bm{\lambda})$ denote the reconstructed MSI and HSI, respectively, obtained by our algorithm using $\bm{\lambda}$. Note that $\mathrm{PSNR}(\widehat{\tY}\pM(\bm{\lambda}), \tY\pM)$ computes the PSNR using the estimated and \textit{observed} HSI and MSI data (no ground-truth SRIs are involved).
For the other parameters, we use $p=0.5$, $\tau=1$, $q=0.5$, and $\epsilon=10^{-3}$ following~\cite{ding2020hyperspectral}. We stop the optimization when relative objective change $<10^{-6}$ or after $1000$ iterations.
For HSR (Eq.~\eqref{eq:diversified_gan_theta}), $\bm f$ is a U-Net and $\bm d_r$ is a 6-layer neural network. 
Detailed architectures for $\bm f$, $\bm g$, and $\bm d_r$ are in the supplementary material. The HSI patch size is set to $B\pH = 12$ in all experiments. We fix $\lambda^{\rm inv} = 10$, $\lambda^{\rm scale} = 15$ and train with batch size 8 for $T_{\max} = 40{,}000$ iterations. 
Learning rate $\eta_0 = 10^{-4}$ decays linearly in the second half:
$
\eta_t = \eta_0 \left(1 - \frac{t - T_{\max}/2}{2T_{\max}} \right), \quad t \geq T_{\max}/2$. Results are averaged over 5 trials.

\begin{figure*}[!t]
    \centering
    \setlength{\tabcolsep}{0pt}
    \begin{subfigure}{0.15\linewidth}
        \centering
        \includegraphics[width=1\linewidth]{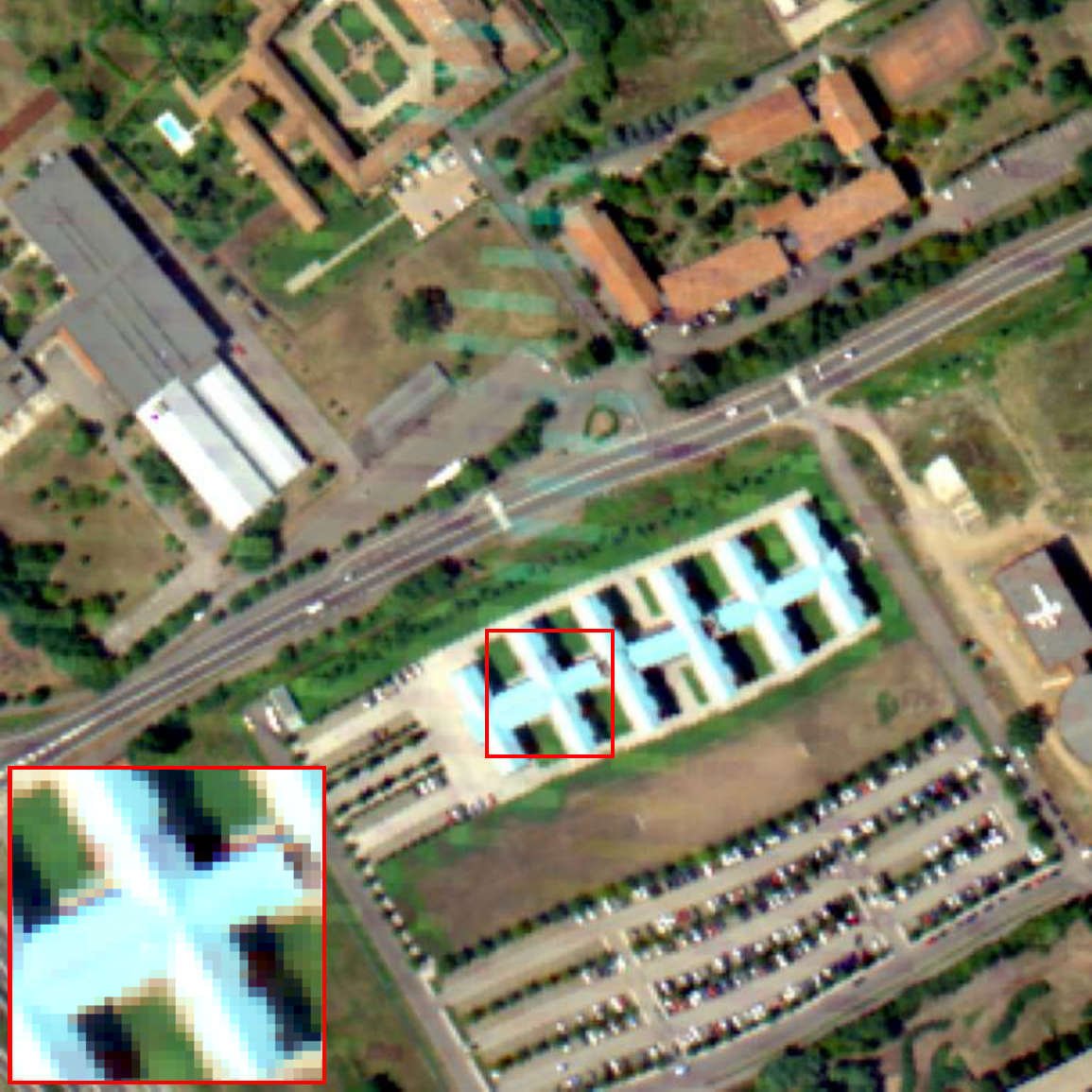}
        \captionsetup{labelformat=empty}
        \caption{\texttt{UHIF-RIM}\cite{ying2021unaligned}}
    \end{subfigure}
    \begin{subfigure}{0.15\linewidth}
        \centering
        \includegraphics[width=\linewidth]{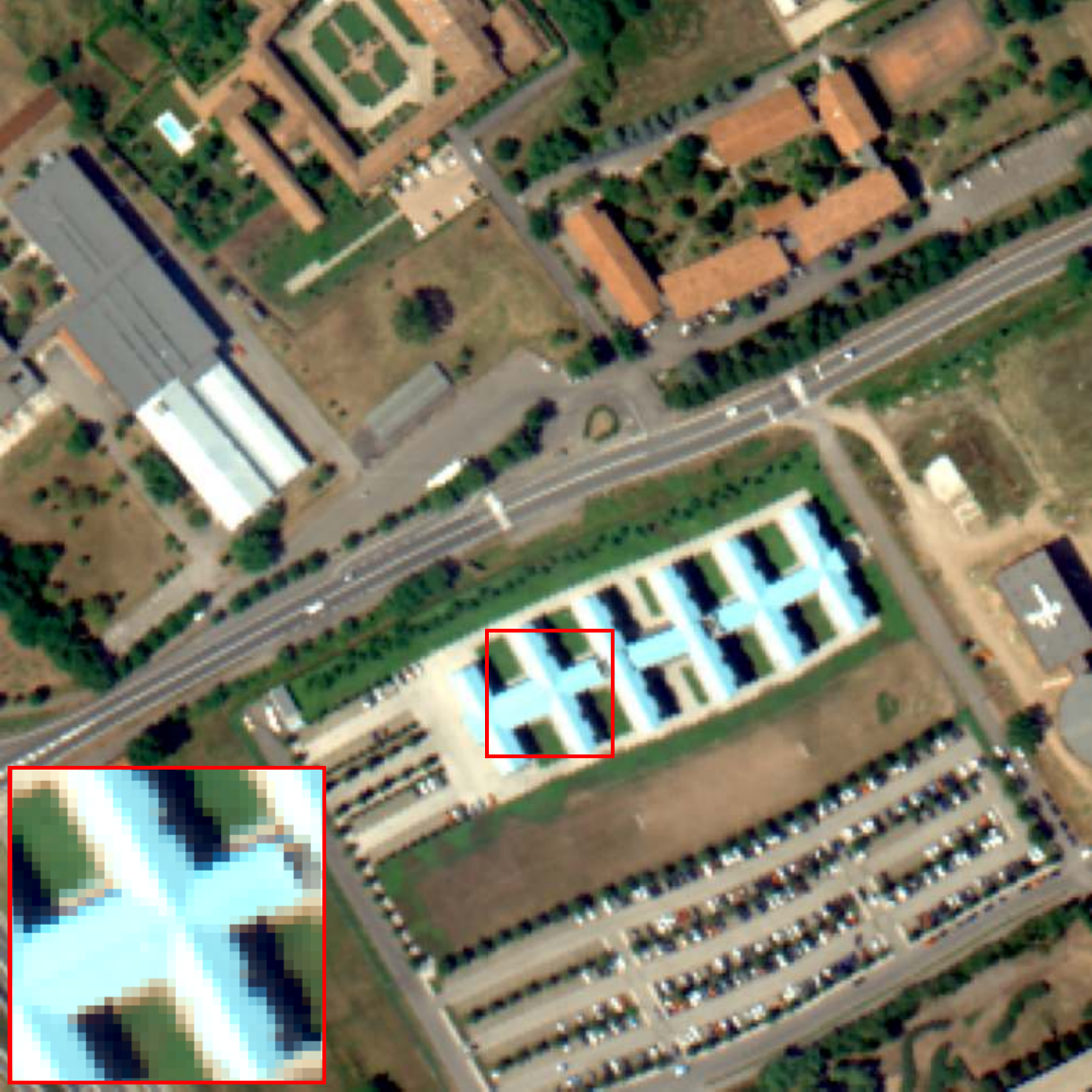}
        \captionsetup{labelformat=empty}
        \caption{\texttt{u2MDN}\cite{qu2021unsupervised}}
    \end{subfigure}
    \begin{subfigure}{0.15\linewidth}
        \centering
        \includegraphics[width=\linewidth]{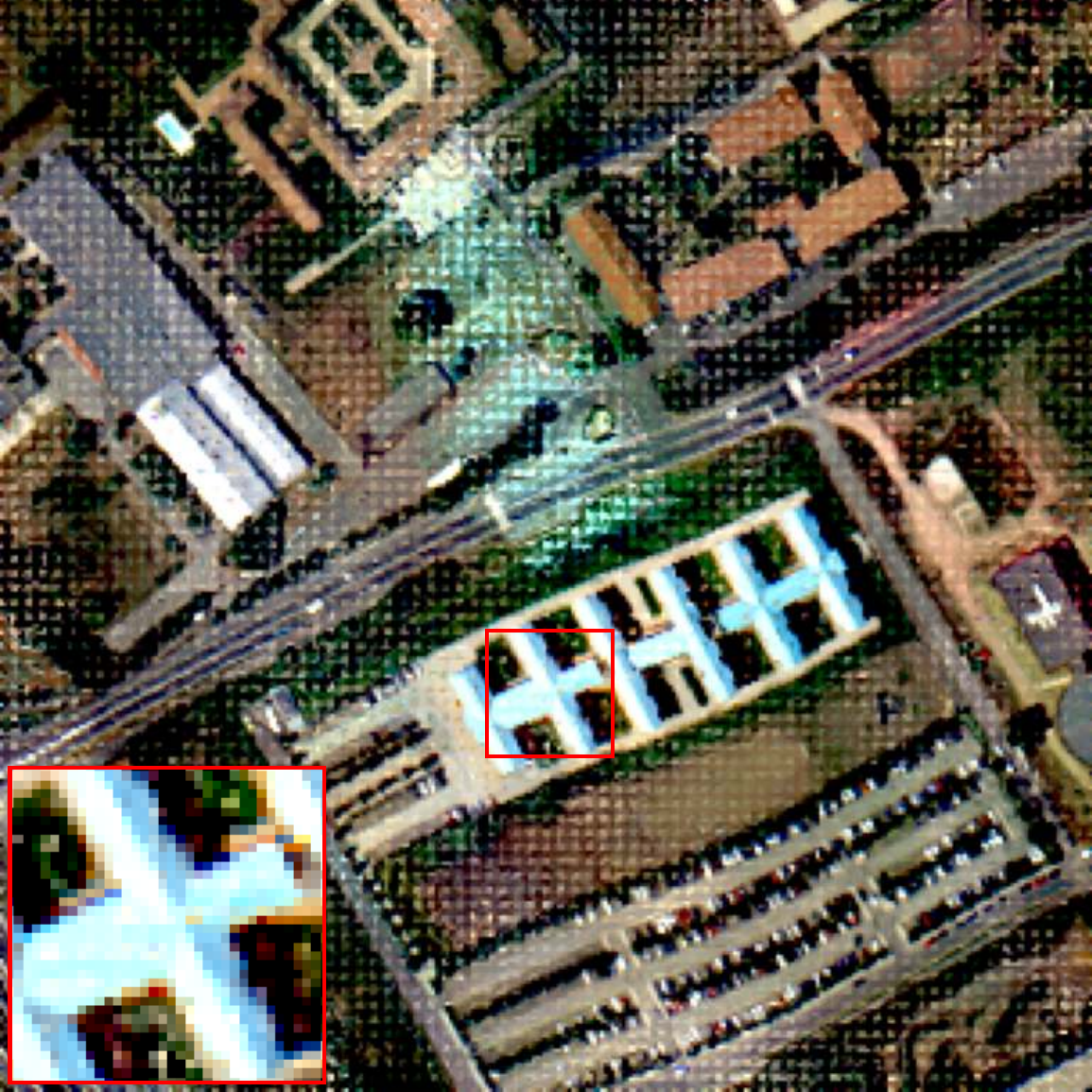}
        \captionsetup{labelformat=empty}
        \caption{\texttt{IARF}\cite{zhou2019integrated}}
    \end{subfigure}
    \begin{subfigure}{0.15\linewidth}
        \centering
        \includegraphics[width=\linewidth]{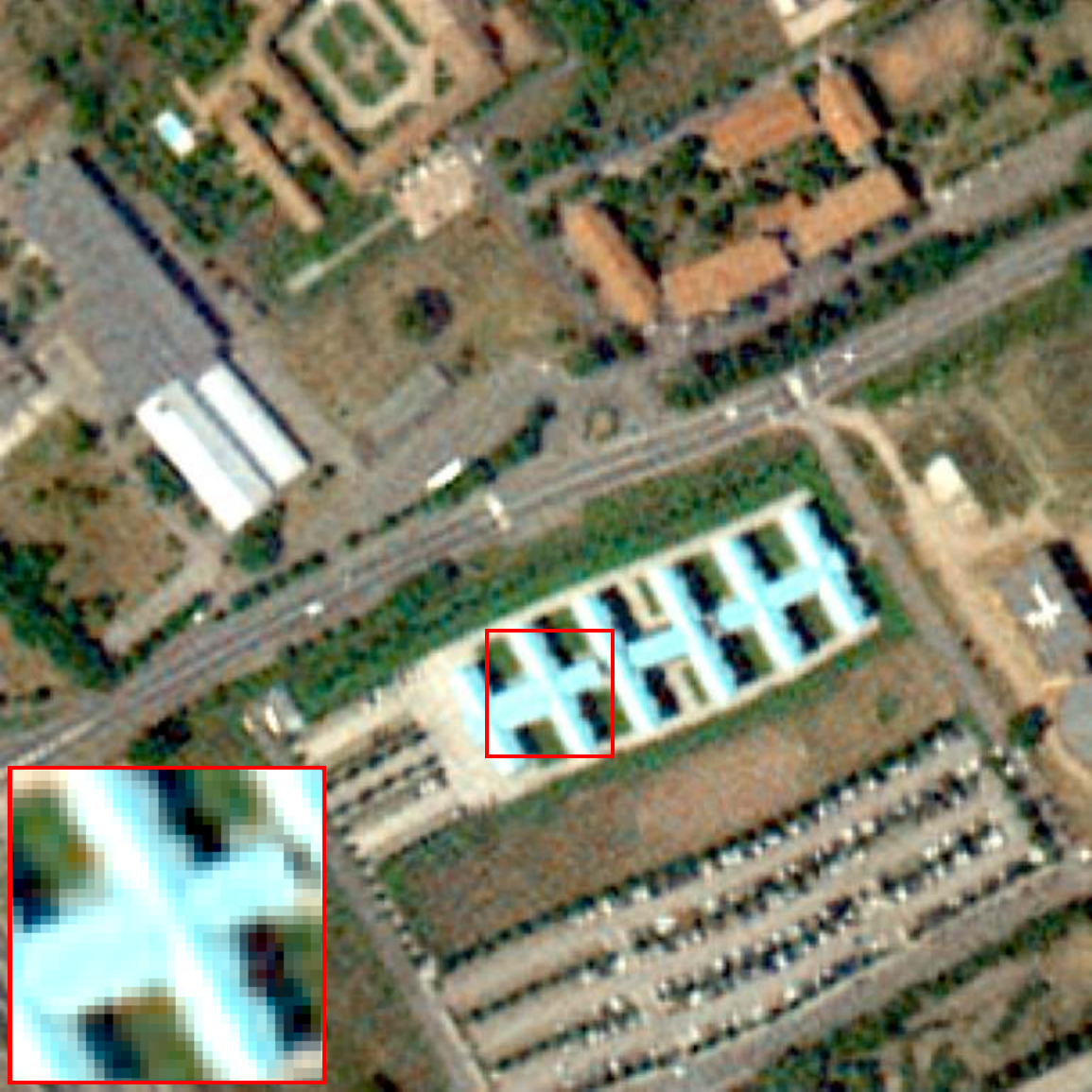}
        \captionsetup{labelformat=empty}
        \caption{\texttt{NBFusion} {\scriptsize \cite{chen2017normalized,kanatsoulis2018hyperspectral}}}
    \end{subfigure}
    \begin{subfigure}{0.15\linewidth}
        \centering
        \includegraphics[width=\linewidth]{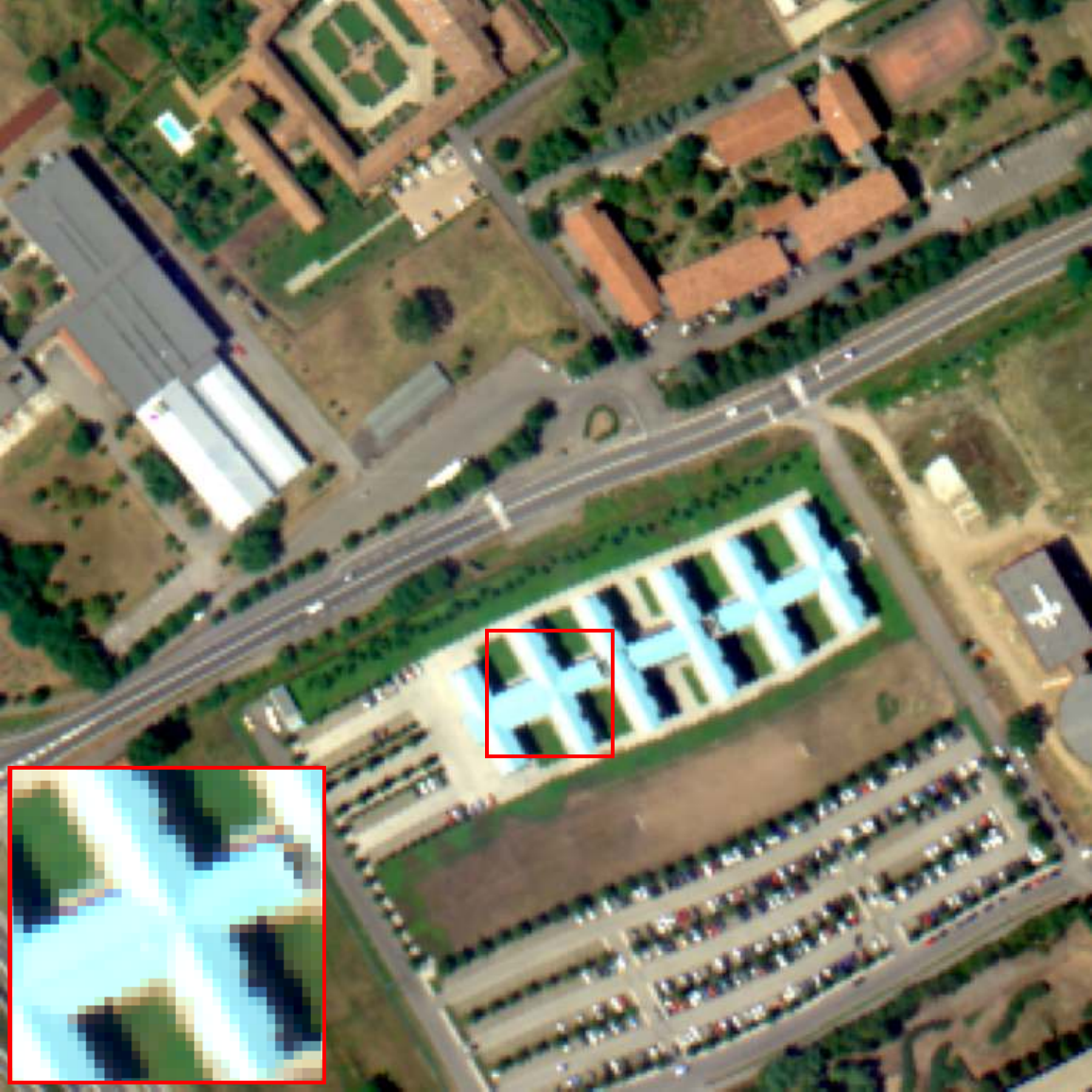}
        \captionsetup{labelformat=empty}
        \caption{\texttt{FRESCO} (\texttt{MSR})}
    \end{subfigure}
    \begin{subfigure}{0.15\linewidth}
        \centering
        \includegraphics[width=\linewidth]{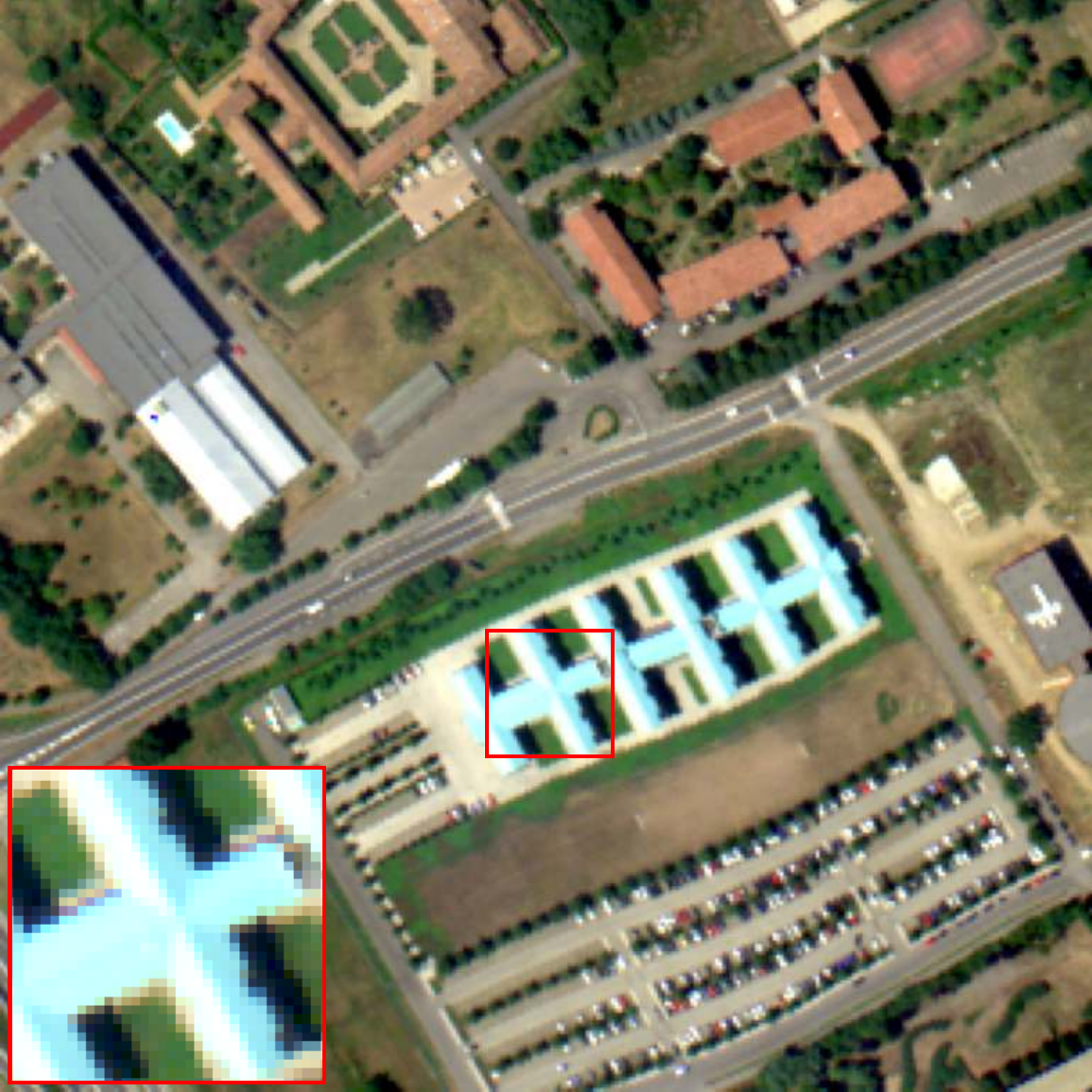}
        \captionsetup{labelformat=empty}
        \caption{Reference: {\scriptsize $\tY\pM_{\rm SRI}$}}
    \end{subfigure} 
    \caption{RGB rendering of recovered $\widehat{\tY}\pM_{\rm SRI}$ (Pavia University); nearest-neighbor downsampling used for degradation.}
    \label{fig:pavia_visual_M}
\end{figure*}

\begin{figure*}[!t]
    \centering
    \setlength{\tabcolsep}{0pt}
    \begin{subfigure}{0.15\linewidth}
        \centering
        \includegraphics[width=1\linewidth]{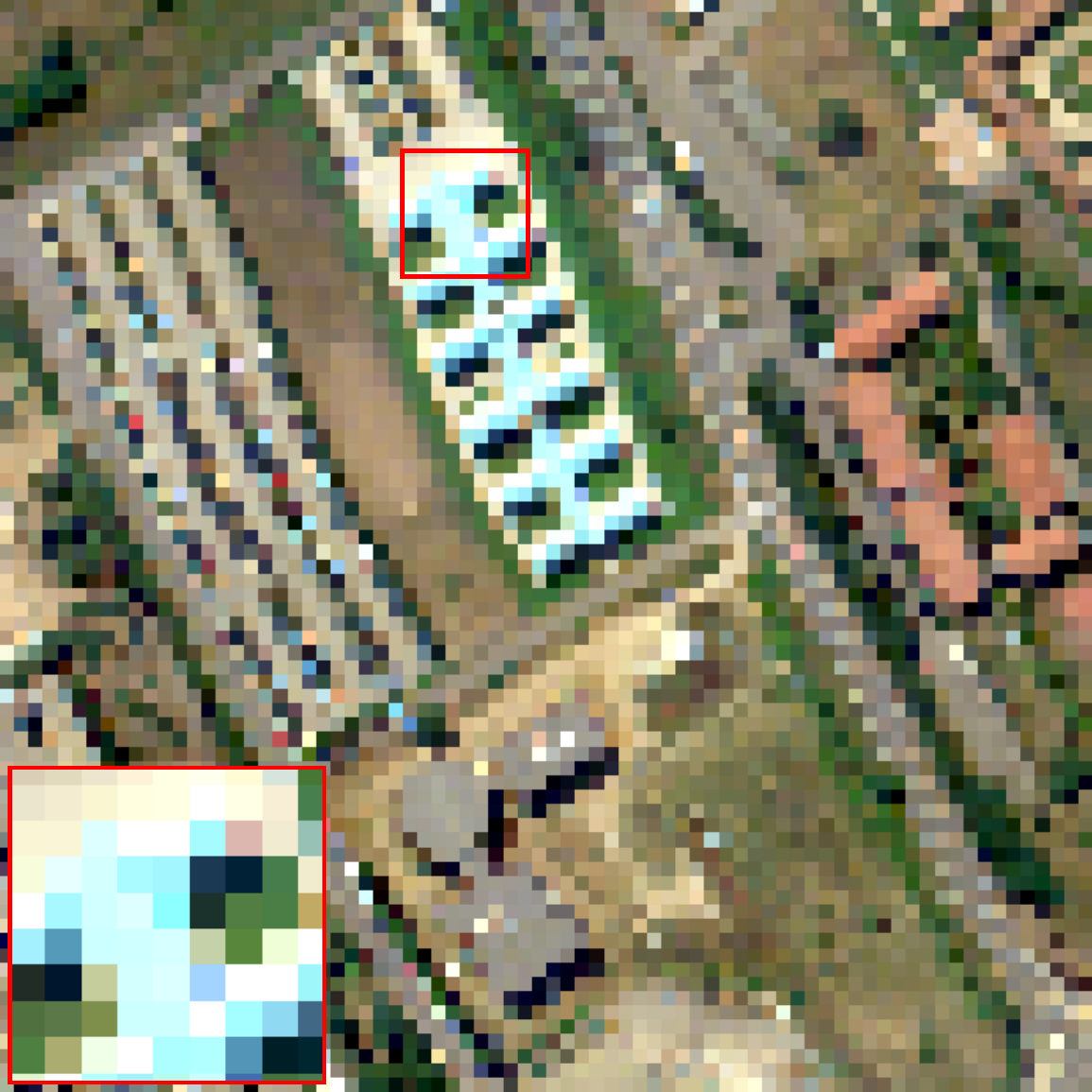}
        \captionsetup{labelformat=empty}
        \caption{HSI}
    \end{subfigure}
    \begin{subfigure}{0.15\linewidth}
        \centering
        \includegraphics[width=\linewidth]{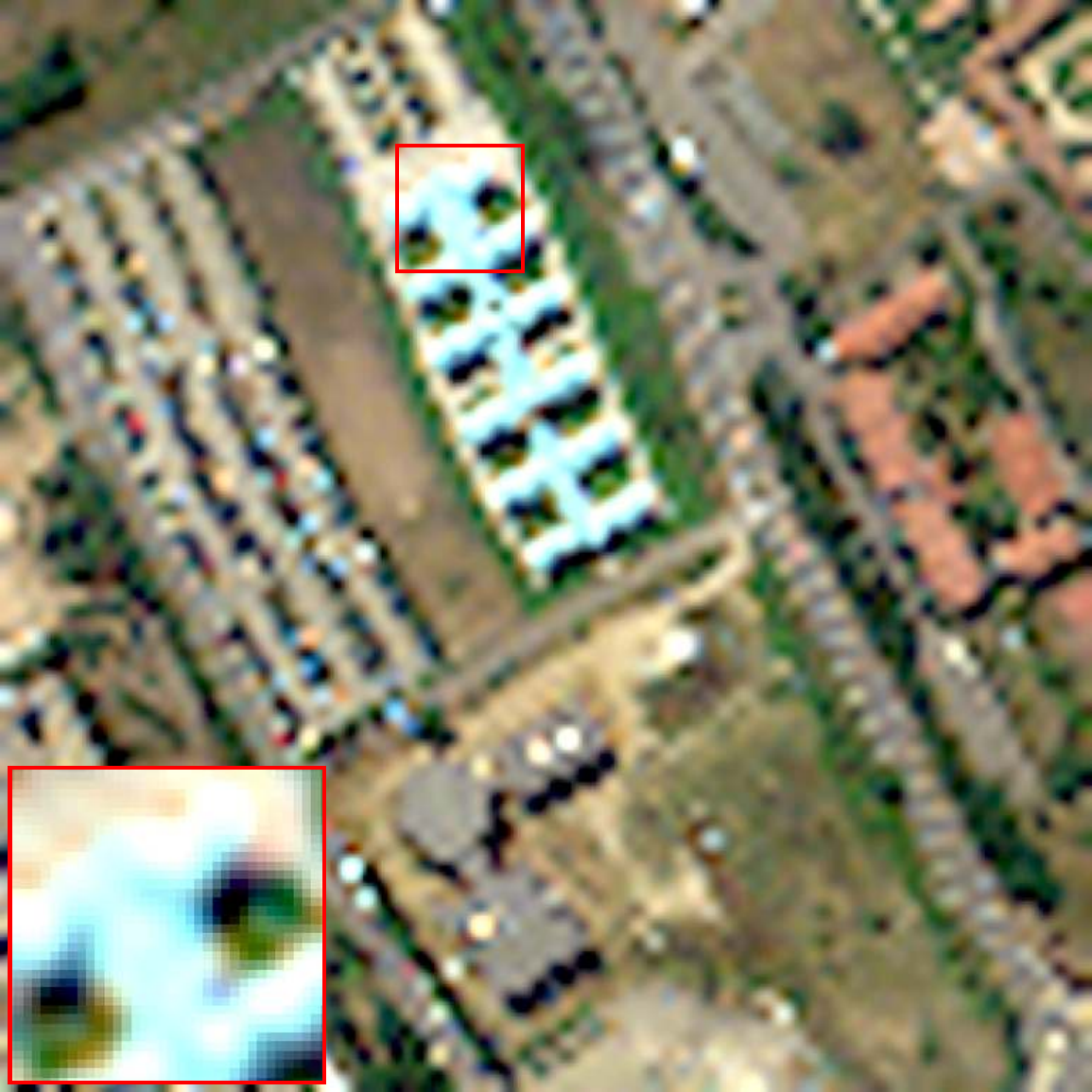}
        \captionsetup{labelformat=empty}
        \caption{\texttt{Lanczos} \cite{duchon1979lanczos}}
    \end{subfigure}
    \begin{subfigure}{0.15\linewidth}
        \centering
        \includegraphics[width=\linewidth]{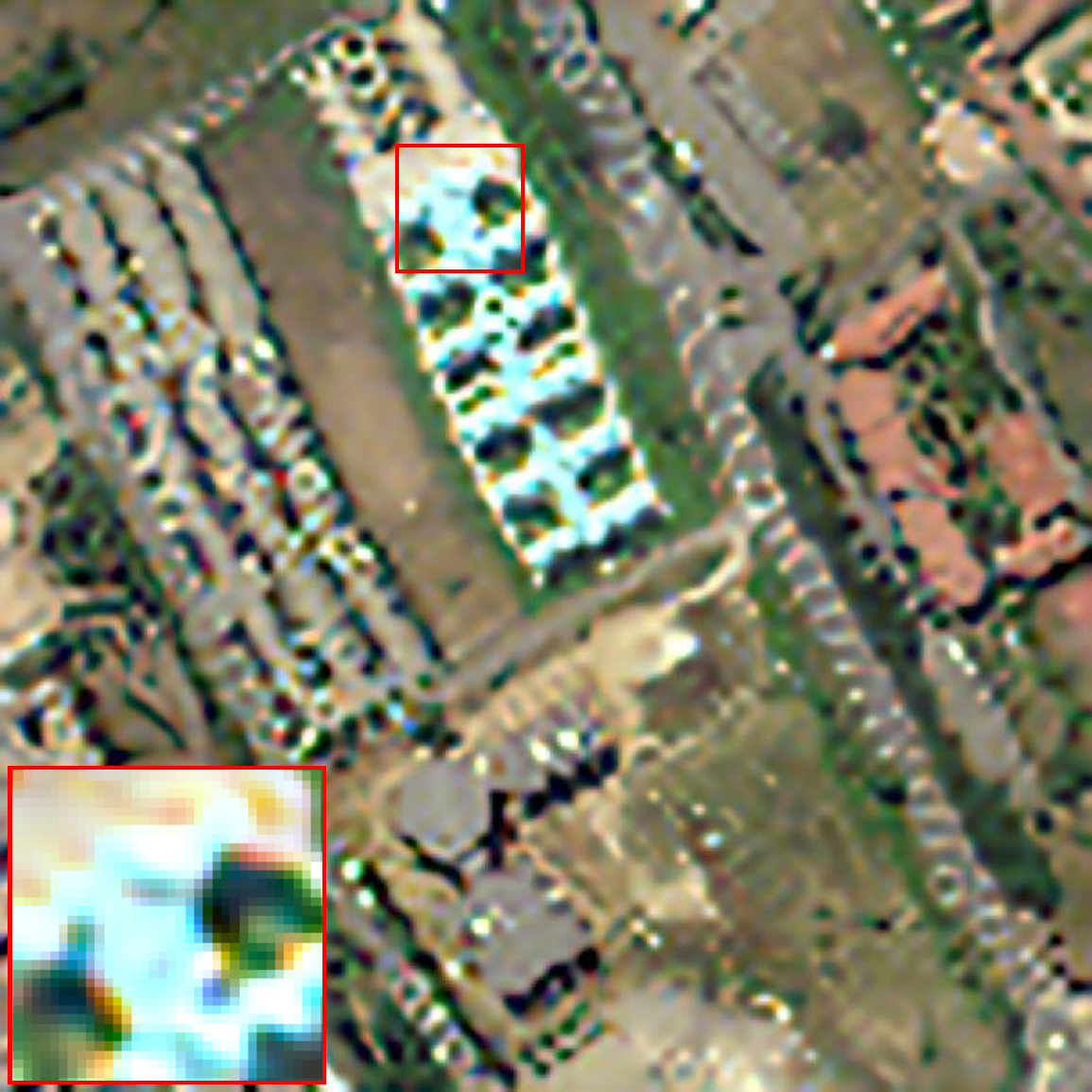}
        \captionsetup{labelformat=empty}
        \caption{\texttt{MC-Net} \cite{li2020mixed}}
    \end{subfigure}
    \begin{subfigure}{0.15\linewidth}
        \centering
        \includegraphics[width=\linewidth]{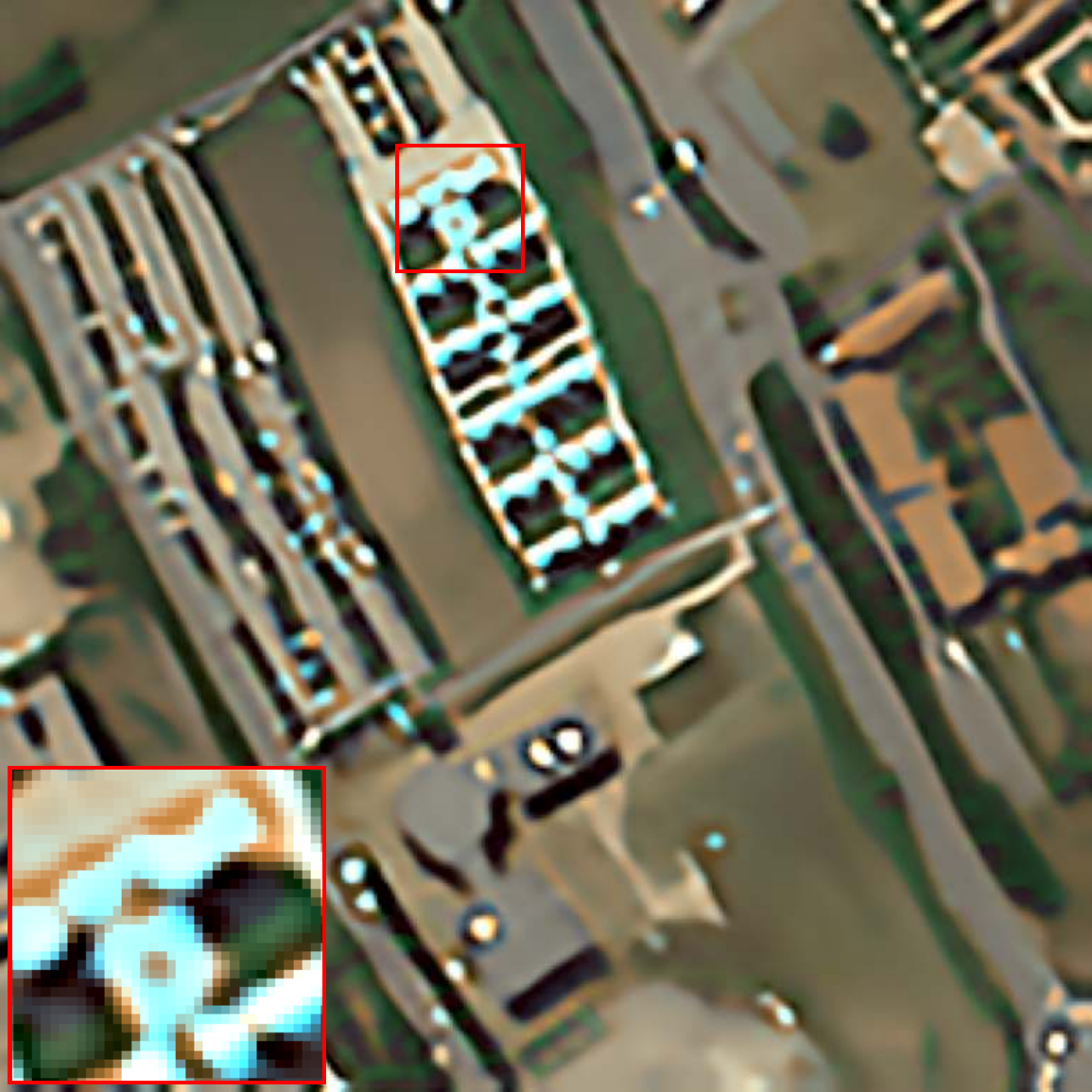}
        \captionsetup{labelformat=empty}
        \caption{\texttt{TSBSR} \cite{liang2023blind}}
    \end{subfigure}
    \begin{subfigure}{0.15\linewidth}
        \centering
        \includegraphics[width=\linewidth]{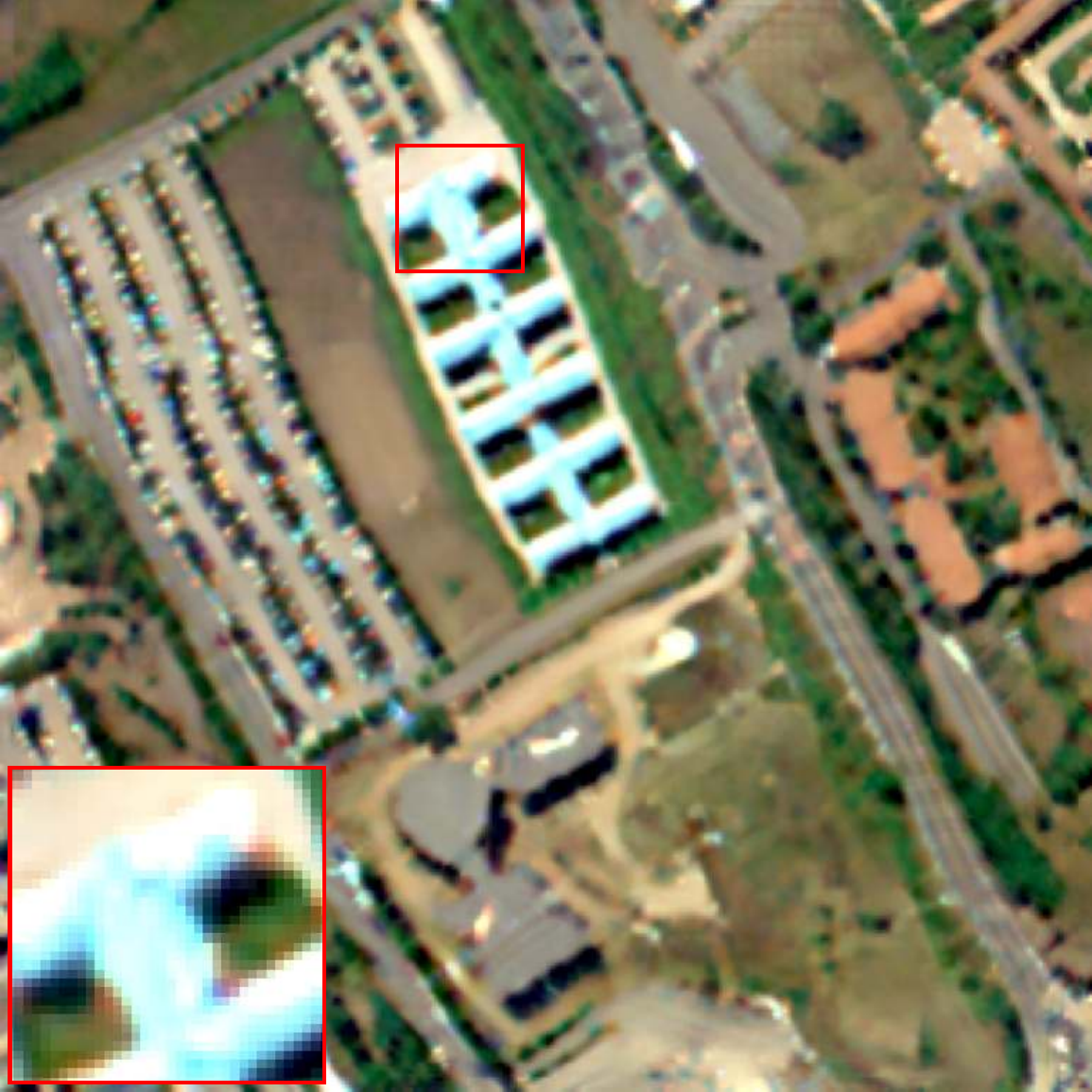}
        \captionsetup{labelformat=empty}
        \caption{\texttt{FRESCO} (\texttt{HSR})}
    \end{subfigure}
    \begin{subfigure}{0.15\linewidth}
        \centering
        \includegraphics[width=\linewidth]{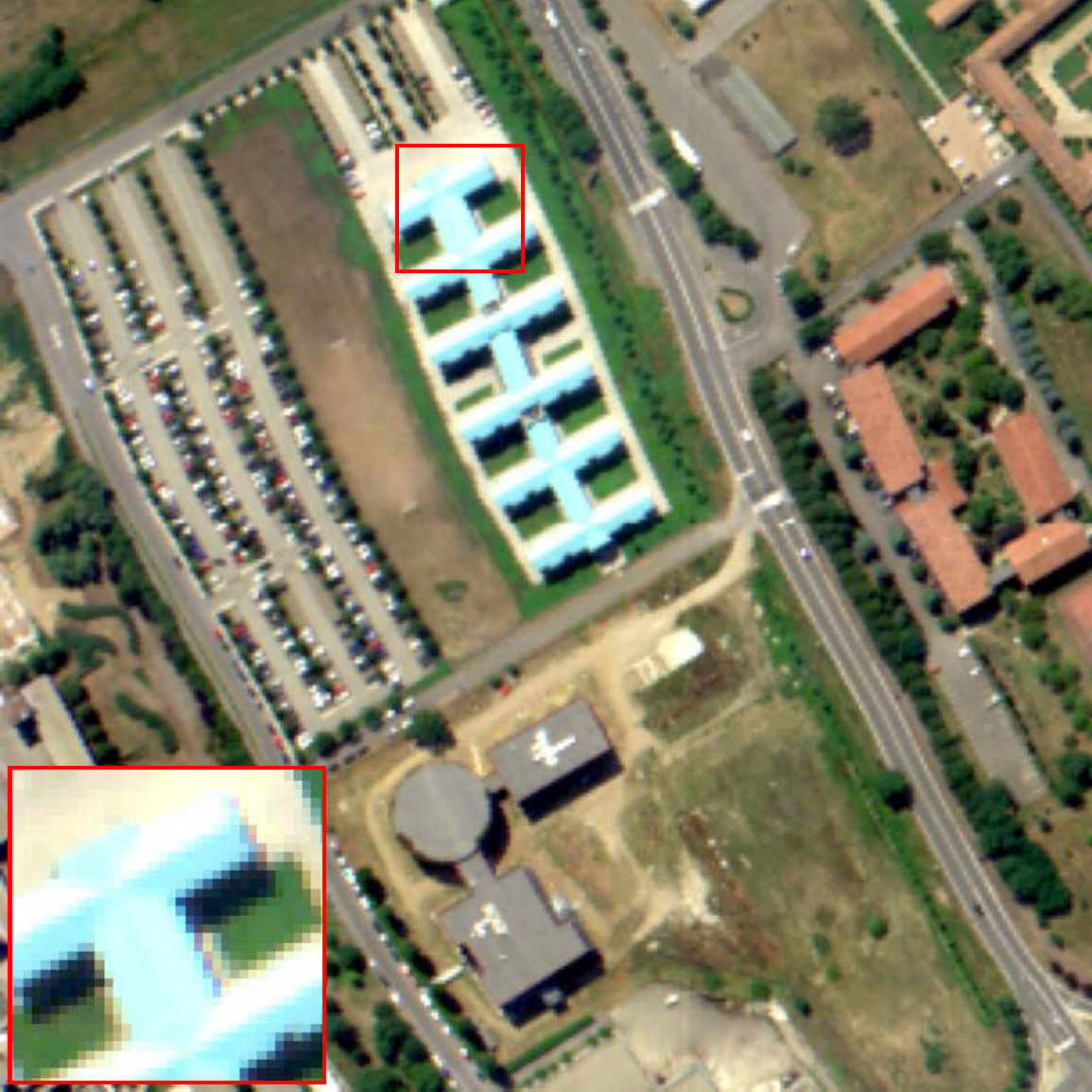}
        \captionsetup{labelformat=empty}
        \caption{Reference: {\scriptsize $\tY\pH_{\rm SRI}$}}
    \end{subfigure} 
    \caption{RGB rendering of recovered $\widehat{\tY}\pH_{\rm SRI}$ (Pavia University); nearest-neighbor downsampling used for degradation.}
    \label{fig:pavia_visual_H}
\end{figure*}

To tackle \eqref{eq:HSRstage}, \texttt{FRESCO} randomly samples patches in each iteration to perform stochastic gradient. 
Specifically, in each iteration, mini-batches of coordinates and rotation angles $(\bm w_\ell,\theta_\ell)$ and $(\bm w'_t,\theta'_t)$ are randomly sampled, where $\theta_\ell,\theta'_t\sim\mathrm{Uniform}(0,360^\circ)$. The corresponding MSI and HSI abundance patches are then extracted on the fly. Since arbitrary rotations may lead to off-grid coordinates on the discrete pixel lattice, the corresponding values are obtained via bilinear interpolation from neighboring pixels whenever the sampling locations are not aligned with pixel centers.

\subsubsection{Results} 
Fig.~\ref{fig:pavia_visual_M} and Fig.~\ref{fig:pavia_visual_H} show results on the Pavia University dataset, visualized using RGB bands of the super-resolved MSI and HSI (not explicitly available in MSI). We extract unregistered SRIs as $\tY_{\mathrm{SRI}}\pM = \tY_{\mathrm{Pavia}}(1\!:\!300,20\!:\!32 0,:)$ and $\tY_{\mathrm{SRI}}\pH =  \tY_{\mathrm{Pavia}}(101\!:\!400,1\!:\!300,:)$ from $\tY_{\mathrm{Pavia}}\in \mathbb{R}^{610 \times 340 \times 103}$, with $\tY_{\mathrm{SRI}}\pM$ rotated $90^\circ$ counterclockwise to introduce misalignment.
We use nearest-neighbor downsampling as the ground-truth spatial degradation operator, which is unknown to all algorithms.  
In Fig.~\ref{fig:pavia_visual_M}, all methods (except \texttt{IARF}) preserve spatial detail reasonably well, with the proposed \texttt{FRESCO} method producing the sharpest MSI reconstruction.  
In Fig.~\ref{fig:pavia_visual_H}, our method \texttt{FRESCO} clearly outperforms others in HSR quality.  
Note that \texttt{MC-Net} and \texttt{TSBSR}, trained using Gaussian downsampling, degrade in performance due to mismatch with the true degradation operator.

Table~\ref{tab:pavia_exp2} presents quantitative results on the Pavia data under three different HSI downsampling operators.  
For MSR, the proposed method consistently outperforms all baselines across all metrics.  
For HSR, it matches the performance of supervised methods when the downsampling operator matches their training setup (i.e., Gaussian kernel). 
When the degradation differs from that used in their training data, our method significantly outperforms the supervised baselines.  
Notably, it achieves the lowest FID and LPIPS scores in all settings, reflecting the best perceptual quality.

\begin{table*}[!t]
\centering
\renewcommand{\arraystretch}{1.3} 
\small 
\caption{Performance of the methods on the Pavia University dataset.}
\begin{adjustbox}{max width=0.9\linewidth}
\begin{tabular}{|c|c|ccccc|cccc|}
\hline
\multirow{2}{*}{\begin{tabular}[c]{@{}c@{}}Underlying Spatial\\ Downsampling Operator \end{tabular}} & \multirow{2}{*}{Metrics} & \multicolumn{5}{c|}{MSR} & \multicolumn{4}{c|}{HSR} \\ \cline{3-11} 
 &  & \texttt{u2MDN} & \texttt{UHIF-RIM} & \texttt{NBFusion} & \texttt{IARF} &\texttt{FRESCO} (\texttt{MSR}) & \texttt{Lanczos} & \texttt{MC-Net} & \texttt{TSBSR} & \multicolumn{1}{c|}{\texttt{FRESCO} (\texttt{HSR})} \\ \hline
\multirow{5}{*}{\begin{tabular}[c]{@{}c@{}}Gaussian Kernel\\ Downsampling\end{tabular}} 
 & PSNR ($\uparrow$) & 42.21 & 34.88 & 32.29 & 20.06 & \textbf{43.59} & 27.75 & \textbf{27.90} & 23.81  & 27.37 \\ 
 & SSIM  ($\uparrow$)& \textbf{0.990}  & 0.941 & 0.877 & 0.520 & 0.989 & 0.745 & \textbf{0.772} & 0.666  & 0.750 \\ 
 & ERGAS ($\downarrow$)& 1.406 & 3.155 & 3.816 & 15.139 & \textbf{1.217} & 6.188 & \textbf{6.147} & 10.242  & 6.42 \\
 & FID ($\downarrow$)& 10.21 & 20.94 & 24.61 & 31.52 & \textbf{9.53} & 73.04 & 62.06 & 69.71  & \textbf{46.17} \\
 & LPIPS  ($\downarrow$)& \textbf{0.003} & 0.046 & 0.124 & 0.261 & 0.003 & 0.443 & 0.289 & 0.304 & \textbf{0.201} \\ \hline 
\multirow{5}{*}{\begin{tabular}[c]{@{}c@{}}Nearest Neighbor \\ Downsampling\end{tabular}} 
 & PSNR ($\uparrow$)& 39.13 & 36.34 & 32.16 & 20.00 & \textbf{43.60} & 25.56 & 22.69 & 21.51 & \textbf{25.69} \\ 
 & SSIM ($\uparrow$)& 0.986 & 0.956 & 0.874 & 0.508 & \textbf{0.990} & \textbf{0.672} & 0.576 & 0.587  & 0.671 \\  
 & ERGAS ($\downarrow$)& 2.066 & 2.619 & 3.884 & 15.233 & \textbf{1.216} & 7.954 & 11.190 & 13.253 & \textbf{7.804} \\ 
 & FID ($\downarrow$)& 12.57 & 13.68 & 25.84 & 32.49 & \textbf{10.12} & 82.58 & 89.16 & 79.93 & \textbf{51.67} \\
  & LPIPS ($\downarrow$)& 0.009 & 0.032 & 0.131 & 0.290 & \textbf{0.003} & 0.472 & 0.388 & 0.342 & \textbf{0.258} \\ \hline
\multirow{5}{*}{\begin{tabular}[c]{@{}c@{}} Uniform \\ Downsampling\end{tabular}} 
 & PSNR ($\uparrow$)& 41.44 & 38.06 & 32.15 & 19.89 & \textbf{43.59} & \textbf{23.99} & 21.51 & 20.12 & 23.71 \\
 & SSIM ($\uparrow$)& 0.989 & 0.965 & 0.874 & 0.500 & \textbf{0.990} & \textbf{0.604} & 0.507 & 0.497 & 0.597 \\
 & ERGAS ($\downarrow$)& 1.549 & 2.091 & 3.894 & 15.431 & \textbf{1.217} & \textbf{9.580} & 12.870 & 15.600 & 9.860 \\
 & FID ($\downarrow$)& 9.81 & 11.72 & 25.99 & 32.93 & \textbf{9.77} & 77.60 & 80.66 & 71.78 & \textbf{52.76} \\
  & LPIPS ($\downarrow$)& 0.004 & 0.018 & 0.135 & 0.291 & \textbf{0.003} & 0.479 & 0.395 & 0.343 & \textbf{0.264} \\ \hline
\end{tabular}
\end{adjustbox}
\label{tab:pavia_exp2}
\end{table*}

\begin{figure*}[!t]
    \centering
    \setlength{\tabcolsep}{0pt}
    \begin{subfigure}{0.15\linewidth}
        \centering
        \includegraphics[width=1\linewidth]{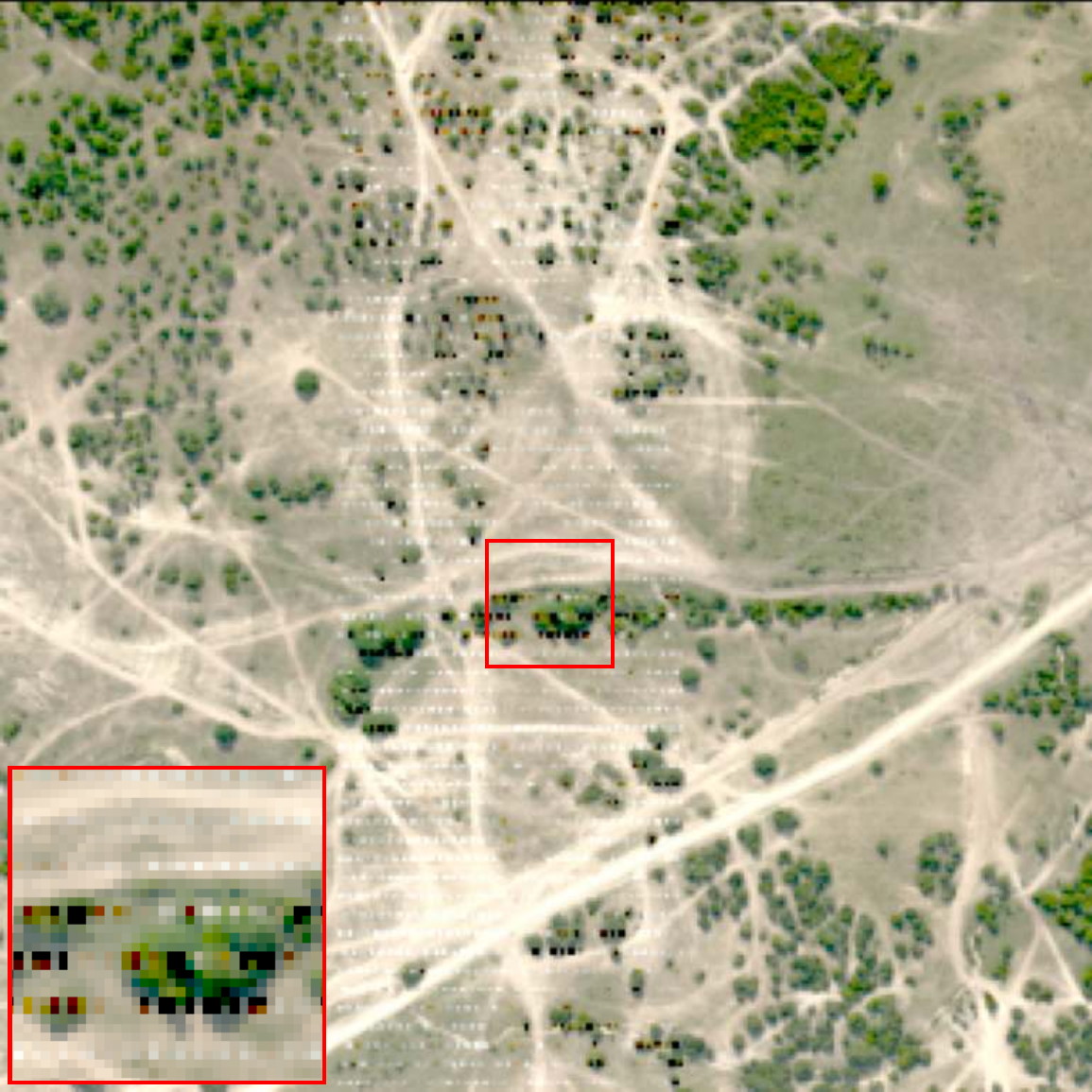}
        \captionsetup{labelformat=empty}
        \caption{\texttt{UHIF-RIM} \cite{ying2021unaligned}}
    \end{subfigure}
    \begin{subfigure}{0.15\linewidth}
        \centering
        \includegraphics[width=\linewidth]{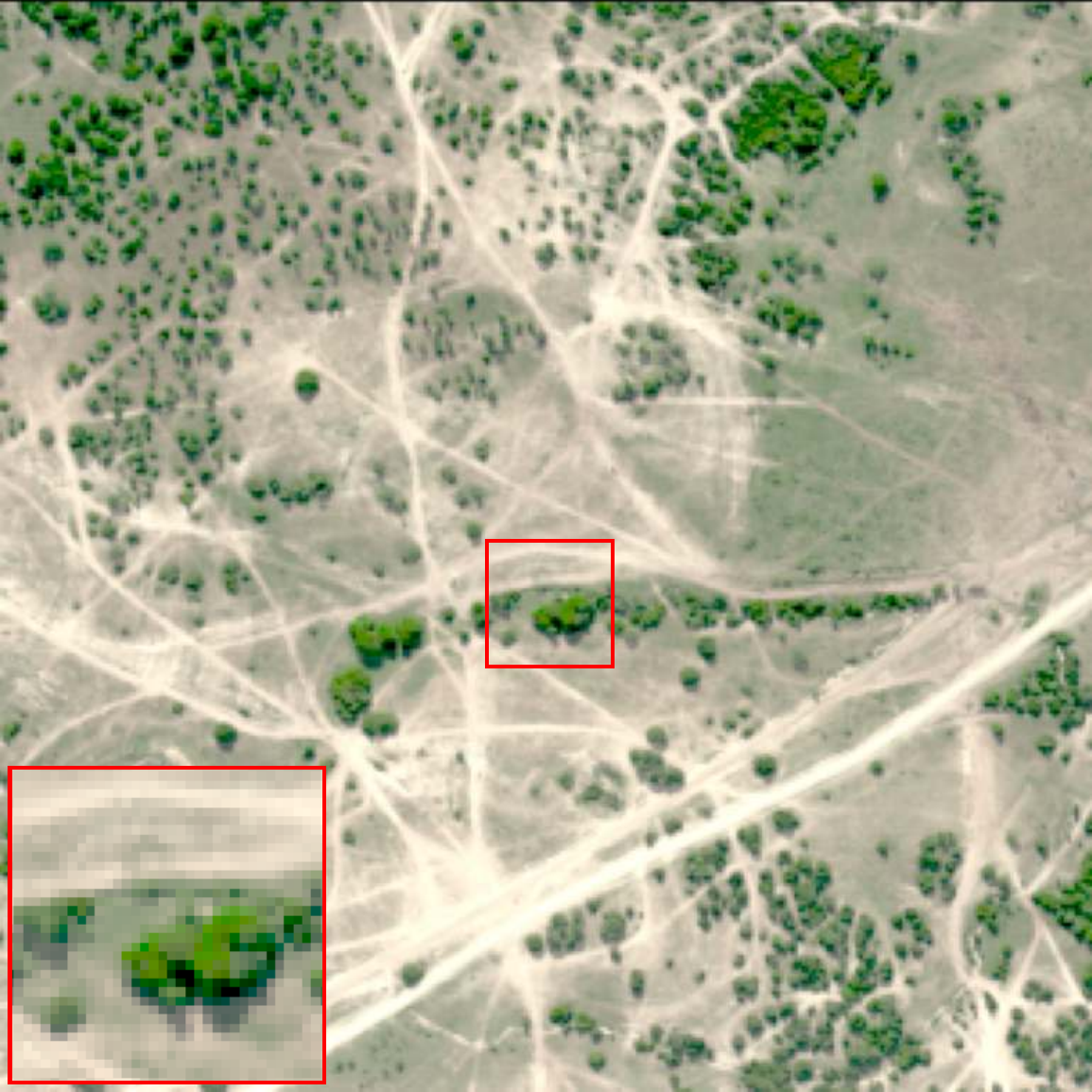}
        \captionsetup{labelformat=empty}
        \caption{\texttt{u2MDN} \cite{qu2021unsupervised}}
    \end{subfigure}
    \begin{subfigure}{0.15\linewidth}
        \centering
        \includegraphics[width=\linewidth]{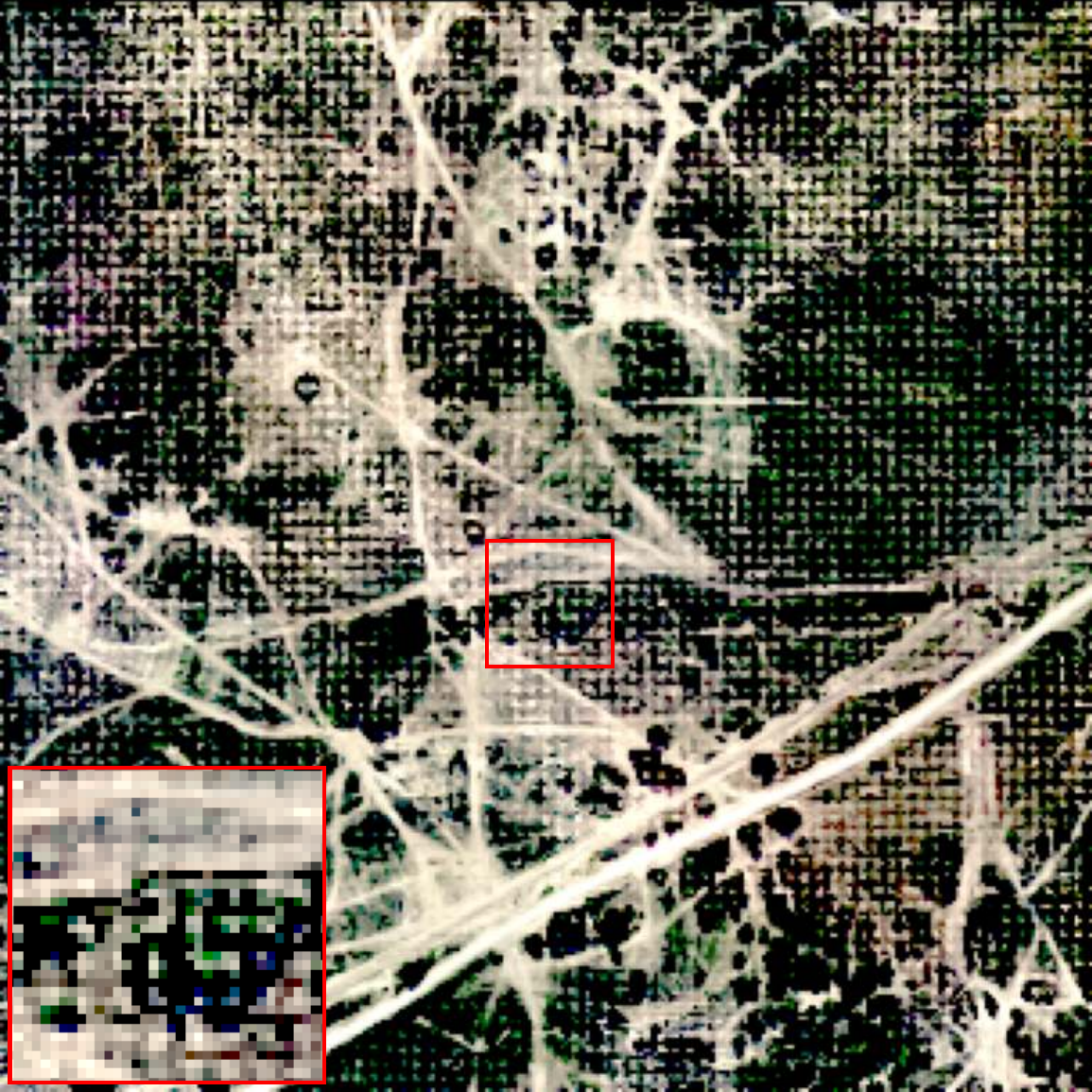}
        \captionsetup{labelformat=empty}
        \caption{\texttt{IARF} \cite{zhou2019integrated}}
    \end{subfigure}
    \begin{subfigure}{0.15\linewidth}
        \centering
        \includegraphics[width=\linewidth]{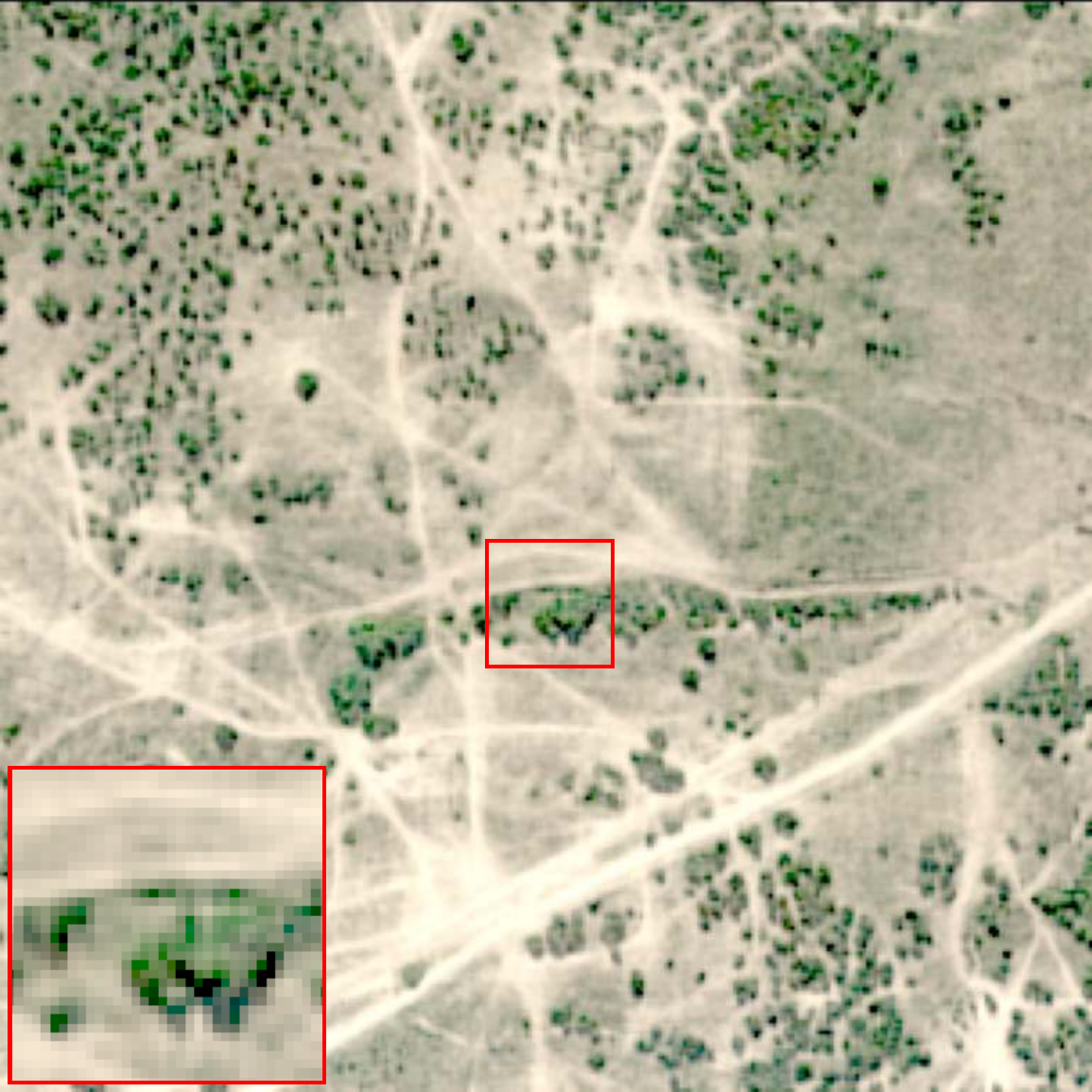}
        \captionsetup{labelformat=empty}
        \caption{\texttt{NBFusion}{\scriptsize \cite{chen2017normalized,kanatsoulis2018hyperspectral}}}
    \end{subfigure}
    \begin{subfigure}{0.15\linewidth}
        \centering
        \includegraphics[width=\linewidth]{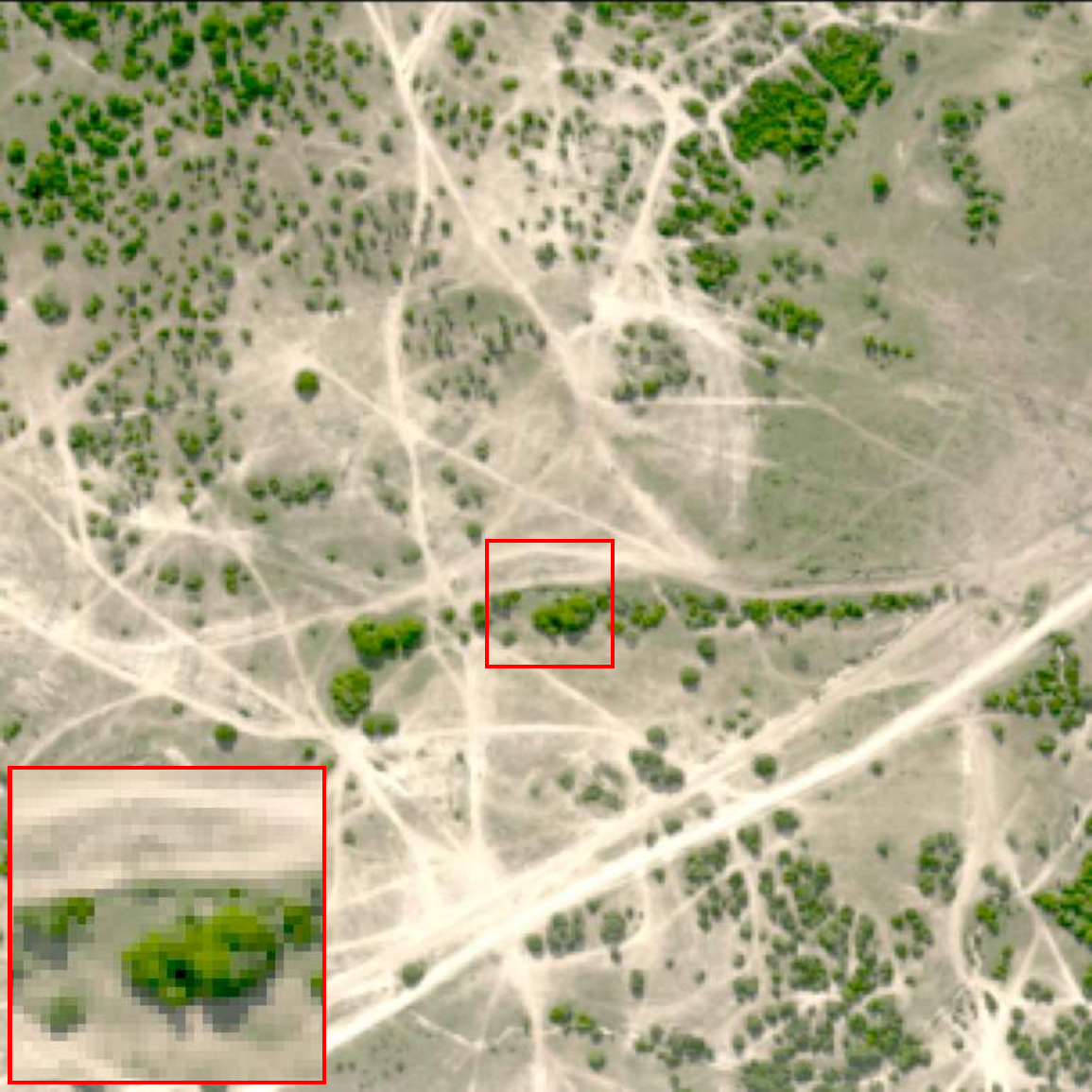}
        \captionsetup{labelformat=empty}
        \caption{\texttt{FRESCO} (\texttt{MSR})}
    \end{subfigure}
    \begin{subfigure}{0.15\linewidth}
        \centering
        \includegraphics[width=\linewidth]{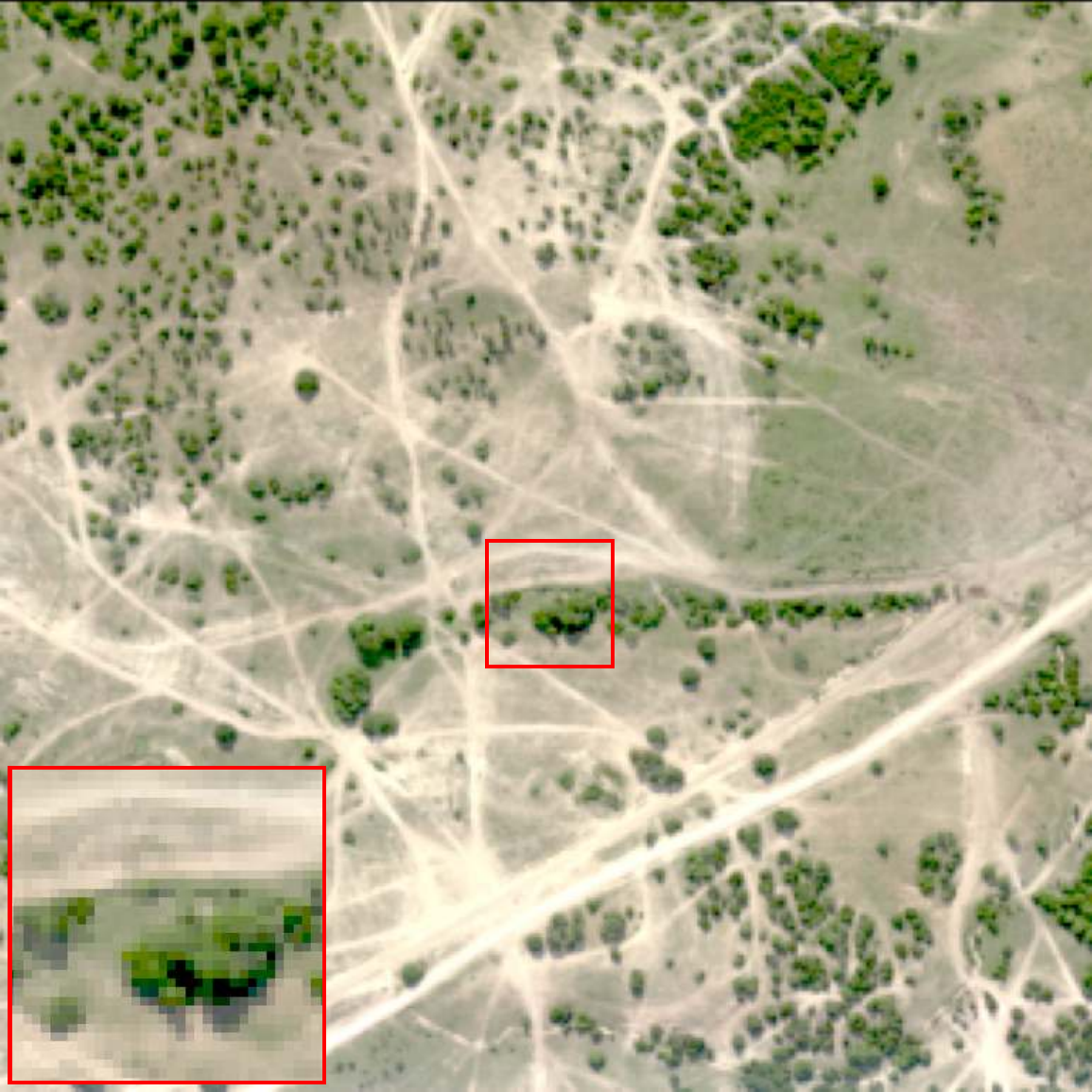}
        \captionsetup{labelformat=empty}
        \caption{Reference: {\scriptsize $\tY\pM_{\rm SRI}$}}
    \end{subfigure} 
    \caption{RGB rendering of recovered $\widehat{\tY}\pM_{\rm SRI}$ (Terrain); uniform downsampling used for degradation.}
    \label{fig:terrain_visual_M}
\end{figure*}

\begin{figure*}[!t]
    \centering
    \setlength{\tabcolsep}{0pt}
    \begin{subfigure}{0.15\linewidth}
        \centering
        \includegraphics[width=\linewidth]{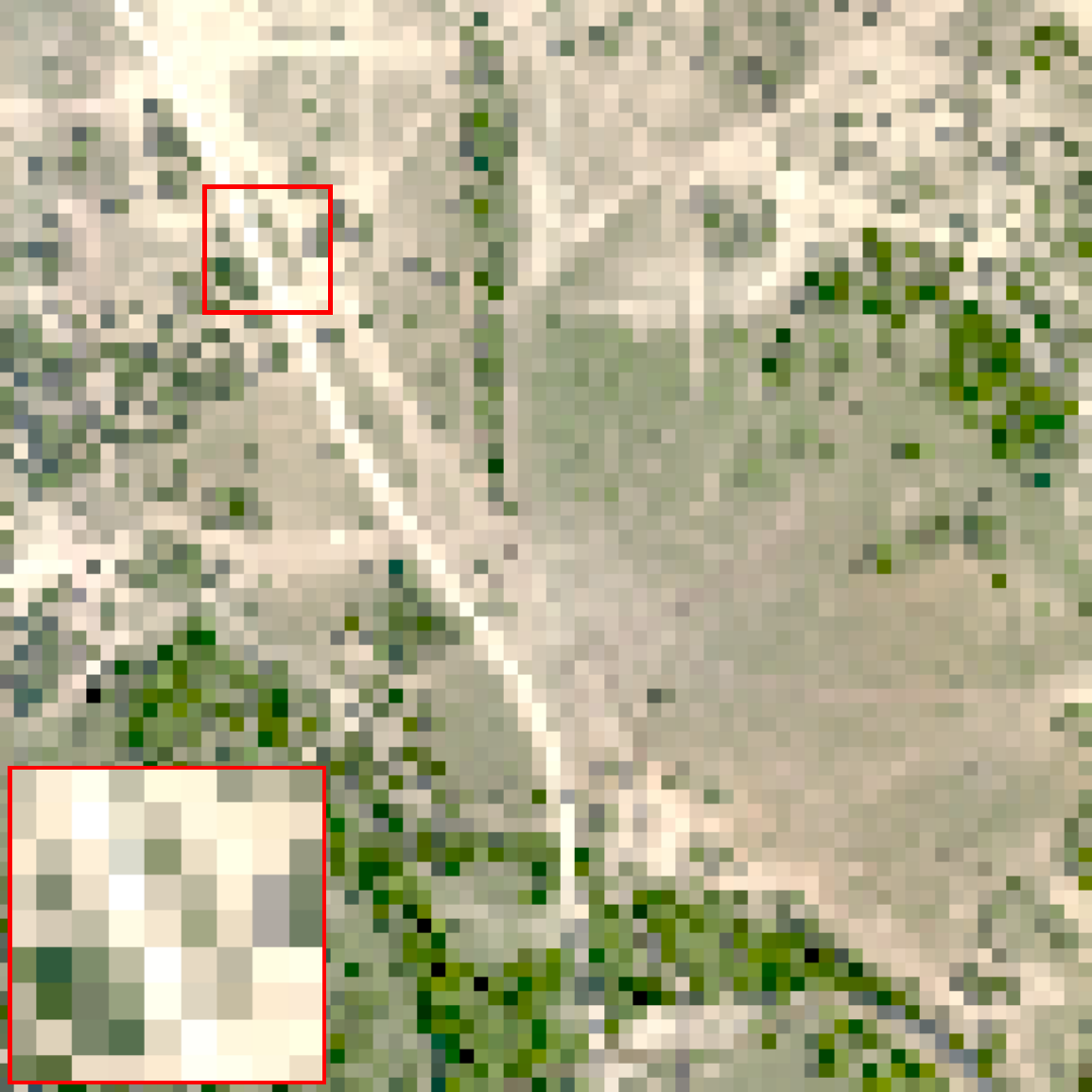}
        \captionsetup{labelformat=empty}
        \caption{HSI}
    \end{subfigure}
    \begin{subfigure}{0.15\linewidth}
        \centering
        \includegraphics[width=\linewidth]{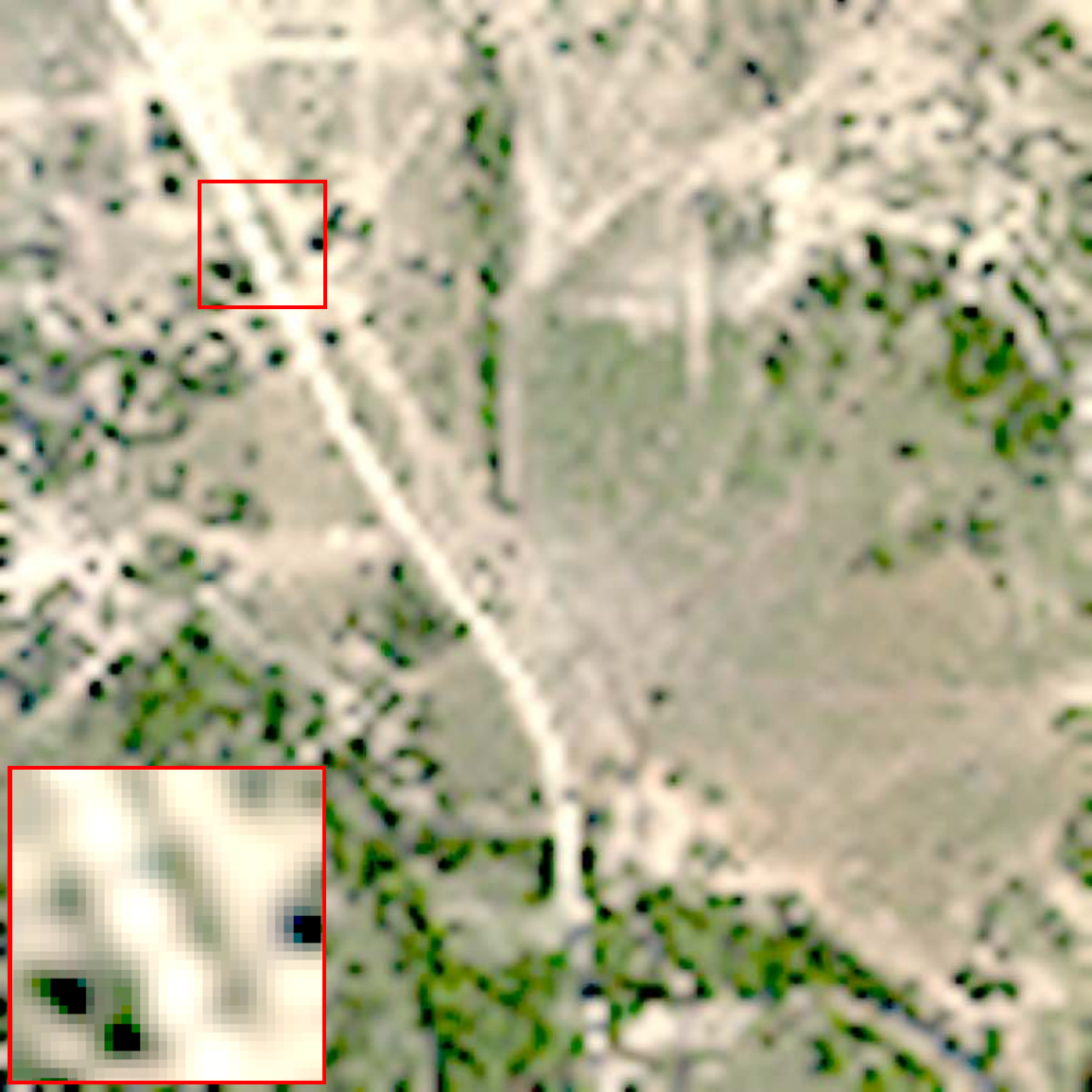}
        \captionsetup{labelformat=empty}
        \caption{\texttt{Lanczos}\cite{duchon1979lanczos}}
    \end{subfigure}
    \begin{subfigure}{0.15\linewidth}
        \centering
        \includegraphics[width=\linewidth]{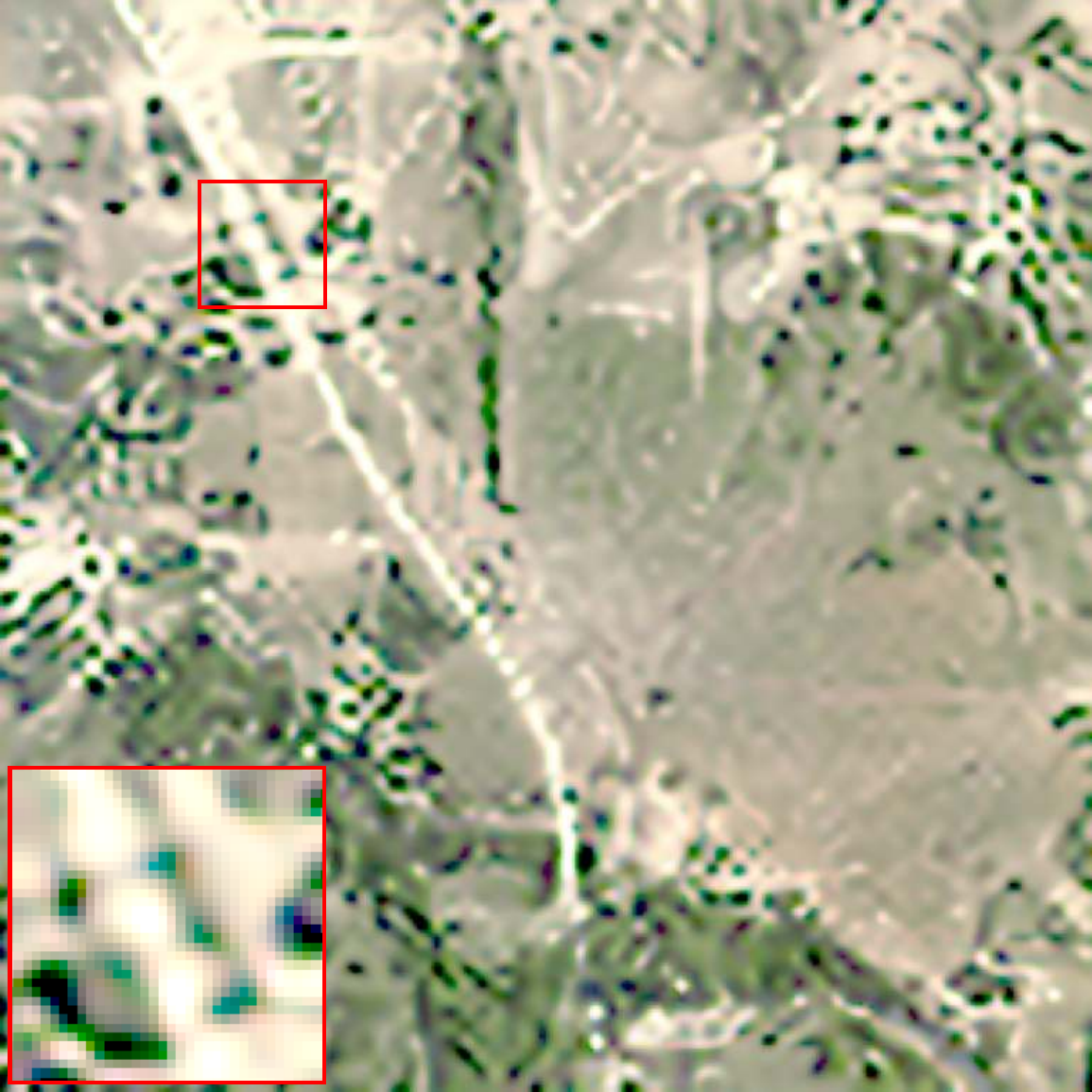}
        \captionsetup{labelformat=empty}
        \caption{\texttt{MC-Net} \cite{li2020mixed}}
    \end{subfigure}
    \begin{subfigure}{0.15\linewidth}
        \centering
        \includegraphics[width=\linewidth]{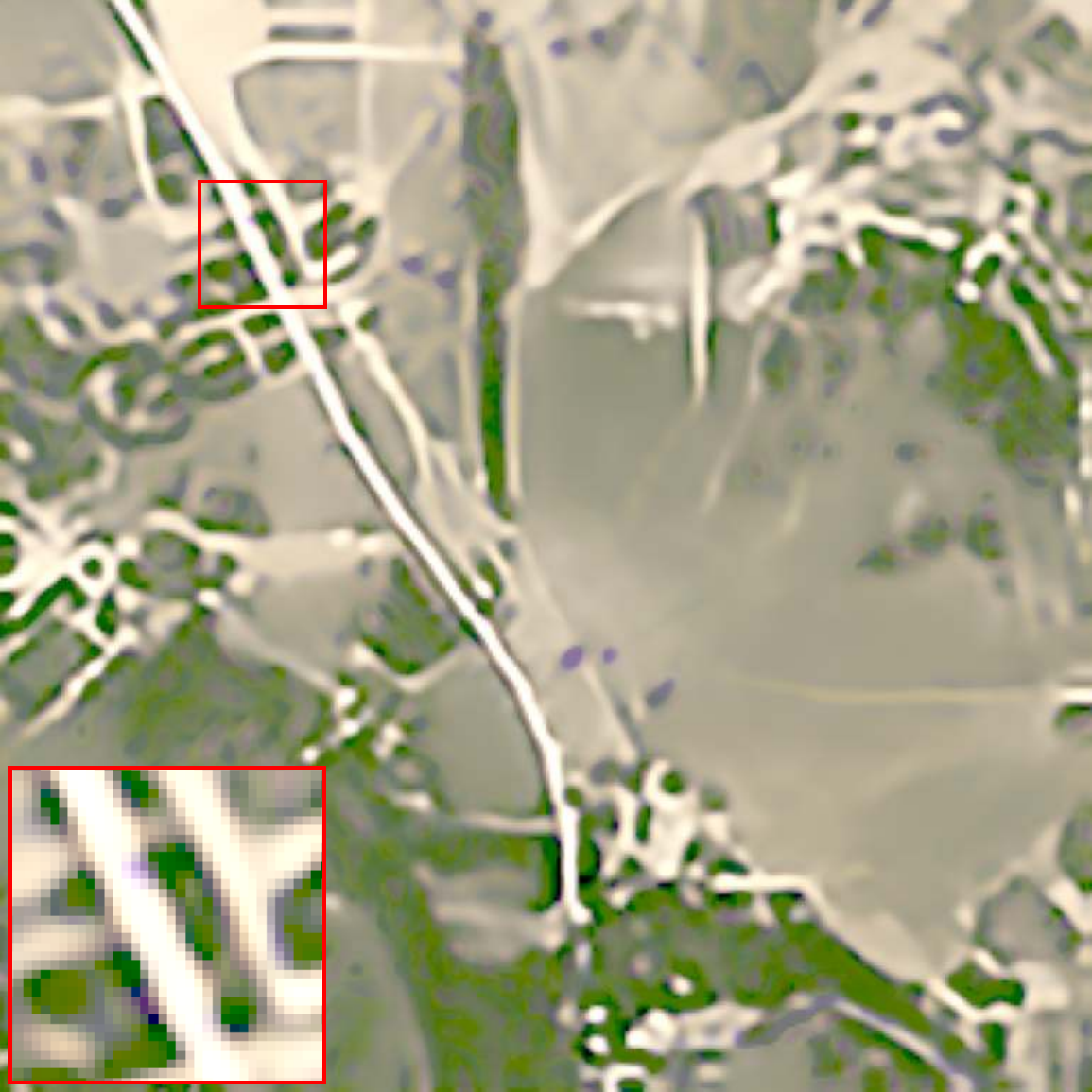}
        \captionsetup{labelformat=empty}
        \caption{\texttt{TSBSR} \cite{liang2023blind}}
    \end{subfigure}
    \begin{subfigure}{0.15\linewidth}
        \centering
        \includegraphics[width=\linewidth]{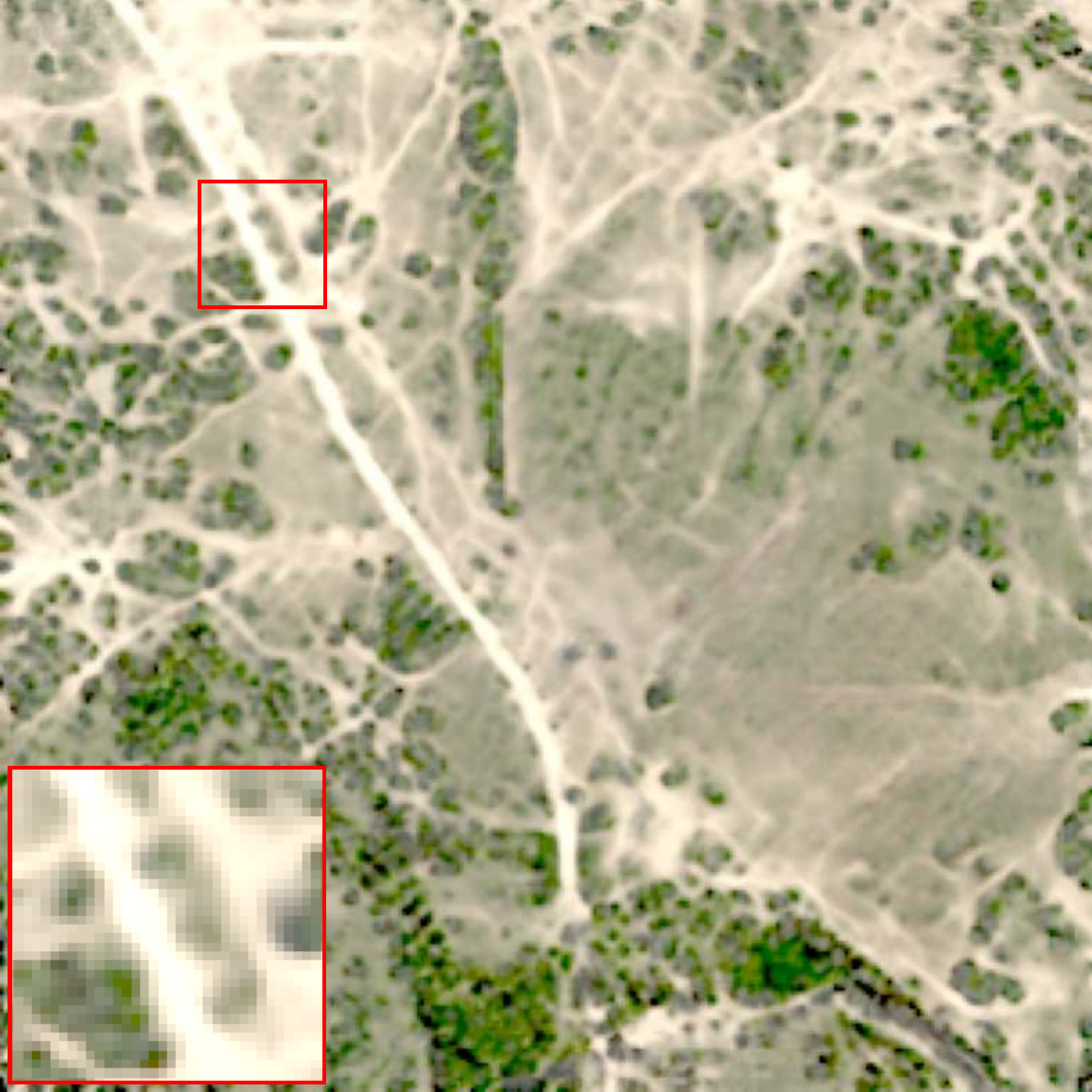}
        \captionsetup{labelformat=empty}
        \caption{\texttt{FRESCO} (\texttt{HSR})}
    \end{subfigure}
    \begin{subfigure}{0.15\linewidth}
        \centering
        \includegraphics[width=\linewidth]{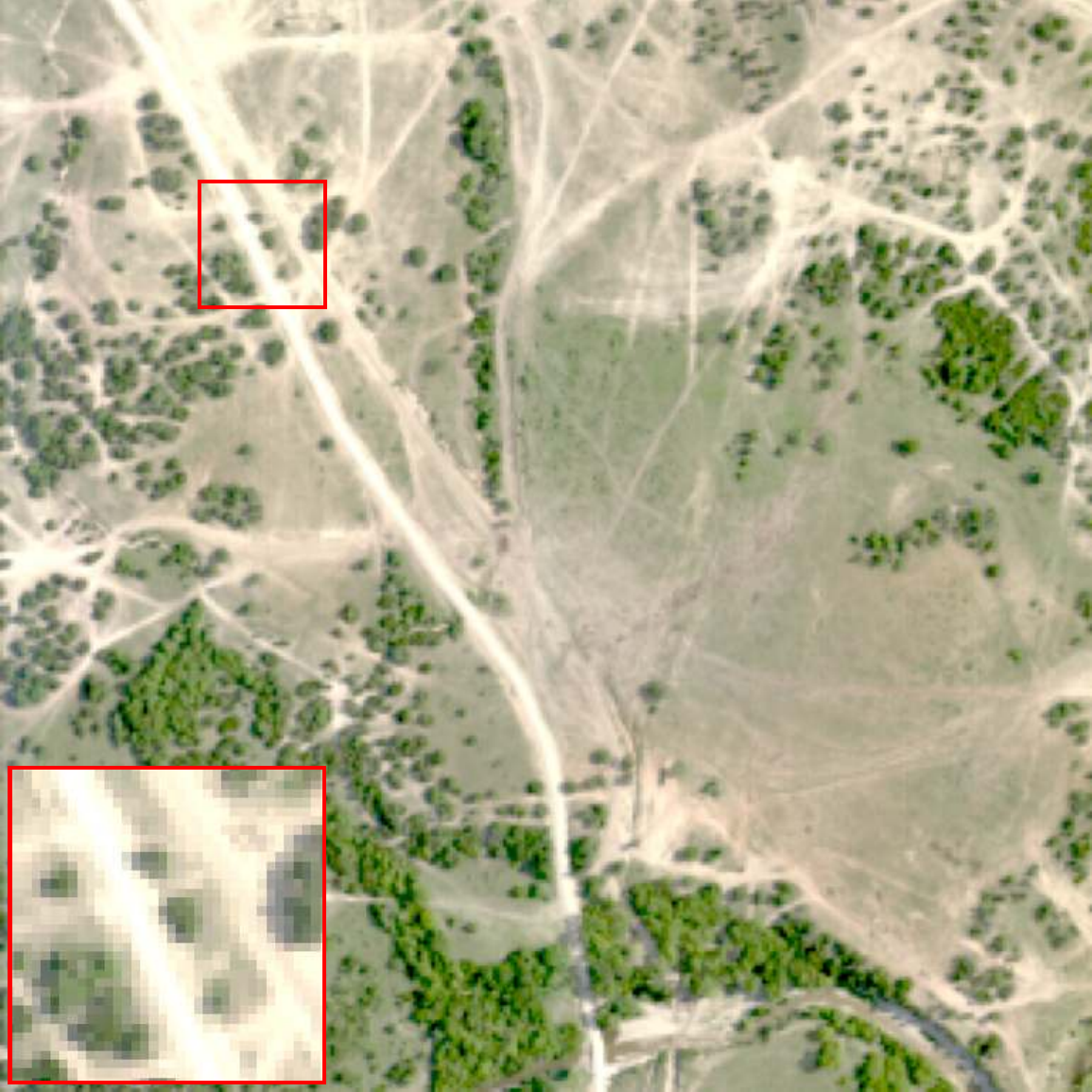}
        \captionsetup{labelformat=empty}
        \caption{Reference: {\scriptsize $\tY\pH_{\rm SRI}$}}
    \end{subfigure}
    \caption{RGB rendering of recovered $\widehat{\tY}\pM_{\rm SRI}$ (Terrain); uniform downsampling used for degradation.}
    \label{fig:terrain_visual_H}
\end{figure*}

Fig.~\ref{fig:terrain_visual_M} and Fig.~\ref{fig:terrain_visual_H} show results on the Terrain dataset, where the unknown downsampling operator is uniform sampling.  
We extract unregistered SRIs as $\tY\pM_{\mathrm{SRI}} = \tY_{\mathrm{Terrain}}(1\!:\!300,1\!:\!300,:)$ and $ \tY\pH_{\mathrm{SRI}} = \tY_{\mathrm{Terrain}}(121\!:\!420,1\!:\!300,:)$ from $ \tY_{\mathrm{Terrain}} \in \mathbb{R}^{500 \times 307 \times 75}$, and rotate $\tY\pM_{\mathrm{SRI}}$ $90^\circ$ counterclockwise to induce misalignment.
Visual trends are similar to those in Fig.~\ref{fig:pavia_visual_M} and Fig.~\ref{fig:pavia_visual_H}.  
Table~\ref{tab:terrain_exp2} reports quantitative results, consistent with the Pavia findings.

\begin{table*}[!t]
\centering
\renewcommand{\arraystretch}{1.3} 
\small 
\caption{Performance of the methods on Terrain.}
\begin{adjustbox}{max width=0.9\linewidth}
\begin{tabular}{|c|c|ccccc|cccc|}
\hline
\multirow{2}{*}{\begin{tabular}[c]{@{}c@{}} Downsampling\\ Methods\end{tabular}} & \multirow{2}{*}{Metrics} & \multicolumn{5}{c|}{MSR} & \multicolumn{4}{c|}{HSR} \\ \cline{3-11} 
 &  & \texttt{u2MDN} & \texttt{UHIF-RIM} &\texttt{NBFusion} & \texttt{IARF} & \texttt{FRESCO} (\texttt{MSR}) & \texttt{Lanczos} & \texttt{MC-Net} & \texttt{TSBSR} & \multicolumn{1}{c|}{\texttt{FRESCO} (\texttt{HSR})} \\ \hline
\multirow{5}{*}{\begin{tabular}[c]{@{}c@{}}Gaussian Kernel\\ Downsampling\end{tabular}} 
 & PSNR ($\uparrow$)& 30.09 & 29.09 & 29.05 & 12.45 & \textbf{44.91} & \textbf{24.08} & 24.03 & 20.88 & 23.64  \\ 
 & SSIM ($\uparrow$) & 0.980 & 0.963 & 0.898 & 0.331 & \textbf{0.998} & 0.744 & \textbf{0.763} & 0.685 & 0.731 \\  
 & ERGAS ($\downarrow$)& 3.075 & 3.690 & 2.949 & 19.856 & \textbf{0.582} & \textbf{5.342} & 5.387 & 7.944 & 5.583\\
 & FID ($\downarrow$)& 4.15 & 6.84 & 16.92 & 30.92 & \textbf{4.02} & 29.35 & 24.55 & 26.49 & \textbf{19.65} \\
 & LPIPS ($\downarrow$)& 0.009 & 0.041 & 0.118 & 0.295 & \textbf{0.006} & 0.372 & 0.277 & 0.320 & \textbf{0.215} \\ \hline 
\multirow{5}{*}{\begin{tabular}[c]{@{}c@{}}Nearest Neighbor \\ Downsampling\end{tabular}} 
 & PSNR ($\uparrow$) & 29.42 & 29.52 & 28.84 & 12.64 & \textbf{44.57} & \textbf{21.76} & 18.76 & 18.16 & 21.70\\ 
 & SSIM ($\uparrow$) & 0.977 & 0.953 & 0.896 & 0.351 & \textbf{0.998} & \textbf{0.679} & 0.589 & 0.612 & 0.659\\ 
 & ERGAS ($\downarrow$)& 3.271 & 3.657 & 3.030 & 19.375 & \textbf{0.604} & \textbf{6.973} & 9.887 & 10.696 & 6.997  \\ 
 & FID  ($\downarrow$)& 4.47 & 7.07 & 17.86 & 28.09 & \textbf{4.01} & 37.68 & 34.44 & 37.67 & \textbf{20.09} \\
  & LPIPS ($\downarrow$)& 0.008 & 0.071 & 0.120 & 0.291 & \textbf{0.006} & 0.393 & 0.348 & 0.371 & \textbf{0.238} \\ \hline
\multirow{5}{*}{\begin{tabular}[c]{@{}c@{}} Uniform \\ Downsampling\end{tabular}} 
 & PSNR ($\uparrow$)& 28.91 & 31.61 & 28.85 & 12.38 & \textbf{44.59} & 20.80 & 18.12 & 17.39  & \textbf{20.94}  \\
 & SSIM ($\uparrow$) & 0.975 & 0.952 & 0.896 & 0.321 & \textbf{0.998} & \textbf{0.625} & 0.538 & 0.534 & 0.615 \\
 & ERGAS ($\downarrow$)& 3.461 & 3.172 & 3.025 & 19.981 & \textbf{0.601} & 7.834 & 10.672 & 11.675 & \textbf{7.672} \\
 & FID ($\downarrow$)& 4.16 & 6.55 & 19.04 & 30.12 & \textbf{4.20} & 38.31 & 33.66 & 34.12 & \textbf{22.76} \\
 & LPIPS ($\downarrow$)& 0.011 & 0.074 & 0.123 & 0.309 & \textbf{0.007} & 0.395 & 0.348 & 0.357& \textbf{0.243}\\ \hline
\end{tabular}
\end{adjustbox}
\label{tab:terrain_exp2}
\end{table*}

\begin{figure}[!t]
    \centering
    \setlength{\tabcolsep}{0pt}
    \begin{subfigure}{0.32\linewidth}
        \centering
        \includegraphics[width=1\linewidth]{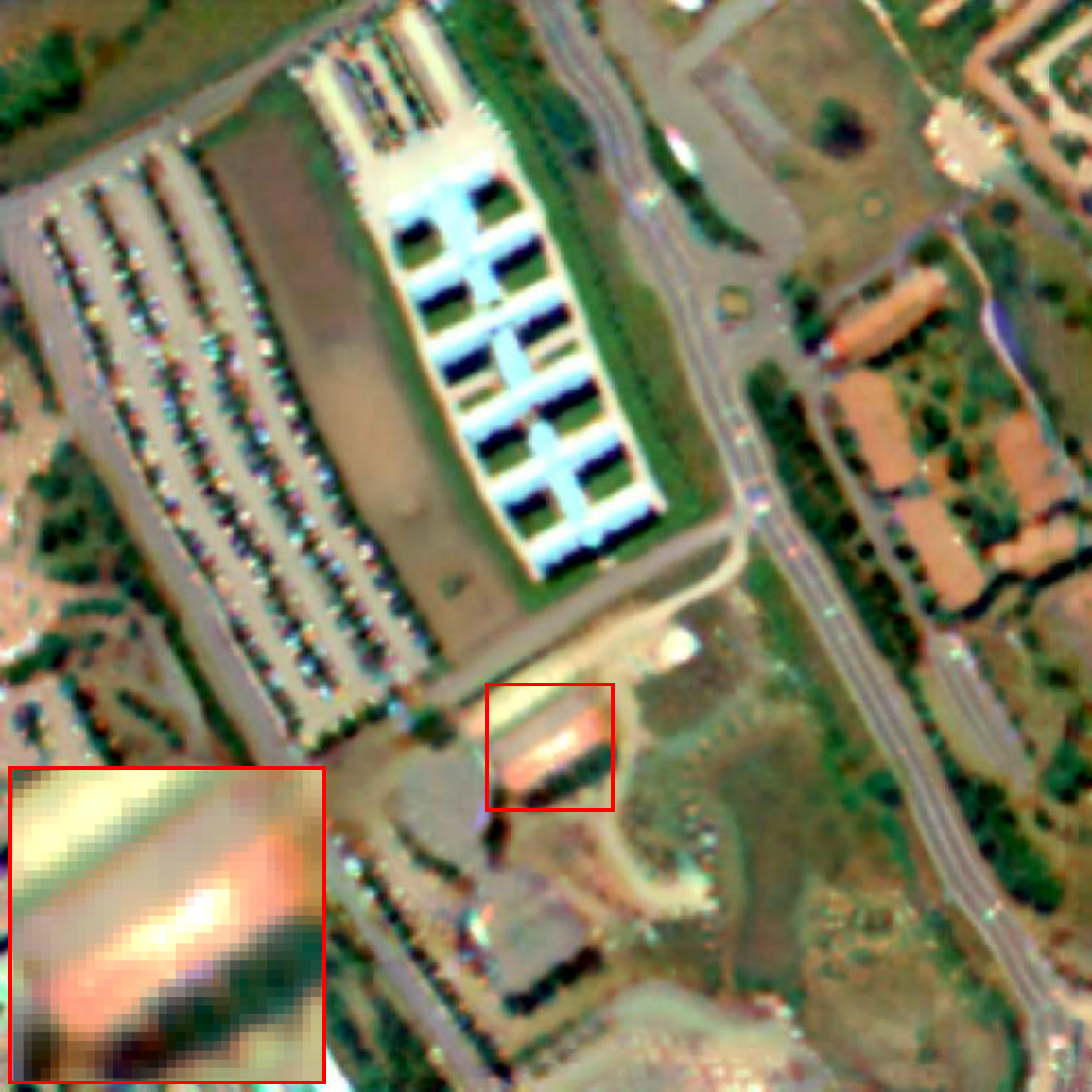}
    \end{subfigure}
    \begin{subfigure}{0.32\linewidth}
        \centering
        \includegraphics[width=\linewidth]{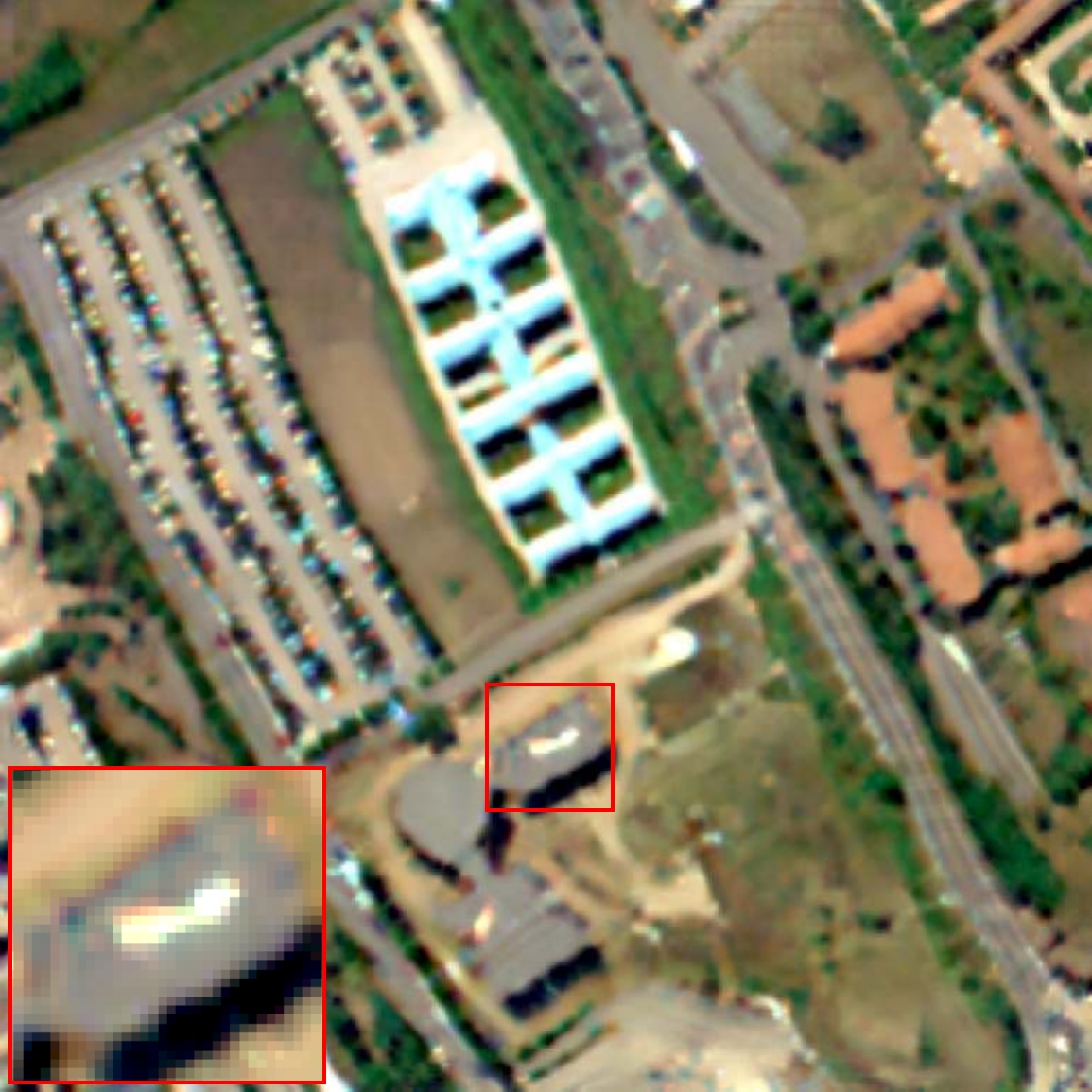}
    \end{subfigure}
    \begin{subfigure}{0.32\linewidth}
        \centering
        \includegraphics[width=\linewidth]{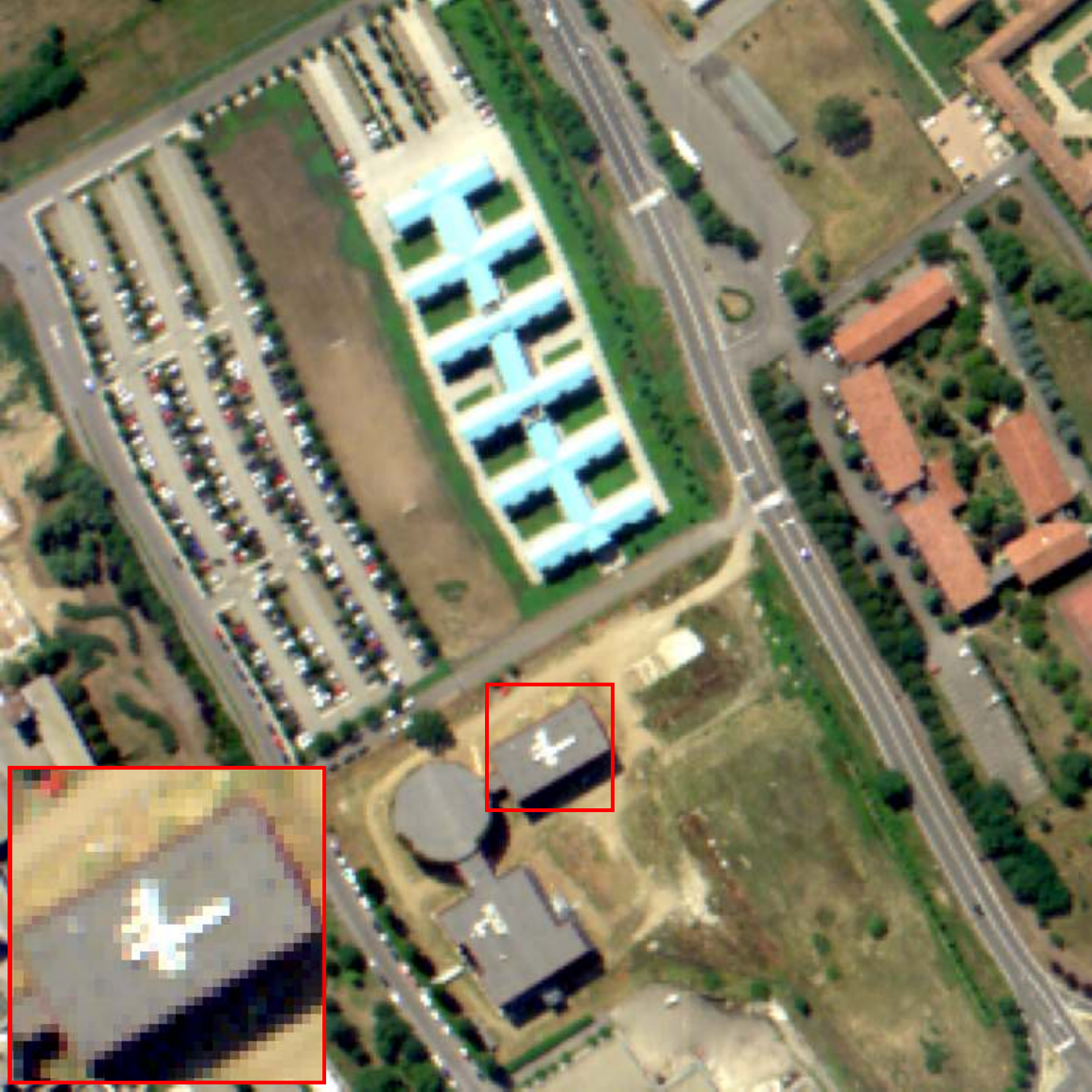}
    \end{subfigure}
    \\
    \vspace{0.5em}
    
    \begin{subfigure}{0.32\linewidth}
        \centering
        \includegraphics[width=\linewidth]{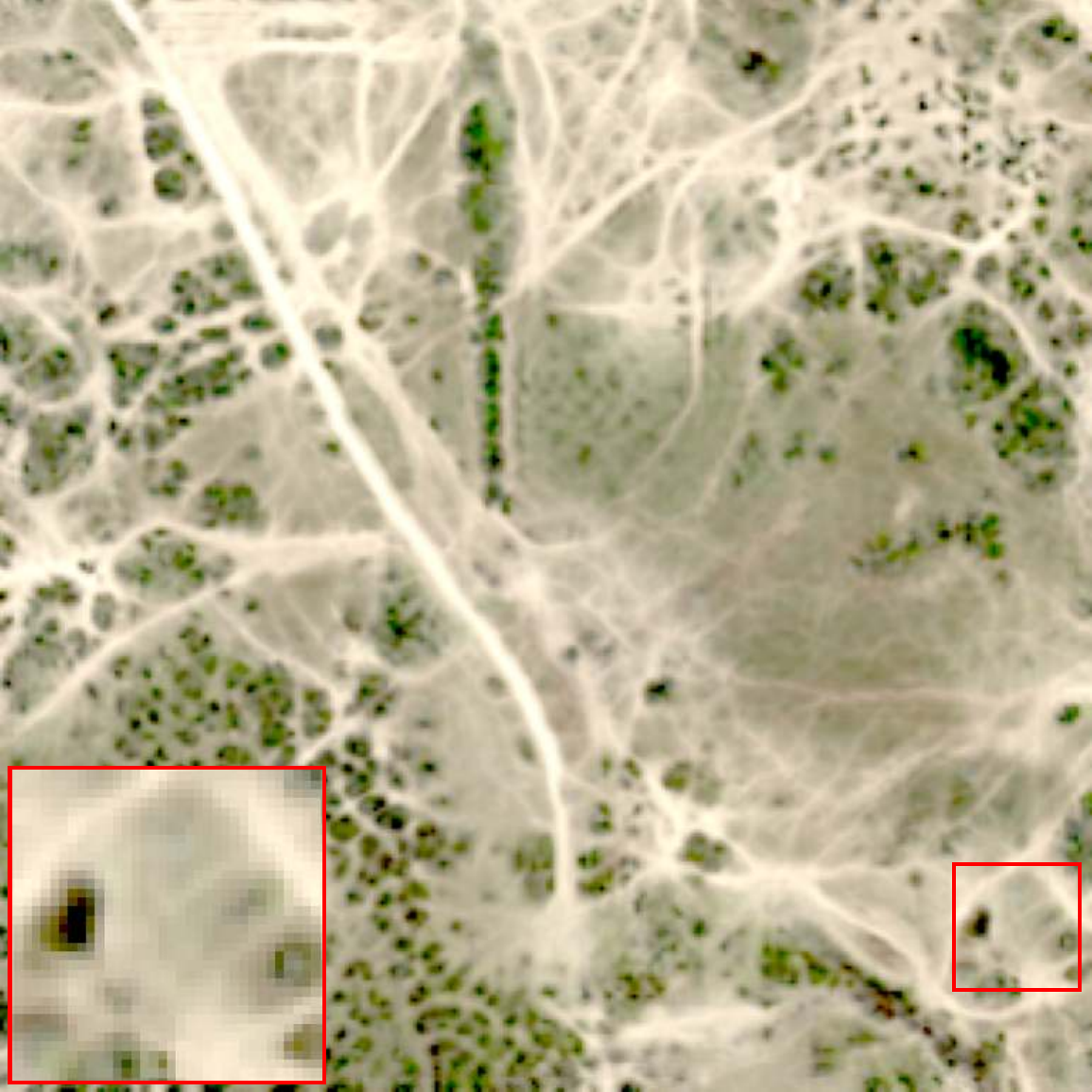}
        \captionsetup{labelformat=empty}
        \caption{\texttt{IAT} (cf. Eq.~\eqref{eq:iat})}
    \end{subfigure}
    \begin{subfigure}{0.32\linewidth}
        \centering
        \includegraphics[width=\linewidth]{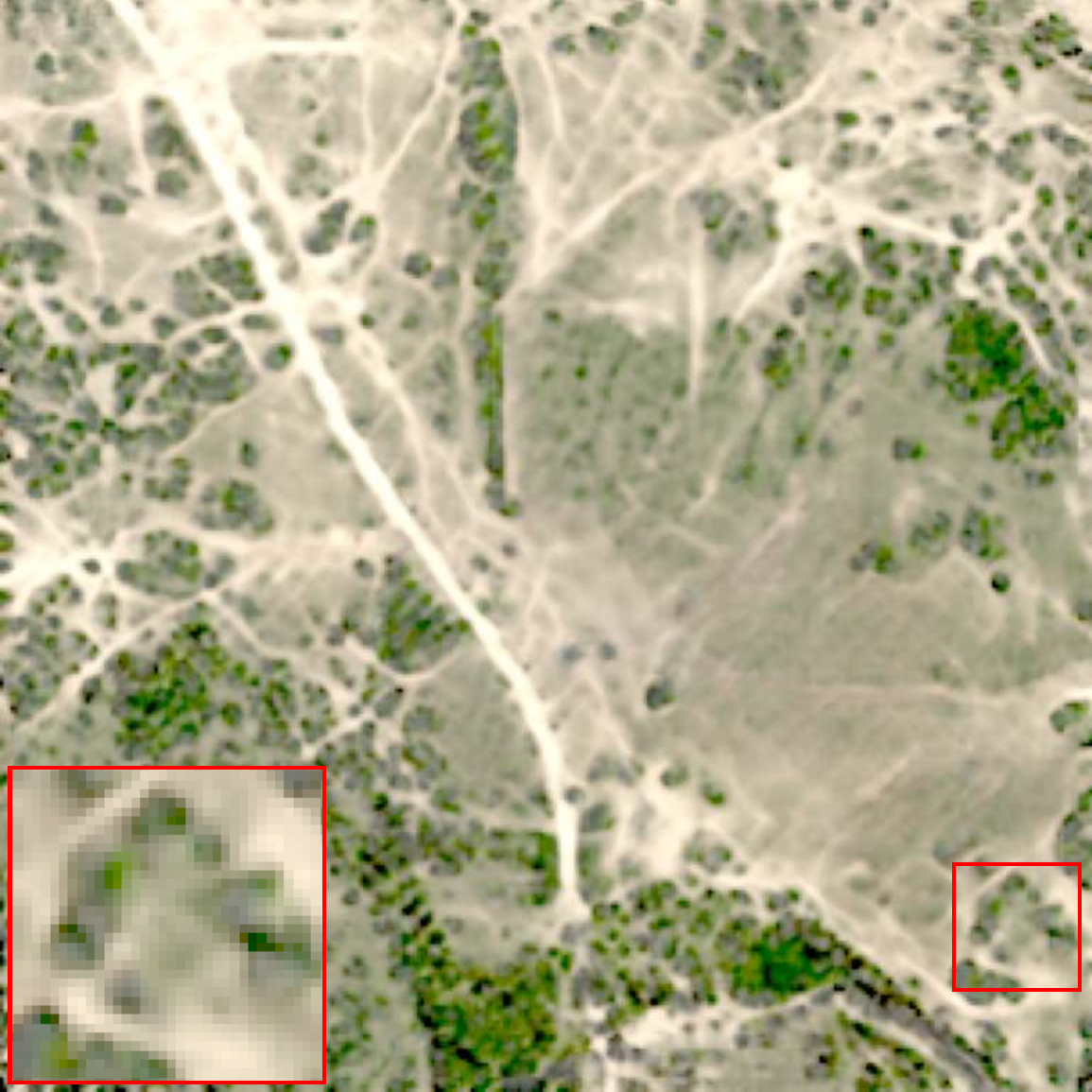}
        \captionsetup{labelformat=empty}
        \caption{\texttt{FRESCO} (\texttt{HSR})}
    \end{subfigure}
    \begin{subfigure}{0.32\linewidth}
        \centering
        \includegraphics[width=\linewidth]{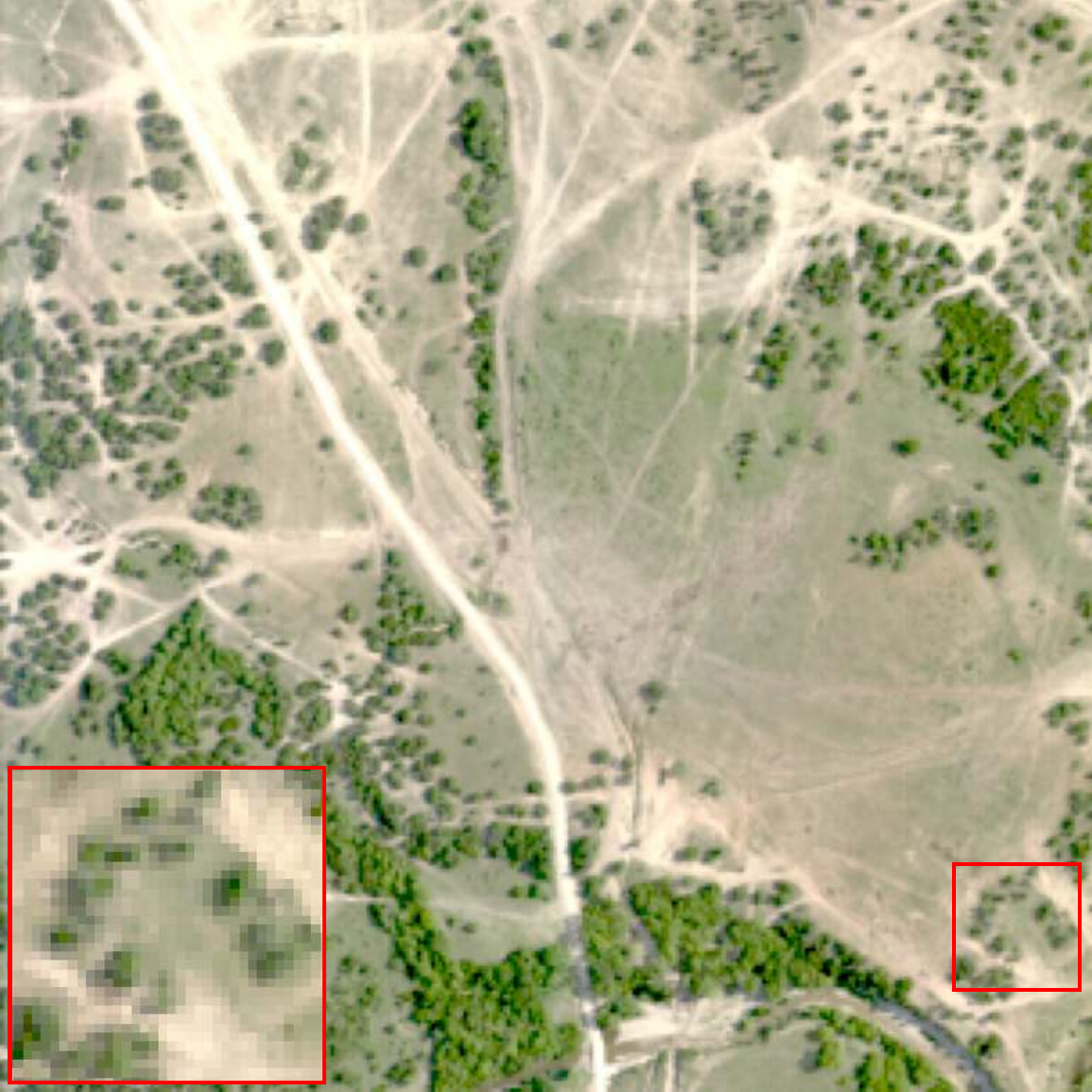}
        \captionsetup{labelformat=empty}
        \caption{Reference: {\scriptsize $\tY\pH_{\rm SRI}$}}
    \end{subfigure}
    \caption{Comparison of \texttt{IAT} and the proposed \texttt{FRESCO}.}
    \label{fig:content_misalignment}
\end{figure}

Fig.~\ref{fig:content_misalignment} compares our method with the \textit{individual abundance translation} (\texttt{IAT}) baseline, which solves  
\begin{align}\label{eq:iat}
[\bm f_r]_{\#} p_r\pH = p_r\pM,
\end{align}
independently for each $r$ without enforcing a shared $\bm f$. This highlights the importance of a unified map across $r$ for recoverability.
The experiment uses the same settings as in Fig.~\ref{fig:pavia_visual_H} and Fig.~\ref{fig:terrain_visual_H}.  
In the first row, \texttt{IAT} misaligns content---e.g., a gray rooftop is translated into orange. In the second, grassy regions are mis-translated into barren land.  
These errors illustrate the lack of identifiability of $\bm f$ in \texttt{IAT}, which permits structurally inconsistent translations.  
Quantitative results comparing our method with \texttt{IAT} appear in Tables~\ref{tab:pavia_exp2} and~\ref{tab:terrain_exp2}.

\begin{figure*}[!t]
    \centering
    \setlength{\tabcolsep}{0pt}
    \colorbox{blue!5}{
    \begin{subfigure}{0.165\linewidth}
        \centering
        \includegraphics[width=1\linewidth]{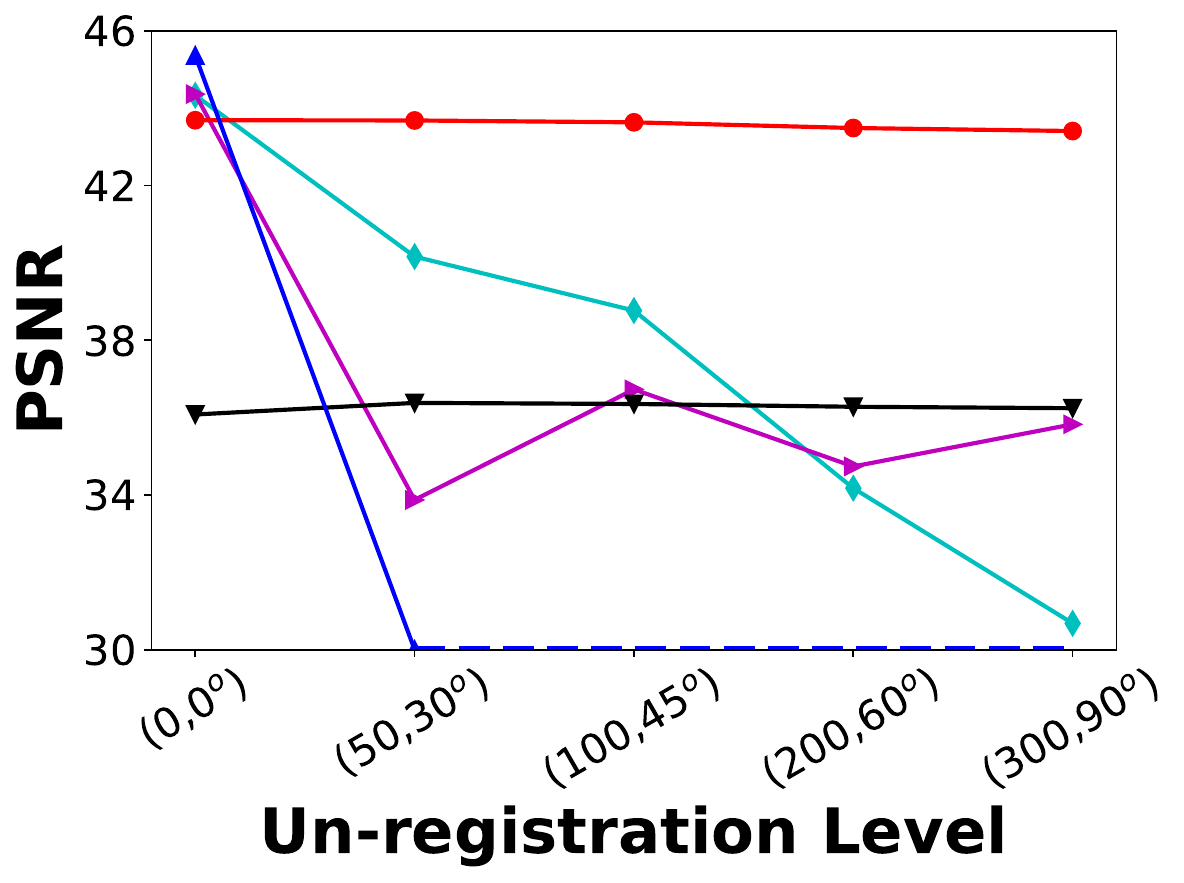}
        \caption{PSNR ($\uparrow$)}
    \end{subfigure}
    \begin{subfigure}{0.165\linewidth}
        \centering
        \includegraphics[width=\linewidth]{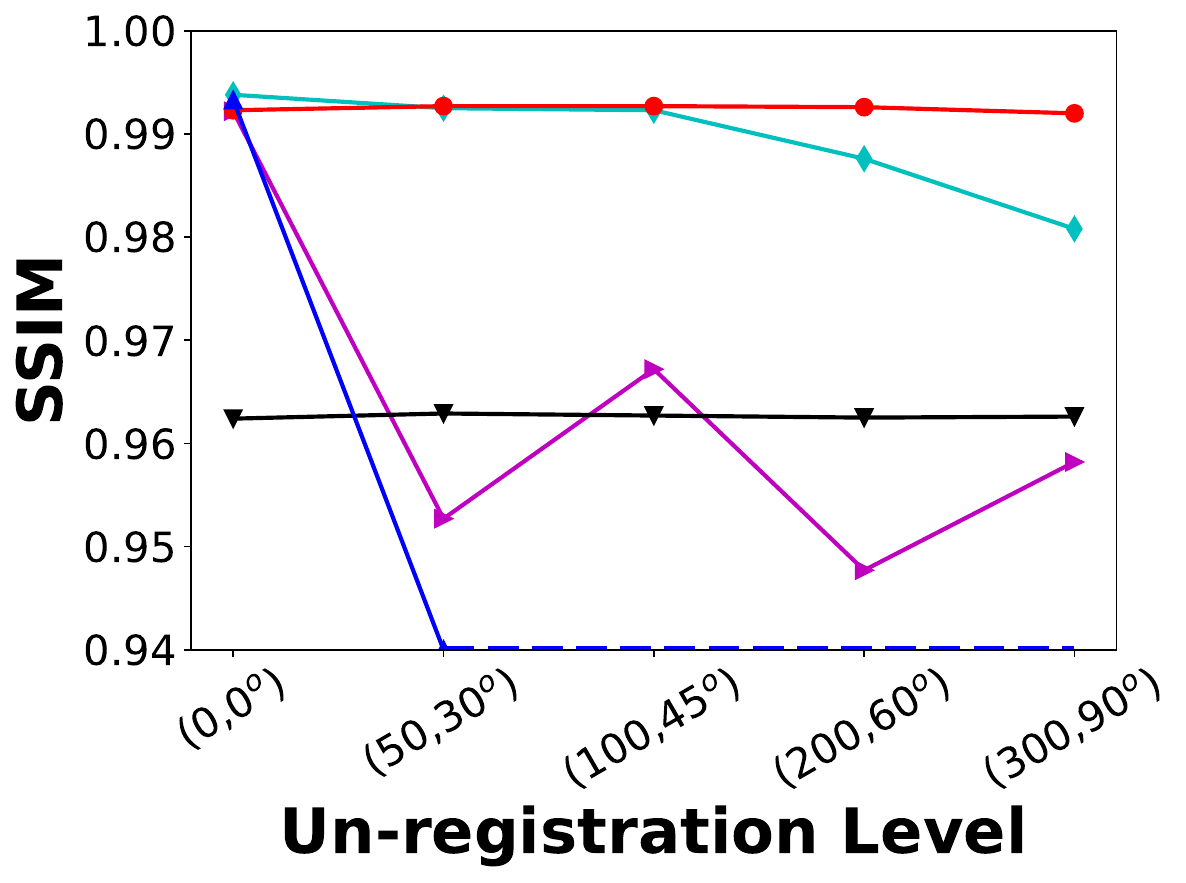}
        \caption{SSIM ($\uparrow$)}
    \end{subfigure}
    \begin{subfigure}{0.165\linewidth}
        \centering
        \includegraphics[width=\linewidth]{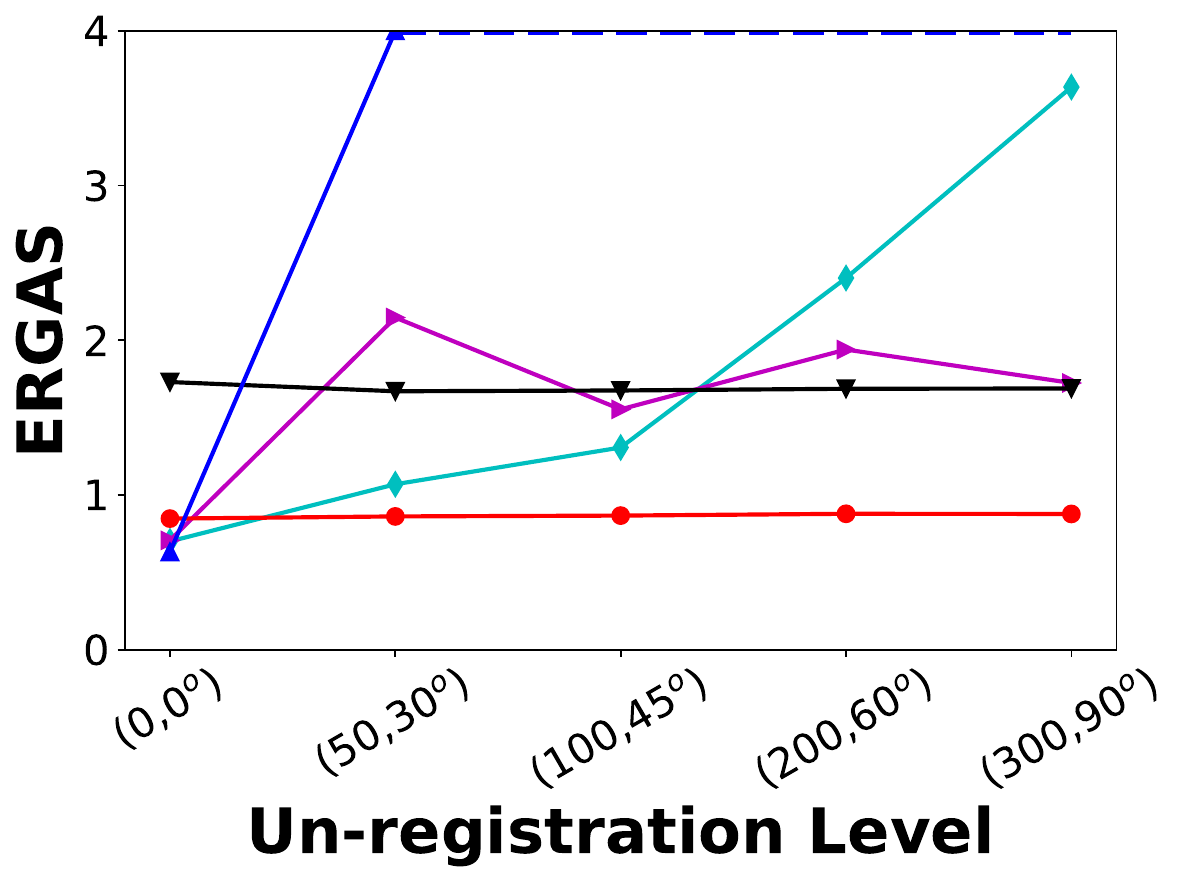}
        \caption{ERGAS ($\downarrow$)}
    \end{subfigure}
    \begin{subfigure}{0.165\linewidth}
        \centering
        \includegraphics[width=\linewidth]{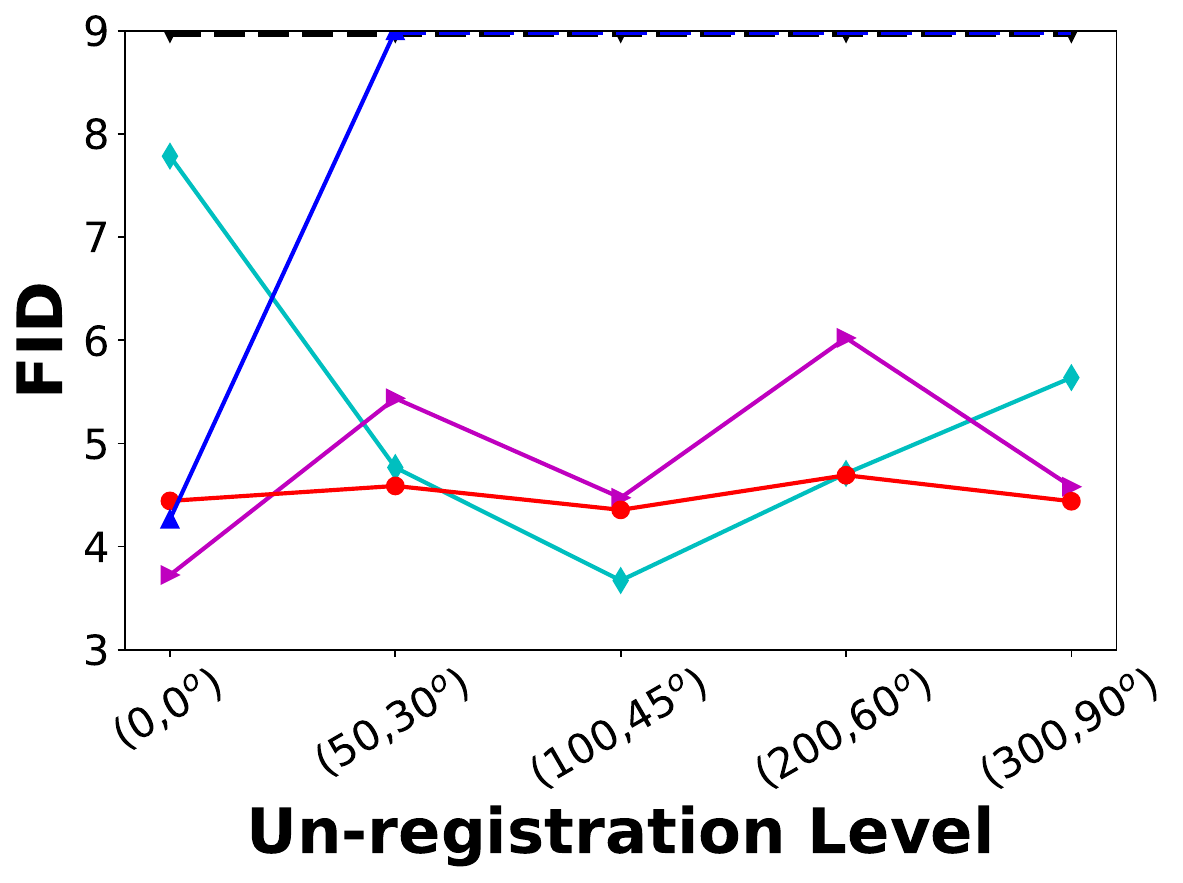}
        \caption{FID ($\downarrow$)}
    \end{subfigure}
    \begin{subfigure}{0.165\linewidth}
        \centering
        \includegraphics[width=\linewidth]{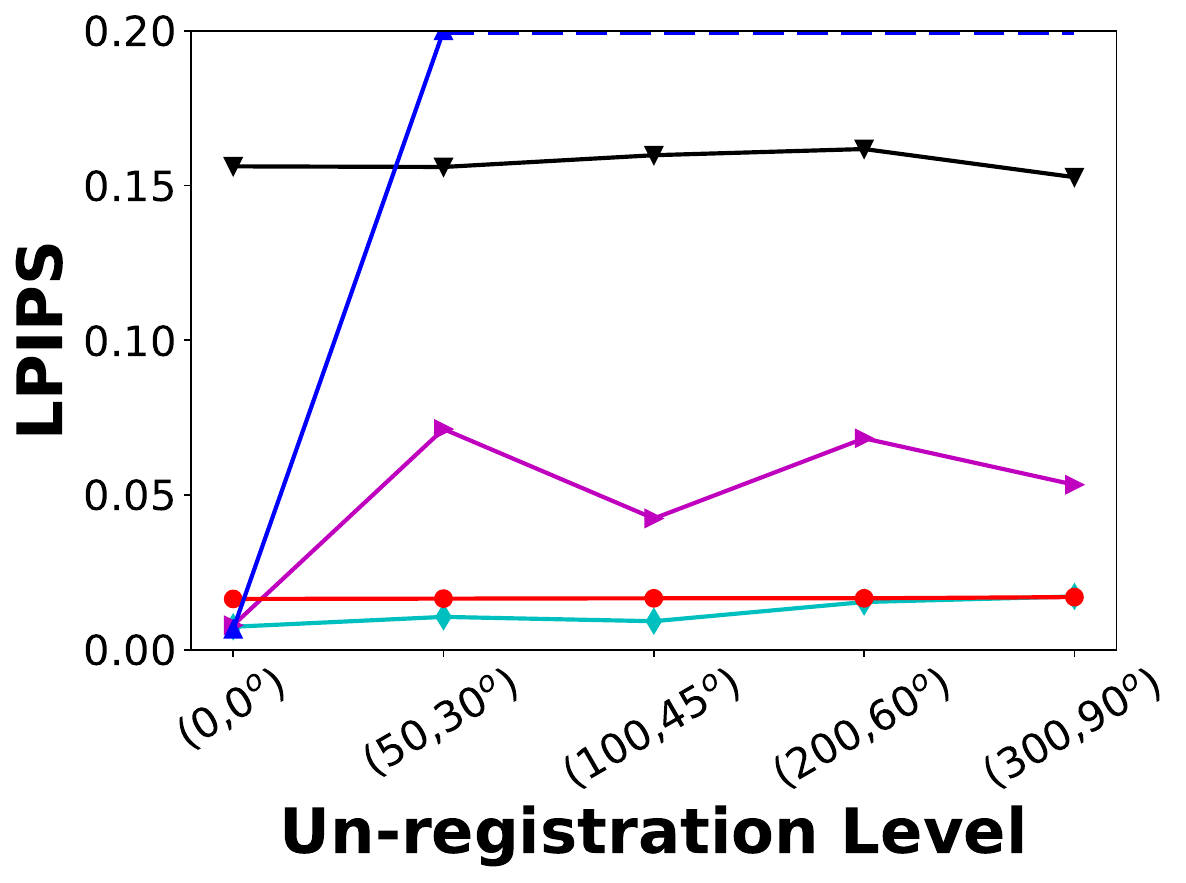}
        \caption{LPIPS ($\downarrow$)}
    \end{subfigure}}
    \begin{subfigure}{0.1\linewidth}
        \centering
        \includegraphics[width=\linewidth,trim = 12 50 12 50,clip]{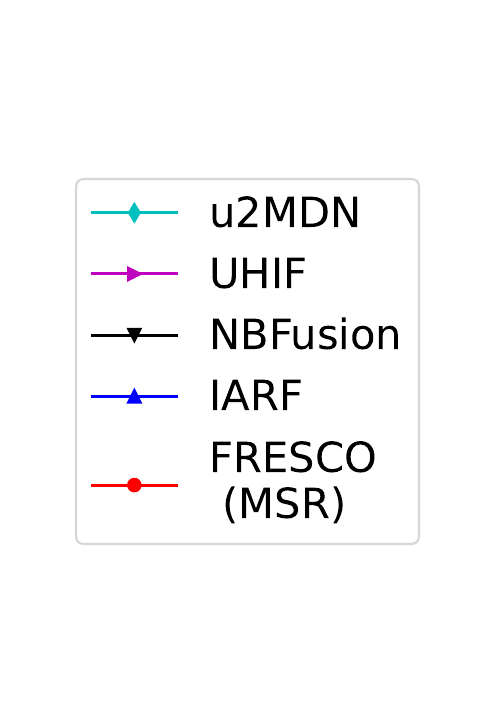}
        \captionsetup{labelformat=empty}
        \caption{}
    \end{subfigure}
    \colorbox{red!5}{
    \begin{subfigure}{0.165\linewidth}
        \centering
        \includegraphics[width=1\linewidth]{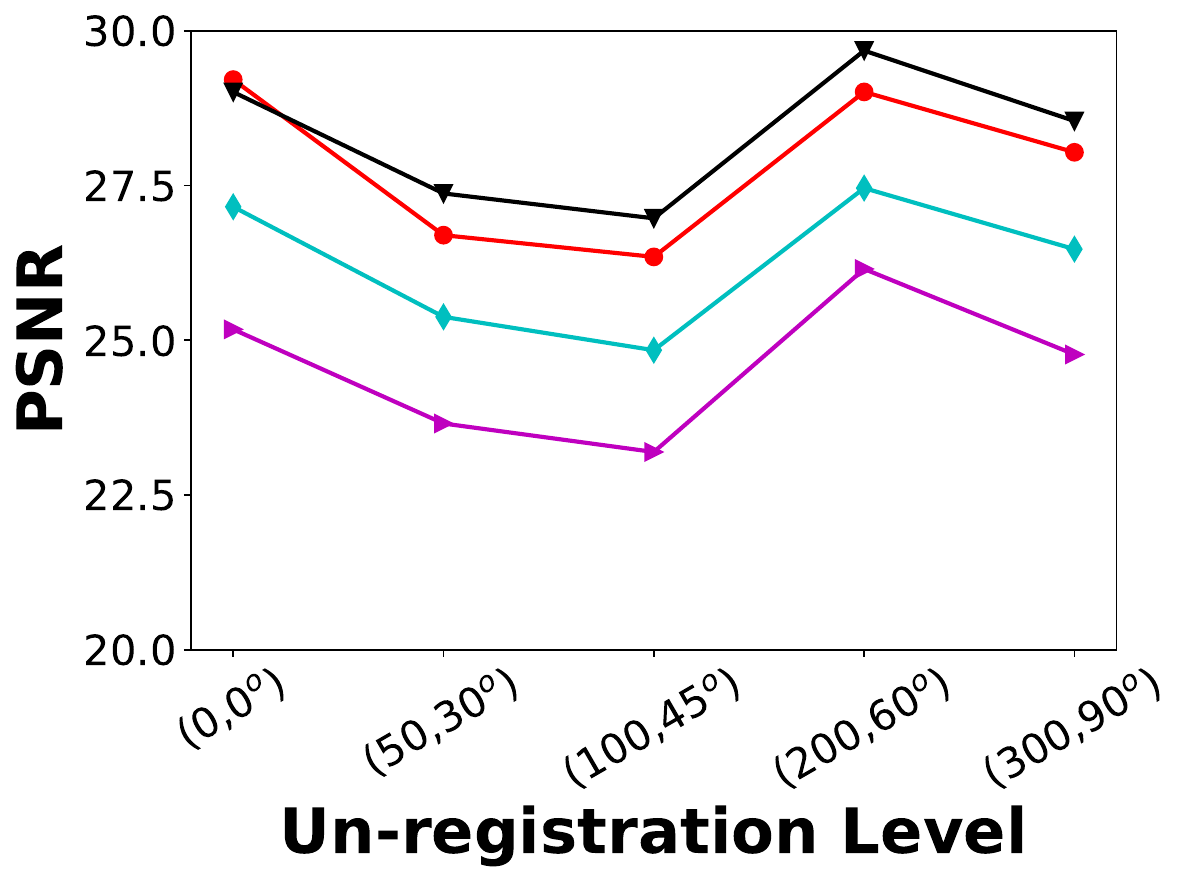}
        \caption{PSNR ($\uparrow$)}
    \end{subfigure}
    \begin{subfigure}{0.165\linewidth}
        \centering
        \includegraphics[width=\linewidth]{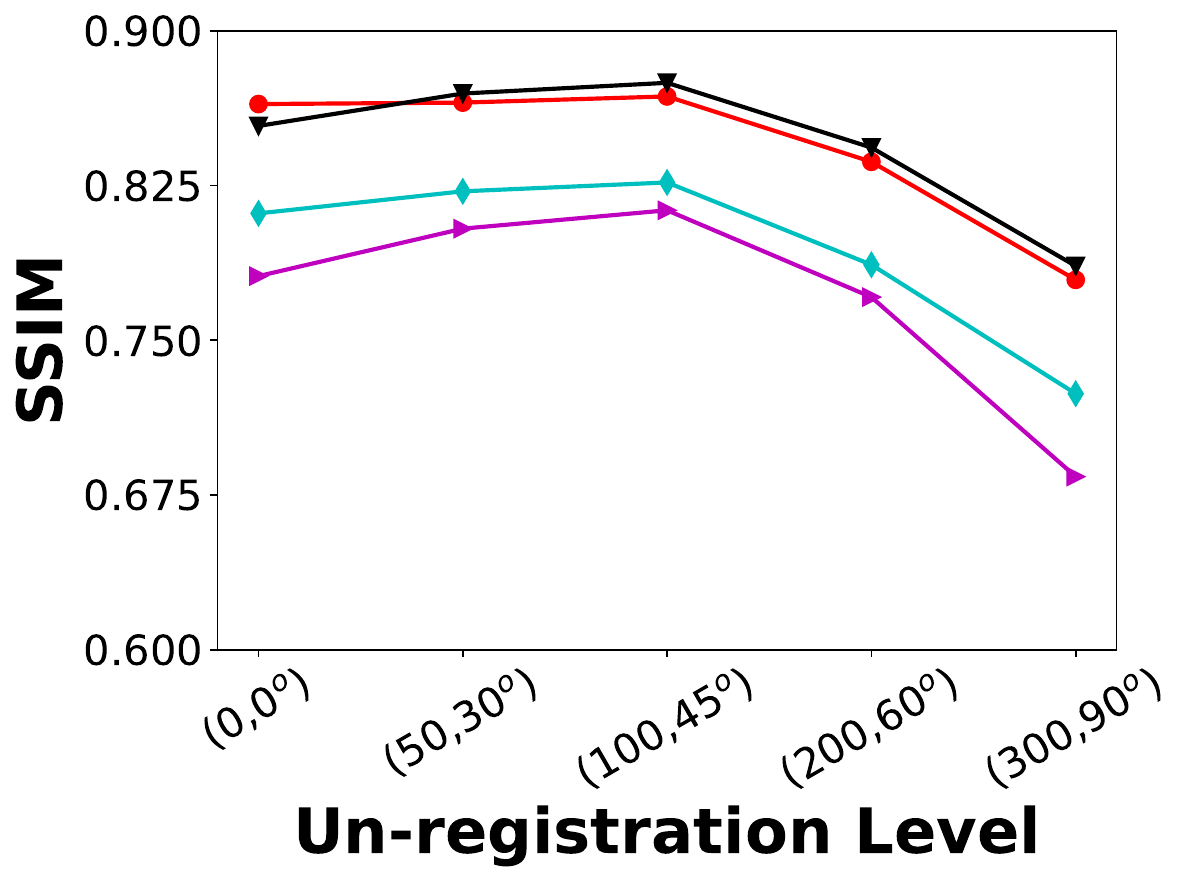}
        \caption{SSIM ($\uparrow$)}
    \end{subfigure}
    \begin{subfigure}{0.165\linewidth}
        \centering
        \includegraphics[width=\linewidth]{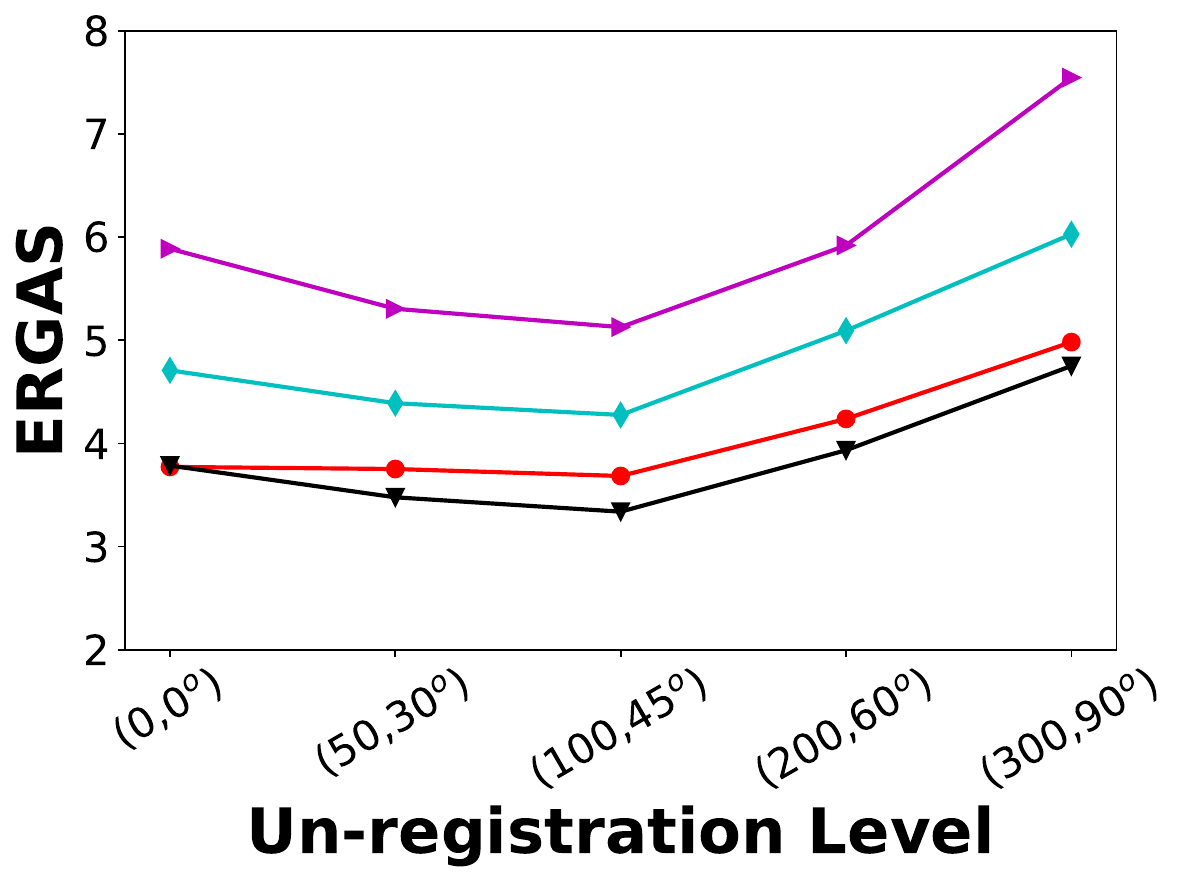}
        \caption{ERGAS ($\downarrow$)}
    \end{subfigure}
    \begin{subfigure}{0.165\linewidth}
        \centering
        \includegraphics[width=\linewidth]{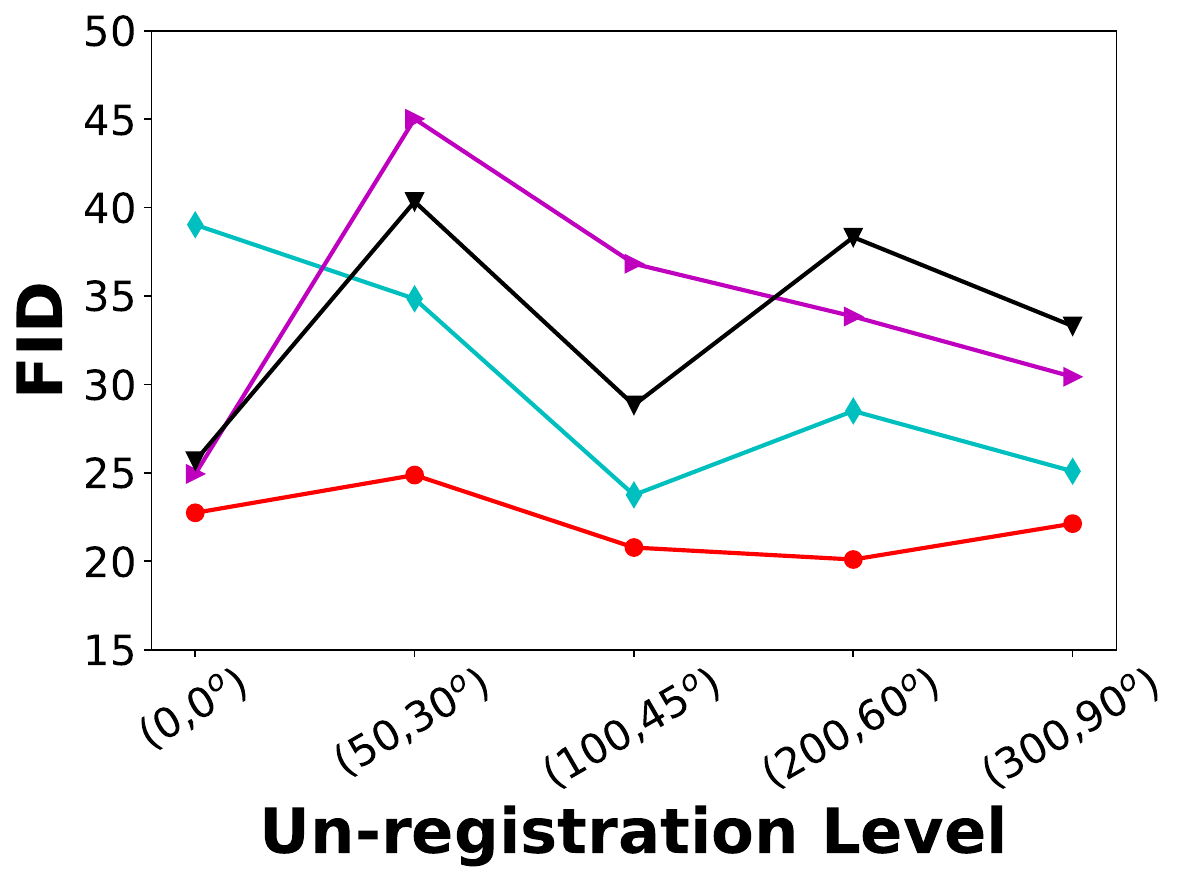}
        \caption{FID ($\downarrow$)}
    \end{subfigure}
    \begin{subfigure}{0.165\linewidth}
        \centering
        \includegraphics[width=\linewidth]{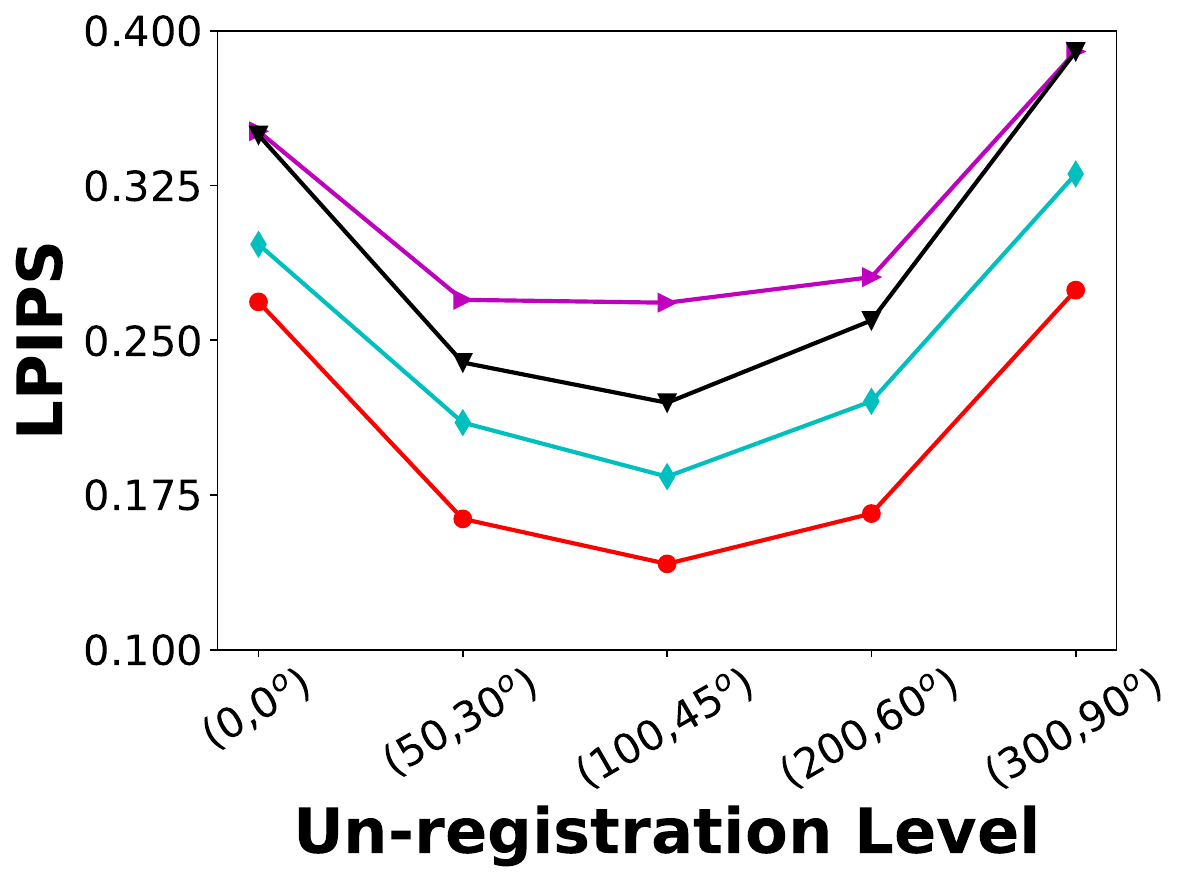}
        \caption{LPIPS ($\downarrow$)}
    \end{subfigure}}
    \begin{subfigure}{0.1\linewidth}
        \includegraphics[width=\linewidth,trim = 12 50 12 50,clip]{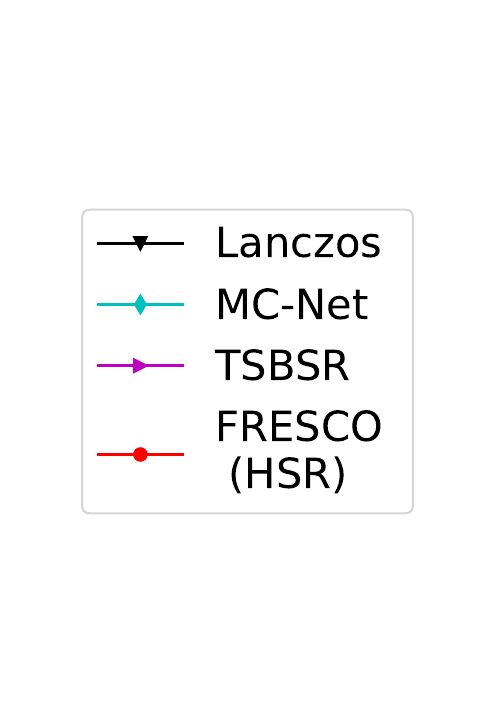}
        \captionsetup{labelformat=empty}
        \caption{}
    \end{subfigure}
    \caption{Performance comparison of MSR (top row) and HSR (bottom row) tasks under various spatial unregistration levels.}
    \label{fig:exp_unregistration_level}
\end{figure*}

Fig.~\ref{fig:exp_unregistration_level} shows performance under increasing spatial unregistration on the Indian Pines dataset.  
We fix $\tY\pM_{\mathrm{SRI}} = \tY_{\mathrm{Pines}}(101\!:\!500,201\!:\!600,:)$ and generate $\tY\pH_{\mathrm{SRI}} = {\sf rotate}_{\theta}(\tY_{\mathrm{Pines}}(101\!:\!500,201\!+\!t\!:\!600\!+\!t,:))$ using shifts $t$ and clockwise rotations $\theta$.  
We test five levels: $(t,\theta) \in \{(0,0^\circ), (50,30^\circ), (100,45^\circ), (200,60^\circ), (300,90^\circ)\}$.  
HSI is created via $5\times5$ Gaussian downsampling ($\sigma=0.3$), which is unknown to the methods.
The proposed method consistently outperforms baselines on both MSR and HSR tasks across all unregistration levels.  
For MSR, while methods like \texttt{IARF} and \texttt{u2MDN} perform well under low misalignment, their performance degrades significantly when $(t,\theta)\geq(50,30^\circ)$. Our method remains robust.  
For HSR, it matches or exceeds \texttt{Lanczos} in PSNR, SSIM, and ERGAS, and achieves the best FID and LPIPS in all cases, indicating strong perceptual quality under severe misalignment.

\begin{figure}[!ht]
    \centering
    \begin{subfigure}[t]{0.48\linewidth}
        \centering
        \includegraphics[width=\linewidth]{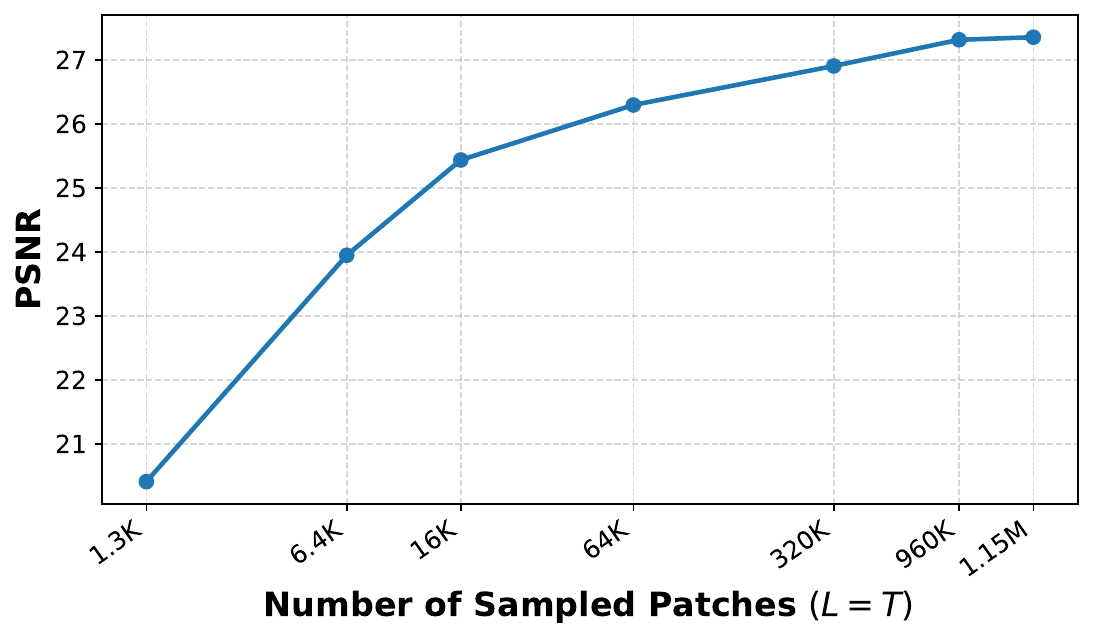}
        \label{fig:pavia_psnr_L}
    \end{subfigure}
    \begin{subfigure}[t]{0.48\linewidth}
        \centering
        \includegraphics[width=\linewidth]{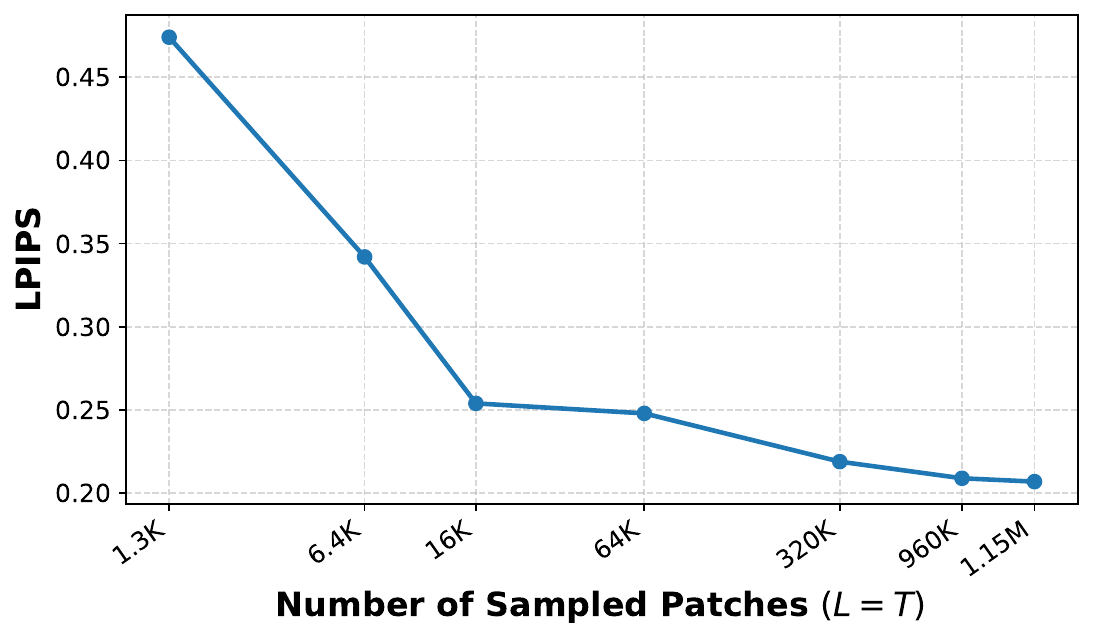}
        \label{fig:pavia_lpips_L}
    \end{subfigure}
    \caption{Performance under various $T(=L)$ on Pavia University.
    }
    \label{fig:pavia_LT}
\end{figure}

\begin{figure}[!ht]
    \centering
    \begin{subfigure}[t]{0.48\linewidth}
        \centering
        \includegraphics[width=\linewidth]{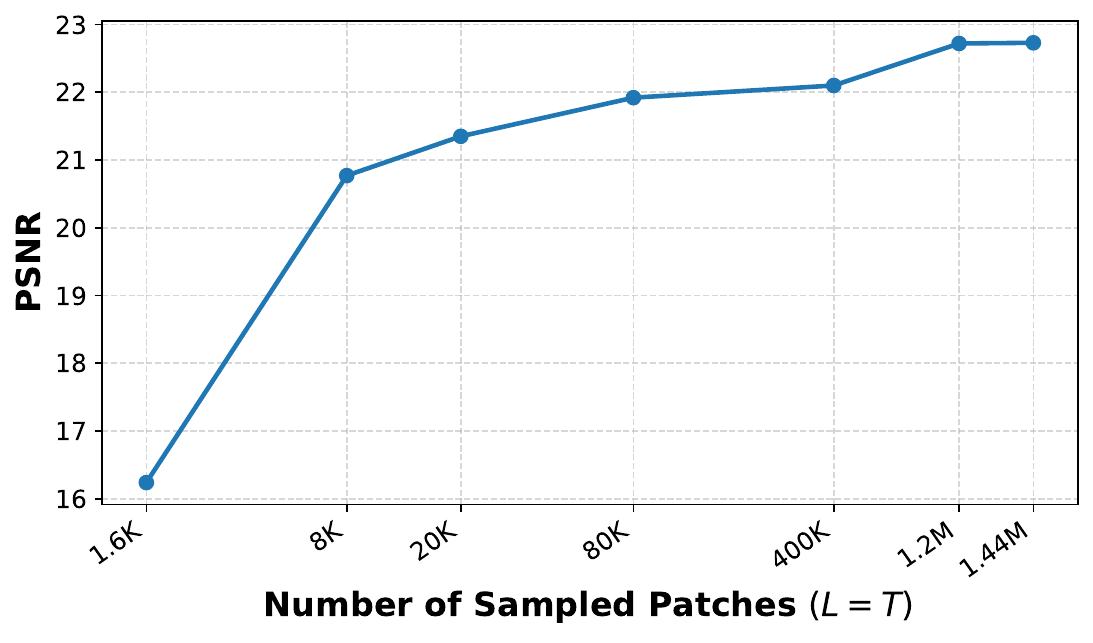}
        \label{fig:terrain_psnr_L}
    \end{subfigure}
    \begin{subfigure}[t]{0.48\linewidth}
        \centering
        \includegraphics[width=\linewidth]{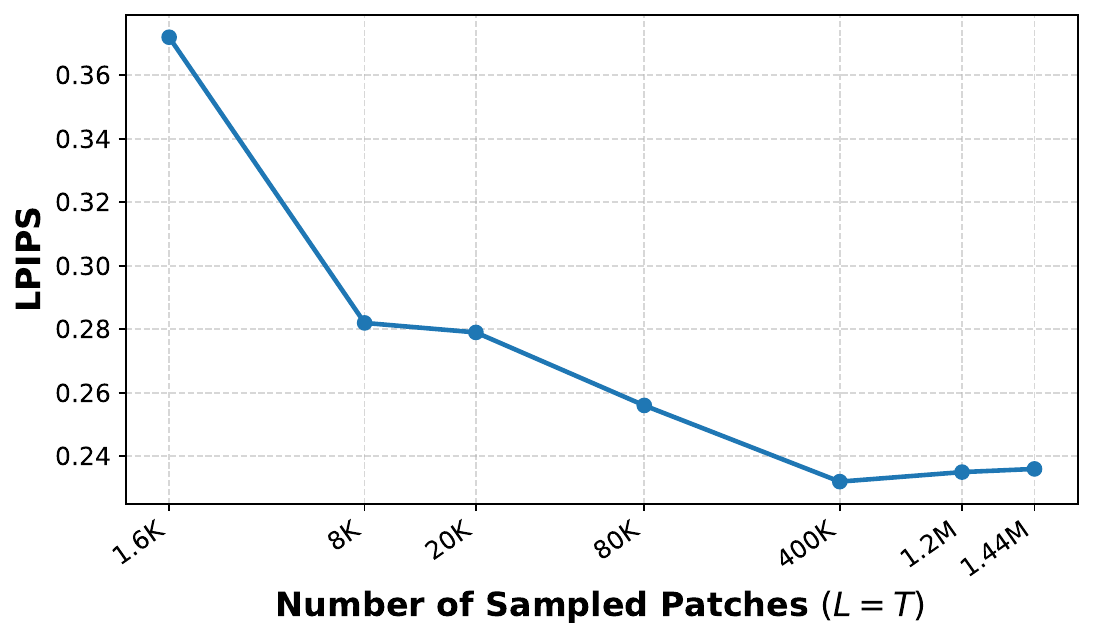}
        \label{fig:terrain_lpips_L}
    \end{subfigure}

    \caption{Performance under various $T(=L)$ on Terrain.}
    \label{fig:terrain_LT}
\end{figure}

\begin{figure}[!t]
    \centering
    \setlength{\tabcolsep}{0pt}
    \begin{subfigure}{0.32\linewidth}
        \centering
        \includegraphics[width=1\linewidth]{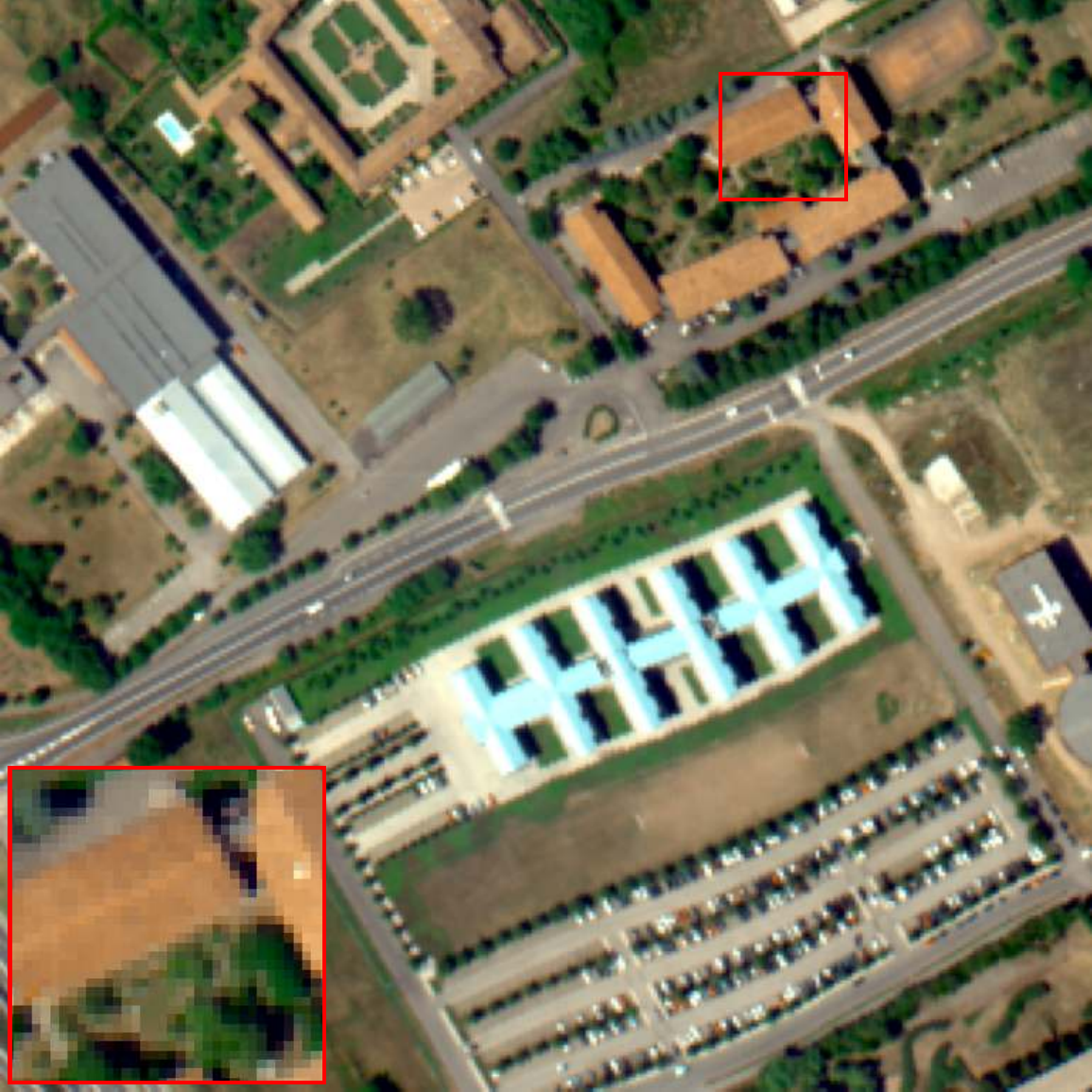}
        \captionsetup{labelformat=empty}
        \caption{\scriptsize \texttt{FRESCO} (\texttt{MSR}) with $\widehat{\bm P}\pM$}
    \end{subfigure}
    \begin{subfigure}{0.32\linewidth}
        \centering
        \includegraphics[width=\linewidth]{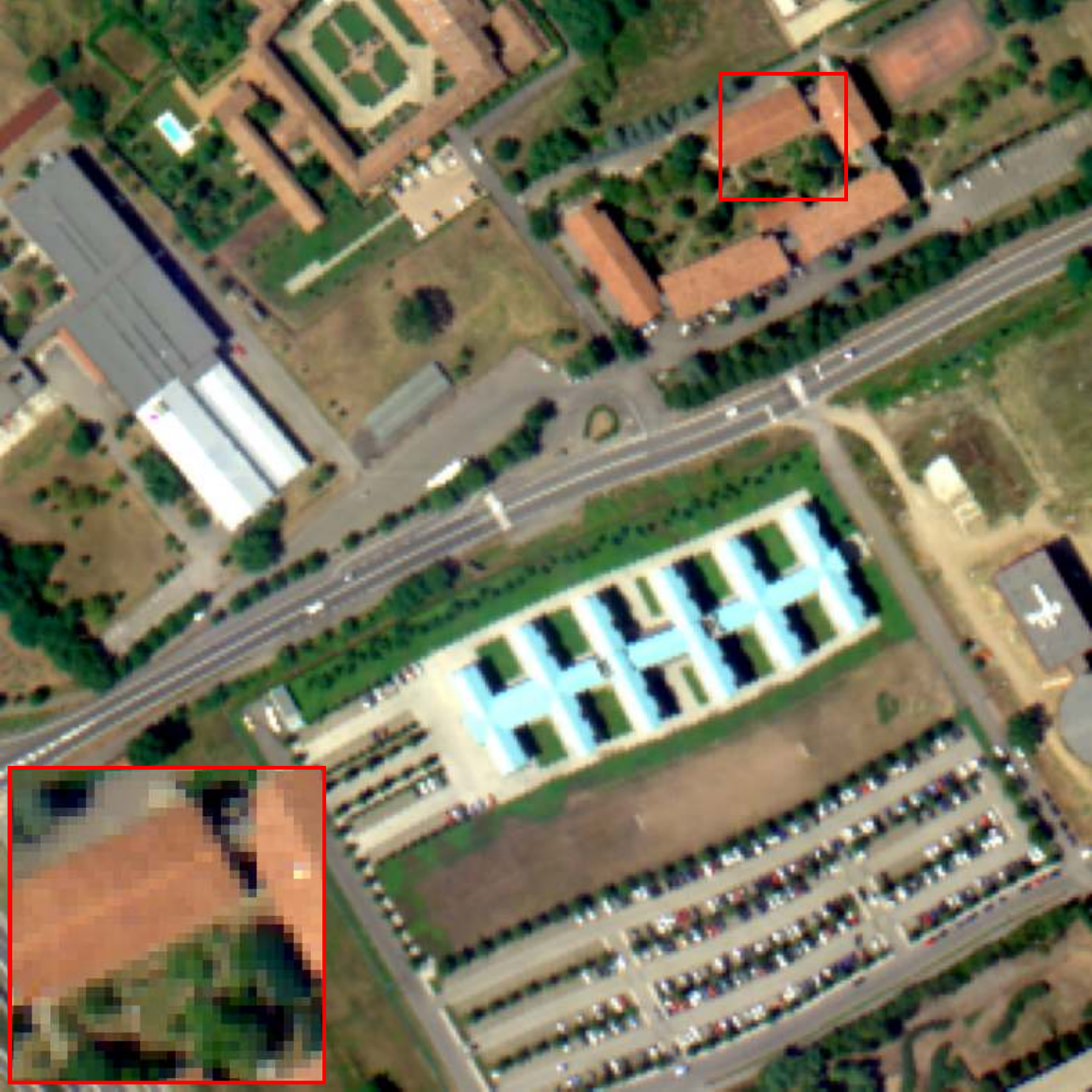}
        \captionsetup{labelformat=empty}
        \caption{\scriptsize \texttt{FRESCO} (\texttt{MSR}) with ${\bm P}\pM$}
    \end{subfigure}
    \begin{subfigure}{0.32\linewidth}
        \centering
        \includegraphics[width=\linewidth]{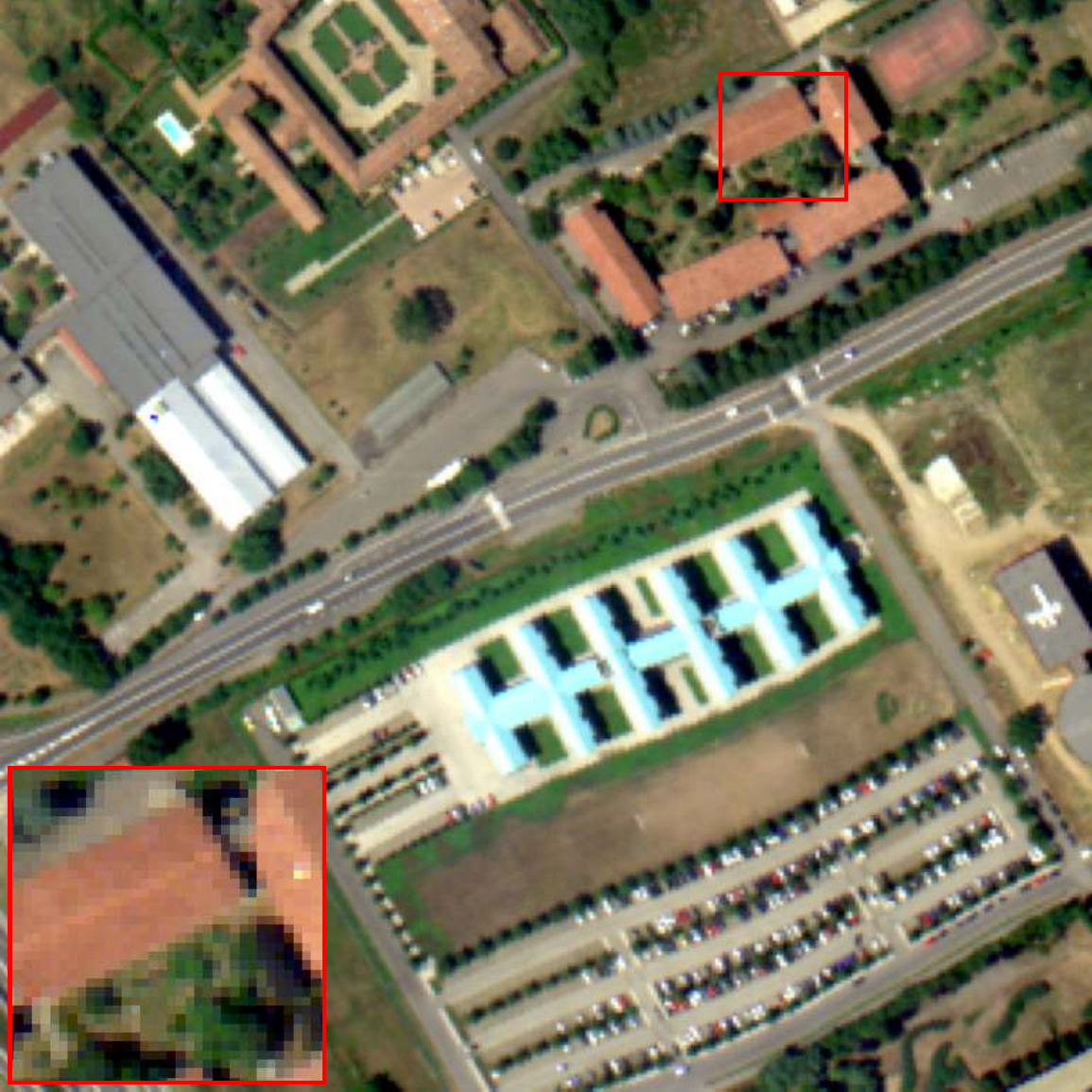}
        \captionsetup{labelformat=empty}
        \caption{\scriptsize Reference: $\tY\pM_{\rm SRI}$ }
    \end{subfigure}
    \\
    \vspace{0.5em}
    
    \begin{subfigure}{0.32\linewidth}
        \centering
        \includegraphics[width=\linewidth]{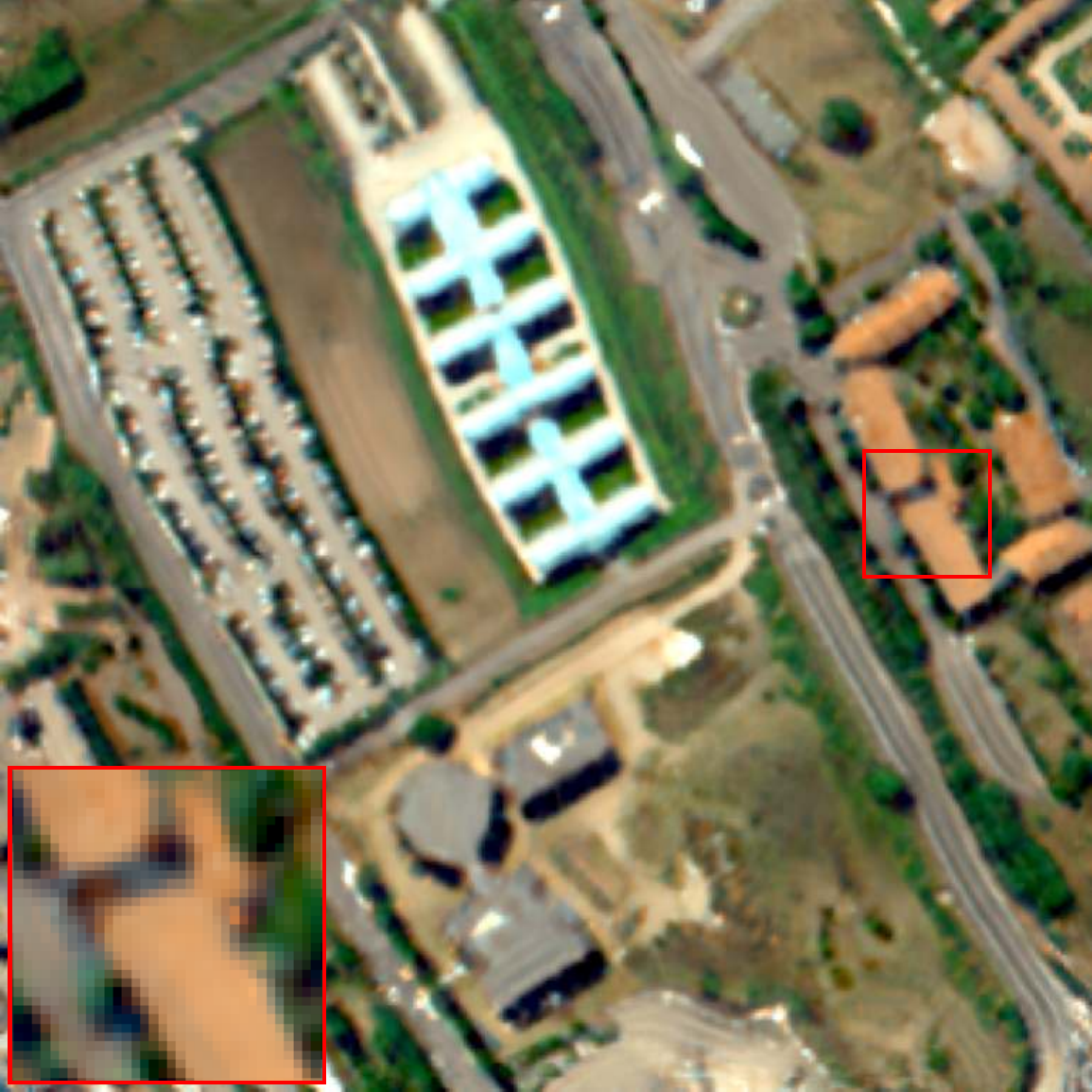}
        \captionsetup{labelformat=empty}
        \caption{\scriptsize \texttt{FRESCO} (\texttt{HSR}) with $\widehat{\bm P}\pM$}
    \end{subfigure}
    \begin{subfigure}{0.32\linewidth}
        \centering
        \includegraphics[width=\linewidth]{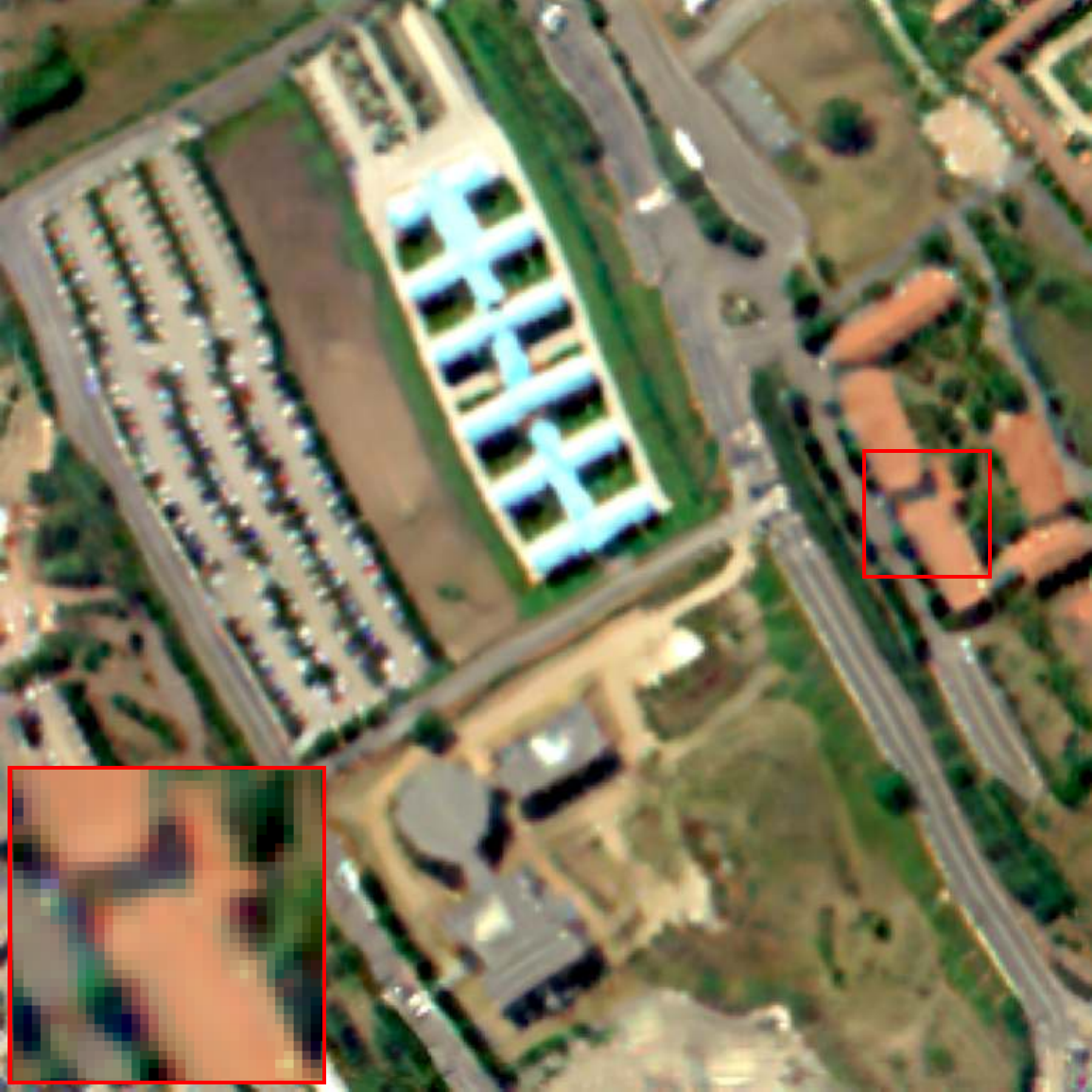}
        \captionsetup{labelformat=empty}
        \caption{\scriptsize \texttt{FRESCO} (\texttt{HSR}) with ${\bm P}\pM$}
    \end{subfigure}
    \begin{subfigure}{0.32\linewidth}
        \centering
        \includegraphics[width=\linewidth]{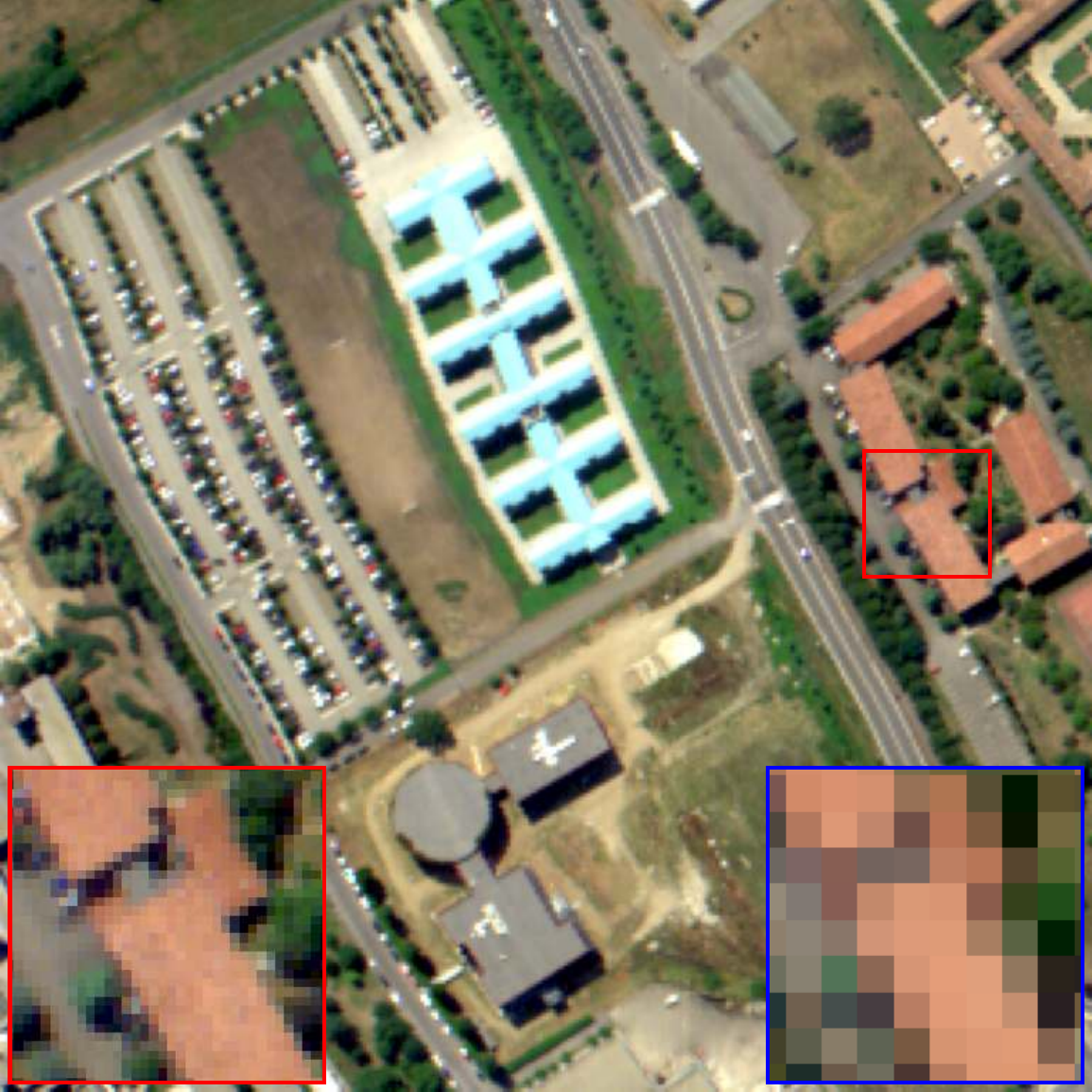}
        \captionsetup{labelformat=empty}
        \caption{\scriptsize Reference: $\tY\pH_{\rm SRI}$ }
    \end{subfigure}
    \caption{RGB renderings of from the recovered SRIs with estimated $\widehat{\bm P}\pM$ and the ground-truth $\bm P\pM$ (Pavia University); Gaussian kernel downsampling with $\sigma = 1.7$.}
    \label{fig:estimated_PM}
\end{figure}

Fig.~\ref{fig:estimated_PM} compares the performance of the proposed \texttt{FRESCO} method using the ground-truth $\bm P\pM$ and the estimated $\widehat{\bm P}\pM$ (obtained via Eq.~\eqref{eq:find_PM_unregistered_HMF}). 
The experiment uses the same Pavia dataset as in Table~\ref{tab:pavia_exp2}, assuming known sensor specifications---that is, the center wavelengths of the MSI bands are available.  
Accordingly, we define $\bm \varOmega$ such that $\bm P\pM(i,j) = 0$ whenever the $j$th HSI band falls outside the spectral response range of the $i$th MSI band.
Visually, the results using $\widehat{\bm P}\pM$ are nearly indistinguishable from those using the ground-truth $\bm P\pM$.  
Table~\ref{tab:estimated_PM} confirms this observation: the performance drop across all metrics is minimal, and \texttt{FRESCO} with $\widehat{\bm P}\pM$ still outperforms all baselines (cf. Table~\ref{tab:pavia_exp2}). 
Similar results are also observed from experiments on the Terrain data; see Table~\ref{tab:estimated_PM_terrain}.

Fig.~\ref{fig:pavia_LT} and Fig.~\ref{fig:terrain_LT} illustrate the performance of \texttt{FRESCO} as the number of sampled patches (i.e., $L$ and $T$ in \eqref{eq:diversified_gan_theta}) increases. Experiments on the Pavia University and Terrain datasets under Gaussian kernel downsampling show that reconstruction performance generally improves with more sampled patches, particularly in the low-sampling regime. The gain gradually saturates once the number of sampled patches reaches the order of $10^6$.

\begin{table}[!t]
\centering
\renewcommand{\arraystretch}{1.4} 
\small 
\caption{Performance of \texttt{FRESCO} Under Estimated $\widehat{\bm P}\pM$ and Ground-Truth $\bm P\pM$ on Pavia University.}
\begin{adjustbox}{max width=\linewidth}
\begin{tabular}{|c|c|cc|cc|}
\hline
\multirow{2}{*}{\begin{tabular}[c]{@{}c@{}} Downsampling\\ Methods\end{tabular}} & \multirow{2}{*}{Metrics} & \multicolumn{2}{c|}{\texttt{FRESCO} (\texttt{MSR})} & \multicolumn{2}{c|}{\texttt{FRESCO} (\texttt{HSR})} \\ \cline{3-6} 
 & & $\widehat{\bm P}\pM$ & ${\bm P}\pM$  &  $\widehat{\bm P}\pM$ & \multicolumn{1}{c|}{${\bm P}\pM$} \\ \hline
\multirow{5}{*}{\begin{tabular}[c]{@{}c@{}}Gaussian Kernel\\ Downsampling\end{tabular}} 
 & PSNR ($\uparrow$)& 41.07 & 43.59 & 27.06 & 27.37\\ 
 & SSIM ($\uparrow$) & 0.986 & 0.989  & 0.738 & 0.750 \\  
 & ERGAS ($\downarrow$)& 1.443 & 1.217 & 6.569 & 6.424\\
 & FID ($\downarrow$)& 12.05 & 9.53 & 49.72 & 46.17\\
 & LPIPS ($\downarrow$)& 0.013 & 0.003 & 0.211 & 0.201\\ \hline 
\multirow{5}{*}{\begin{tabular}[c]{@{}c@{}}Nearest Neighbor \\ Downsampling\end{tabular}} 
 & PSNR ($\uparrow$) & 41.10 & 43.60 & 25.55 & 25.69\\ 
 & SSIM ($\uparrow$) & 0.986 & 0.990 & 0.658 & 0.671\\ 
 & ERGAS ($\downarrow$)& 1.436 & 1.216 & 7.813 & 7.804\\ 
 & FID  ($\downarrow$)& 13.25 & 10.12 & 54.89 & 51.67\\
  & LPIPS ($\downarrow$)& 0.013 & 0.003 & 0.272 & 0.258\\ \hline
\multirow{5}{*}{\begin{tabular}[c]{@{}c@{}} Uniform \\ Downsampling\end{tabular}} 
 & PSNR ($\uparrow$)& 41.02 & 43.59 & 23.42 & 23.71\\
 & SSIM ($\uparrow$) & 0.986 & 0.990 & 0.591 & 0.597\\
 & ERGAS ($\downarrow$)& 1.446 & 1.217 & 9.959 & 9.860\\
 & FID ($\downarrow$)& 12.23 & 9.77 & 56.25 & 52.76\\
 & LPIPS ($\downarrow$)& 0.014 & 0.003 & 0.275 & 0.264\\ \hline
\end{tabular}
\end{adjustbox}
\label{tab:estimated_PM}
\end{table}

\begin{table}[!t]
\centering
\renewcommand{\arraystretch}{1.4} 
\small 
\caption{Performance of \texttt{FRESCO} Under Estimated $\widehat{\bm P}\pM$ and Ground-Truth $\bm P\pM$ on Terrain.}
\begin{adjustbox}{max width=\linewidth}
\begin{tabular}{|c|c|cc|cc|}
\hline
\multirow{2}{*}{\begin{tabular}[c]{@{}c@{}} Downsampling\\ Methods\end{tabular}} & \multirow{2}{*}{Metrics} & \multicolumn{2}{c|}{\texttt{FRESCO} (\texttt{MSR})} & \multicolumn{2}{c|}{\texttt{FRESCO} (\texttt{HSR})} \\ \cline{3-6} 
 & & $\widehat{\bm P}\pM$ & ${\bm P}\pM$  &  $\widehat{\bm P}\pM$ & \multicolumn{1}{c|}{${\bm P}\pM$} \\ \hline
\multirow{5}{*}{\begin{tabular}[c]{@{}c@{}}Gaussian Kernel\\ Downsampling\end{tabular}} 
 & PSNR ($\uparrow$)& 39.83 & 44.91 & 22.74 & 23.64\\ 
 & SSIM ($\uparrow$) & 0.996 & 0.998 & 0.726 & 0.731 \\  
 & ERGAS ($\downarrow$)& 0.982 & 0.582 & 6.196 & 5.583 \\
 & FID ($\downarrow$)& 4.09 & 4.02 & 20.06 & 19.65\\
 & LPIPS ($\downarrow$)& 0.010 & 0.006 & 0.224 & 0.215 \\ \hline 
\multirow{5}{*}{\begin{tabular}[c]{@{}c@{}}Nearest Neighbor \\ Downsampling\end{tabular}} 
 & PSNR ($\uparrow$)& 40.12 & 44.57 & 21.31 & 21.70 \\ 
 & SSIM ($\uparrow$) & 0.996 & 0.998 & 0.653 & 0.659\\ 
 & ERGAS ($\downarrow$)& 0.967 & 0.604 & 7.315 & 6.997 \\ 
 & FID  ($\downarrow$)& 4.10 & 4.01 & 21.43 & 20.09\\
  & LPIPS ($\downarrow$)& 0.007 & 0.006 & 0.240 & 0.238 \\ \hline
\multirow{5}{*}{\begin{tabular}[c]{@{}c@{}} Uniform \\ Downsampling\end{tabular}} 
 & PSNR ($\uparrow$)& 39.33 & 44.59 & 20.81 & 20.94\\
 & SSIM ($\uparrow$) & 0.997 & 0.998 & 0.602 & 0.615 \\
 & ERGAS ($\downarrow$)& 0.922 & 0.601 & 8.019 & 7.672 \\
 & FID ($\downarrow$)& 4.34 & 4.20 & 23.62 & 22.76 \\
 & LPIPS ($\downarrow$)& 0.009 & 0.007 & 0.253 & 0.243 \\ \hline
\end{tabular}
\end{adjustbox}
\label{tab:estimated_PM_terrain}
\end{table}

\subsection{Ablation Study}\label{sec:ablation}
Table~\ref{tab:msr_ablation} and Table~\ref{tab:hsr_ablation} report results from ablation studies for the MSR and HSR settings, respectively, where the Pavia University dataset under Gaussian kernel downsampling is used.

For MSR, the fitting term $\mathcal{L}^{\rm fitting}$ is always enabled, and we evaluate the effect of each additional regularization term.
For HSR, the distribution matching loss $\mathcal{L}^{\rm DM}$ is always kept active, while the invertiability regularization $\mathcal{L}^{\rm inv}$ and the scaling regularization $\mathcal{L}^{\rm scale}$ are ablated individually.
In both tables, a checkmark indicates that the corresponding loss term is included in training, whereas a cross denotes its removal.

The results show that enabling all relevant loss components consistently leads to the best overall performance across almost all evaluation metrics, confirming the effectiveness of the proposed loss design.

\subsection{Run time Analysis}
Table~\ref{tab:runtime_analysis} reports the runtime comparison on the Pavia University dataset under Gaussian-kernel downsampling. We note that the compared methods employ different training and implementation settings. In particular, \texttt{MC-Net}~\cite{li2020mixed} is a supervised method trained offline once on a dataset and then applied to new test images through fast inference. Therefore, Table~\ref{tab:runtime_analysis} reports its offline training time for reference. For the other unsupervised or optimization-based methods, the reported runtime corresponds to the total per-pair processing time required for a given HSI-MSI pair.
All runtimes were measured on a server equipped with two Intel Xeon Gold 6148 CPUs @ 2.40GHz and an NVIDIA Tesla V100-SXM3 GPU with 32GB memory.

\begin{table}[!t]
\centering
\renewcommand{\arraystretch}{1.25}
\caption{MSR ablation study on Pavia University.}
\begin{adjustbox}{max width=\linewidth}
\begin{tabular}{|c|c|c|c|c|c|c|c|}
\hline
\multicolumn{3}{|c|}{Experiment Setup} & \multicolumn{5}{c|}{Results} \\ \hline
\begin{tabular}[c]{@{}c@{}}$\mathcal{L}^{\rm LR}$\end{tabular}
&
\begin{tabular}[c]{@{}c@{}}$\mathcal{L}^{\rm TV}$\end{tabular}
&
\begin{tabular}[c]{@{}c@{}}$\mathcal{L}^{\rm sto}$\end{tabular}
& PSNR ($\uparrow$) & SSIM ($\uparrow$) & ERGAS ($\downarrow$) & FID ($\downarrow$) & LPIPS ($\downarrow$)\\ \hline
\cmark & \cmark & \cmark  & \textbf{43.59} & \textbf{0.989} & 1.22 & \textbf{9.53} & \textbf{0.003} \\ \hline
\xmark & \cmark & \cmark  & 43.48 & 0.989 & 1.09 & 10.27 & 0.004 \\ \hline
\cmark & \xmark & \cmark  & 43.28 & 0.980 & 1.12 & 9.89 & 0.003 \\ \hline
\cmark & \cmark & \xmark  & 43.42 & 0.989 & \textbf{1.09} & 10.18 & 0.004 \\ \hline
\end{tabular}
\label{tab:msr_ablation}
\end{adjustbox}
\end{table}

\begin{table}[!t]
\centering
\renewcommand{\arraystretch}{1.25}
\caption{HSR ablation study on  Pavia University.}
\begin{adjustbox}{max width=\linewidth}
\begin{tabular}{|c|c|c|c|c|c|c|}
\hline
\multicolumn{2}{|c|}{Experiment Setup} & \multicolumn{5}{c|}{Results} \\ \hline
\begin{tabular}[c]{@{}c@{}}$\mathcal{L}^{\rm inv}$\end{tabular}
&
\begin{tabular}[c]{@{}c@{}}$\mathcal{L}^{\rm scale}$\end{tabular}
& PSNR ($\uparrow$) & SSIM ($\uparrow$) & ERGAS ($\downarrow$) & FID ($\downarrow$) & LPIPS ($\downarrow$) \\ \hline

\cmark & \cmark & \textbf{27.37} & \textbf{0.751}  & \textbf{6.42}  & \textbf{46.17} & \textbf{0.201} \\ \hline
\xmark & \cmark & 25.04 & 0.621 & 8.38  & 51.54 & 0.245 \\ \hline
\cmark & \xmark & 22.72 & 0.443  & 13.63 & 67.12 & 0.297 \\ \hline
\end{tabular}
\label{tab:hsr_ablation}
\end{adjustbox}
\end{table}

\begin{table}[!ht]
\centering
\caption{Runtime comparison on the Pavia University dataset.}
\label{tab:runtime_analysis}
\renewcommand{\arraystretch}{1.2}
\setlength{\tabcolsep}{3.2pt}
\resizebox{.8\columnwidth}{!}{
\begin{tabular}{|c|c|c|c|c|}
\hline
Stage 
& Method 
& Setting 
& Device 
& Runtime (min) \\
\hline
\multirow{9}{*}{\begin{tabular}[c]{@{}c@{}}MSR\end{tabular}}
& \texttt{UHIF}          
& \begin{tabular}[c]{@{}c@{}}Per-pair\\optimization\end{tabular}
& CPU & 0.32 \\
\cline{2-5}
& \texttt{u2MDN}         
& \begin{tabular}[c]{@{}c@{}}Per-pair\\optimization\end{tabular}
& GPU & 28.49 \\
\cline{2-5}
& \texttt{IARF}          
& \begin{tabular}[c]{@{}c@{}}Per-pair\\optimization\end{tabular}
& CPU & 1.65 \\
\cline{2-5}
& \texttt{NBFusion}      
& \begin{tabular}[c]{@{}c@{}}Per-pair\\optimization\end{tabular}
& CPU & 6.83 \\
\cline{2-5}
& \begin{tabular}[c]{@{}c@{}}\texttt{FRESCO}\\\texttt{(MSR)}\end{tabular}
& \begin{tabular}[c]{@{}c@{}}Per-pair\\optimization\end{tabular}
& CPU & 2.56 \\
\hline

\multirow{7}{*}{\begin{tabular}[c]{@{}c@{}}HSR\end{tabular}}
& \texttt{Lanczos}       
& \begin{tabular}[c]{@{}c@{}}Direct\\interpolation\end{tabular}
& CPU & 0.007 \\
\cline{2-5}
& \texttt{MC-Net}        
& \begin{tabular}[c]{@{}c@{}}Offline\\model training\end{tabular}
& GPU & 487.80 \\
\cline{2-5}
& \texttt{TSBSR}         
& \begin{tabular}[c]{@{}c@{}}Per-pair\\optimization\end{tabular}
& GPU & 204.60 \\
\cline{2-5}
& \begin{tabular}[c]{@{}c@{}}\texttt{FRESCO}\\\texttt{(HSR)}\end{tabular}
& \begin{tabular}[c]{@{}c@{}}Per-pair\\optimization\end{tabular}
& GPU & 142.20 \\
\hline
\end{tabular}
}
\end{table}

\subsection{Real-Data Experiment}

\begin{figure*}[!t]
    \centering
    \setlength{\tabcolsep}{0pt}
    \begin{subfigure}{0.15\linewidth}
    \centering
    \includegraphics[width=1\linewidth]{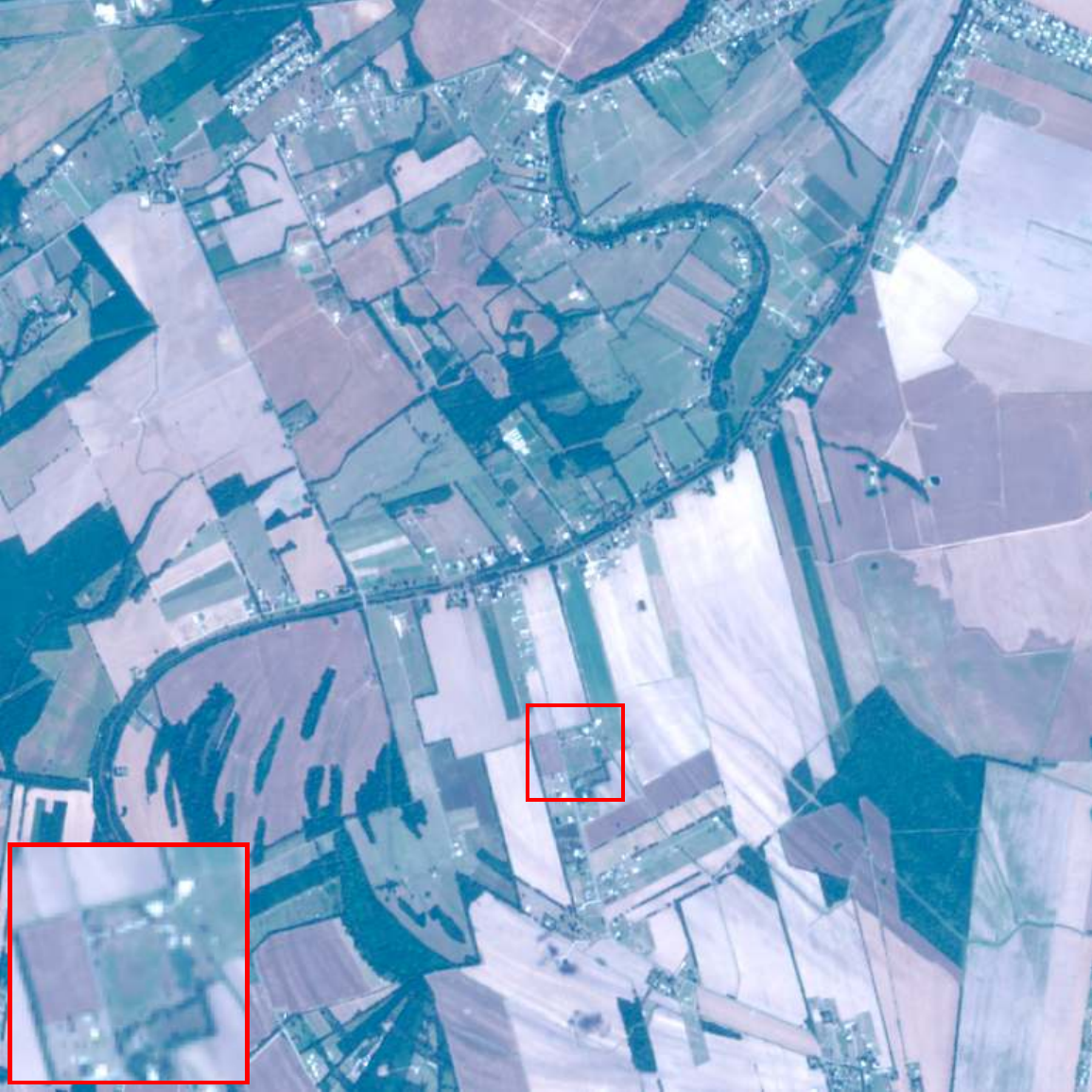}
    \captionsetup{labelformat=empty}
    \caption{MSI}
    \end{subfigure}
    \begin{subfigure}{0.15\linewidth}
        \centering
        \includegraphics[width=1\linewidth]{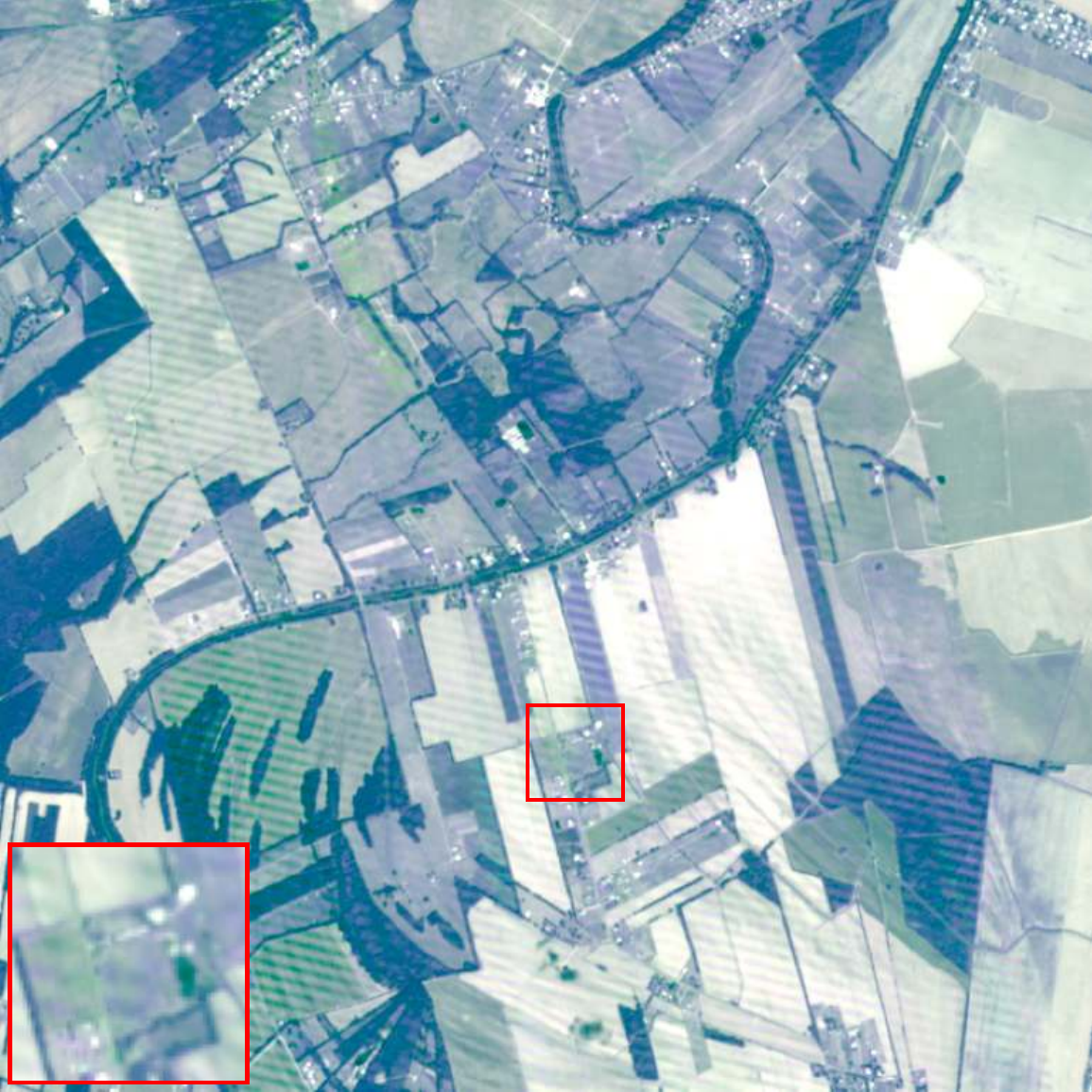}
        \captionsetup{labelformat=empty}
        \caption{\texttt{UHIF-RIM} \cite{ying2021unaligned}}
    \end{subfigure}
    \begin{subfigure}{0.15\linewidth}
        \centering
        \includegraphics[width=\linewidth]{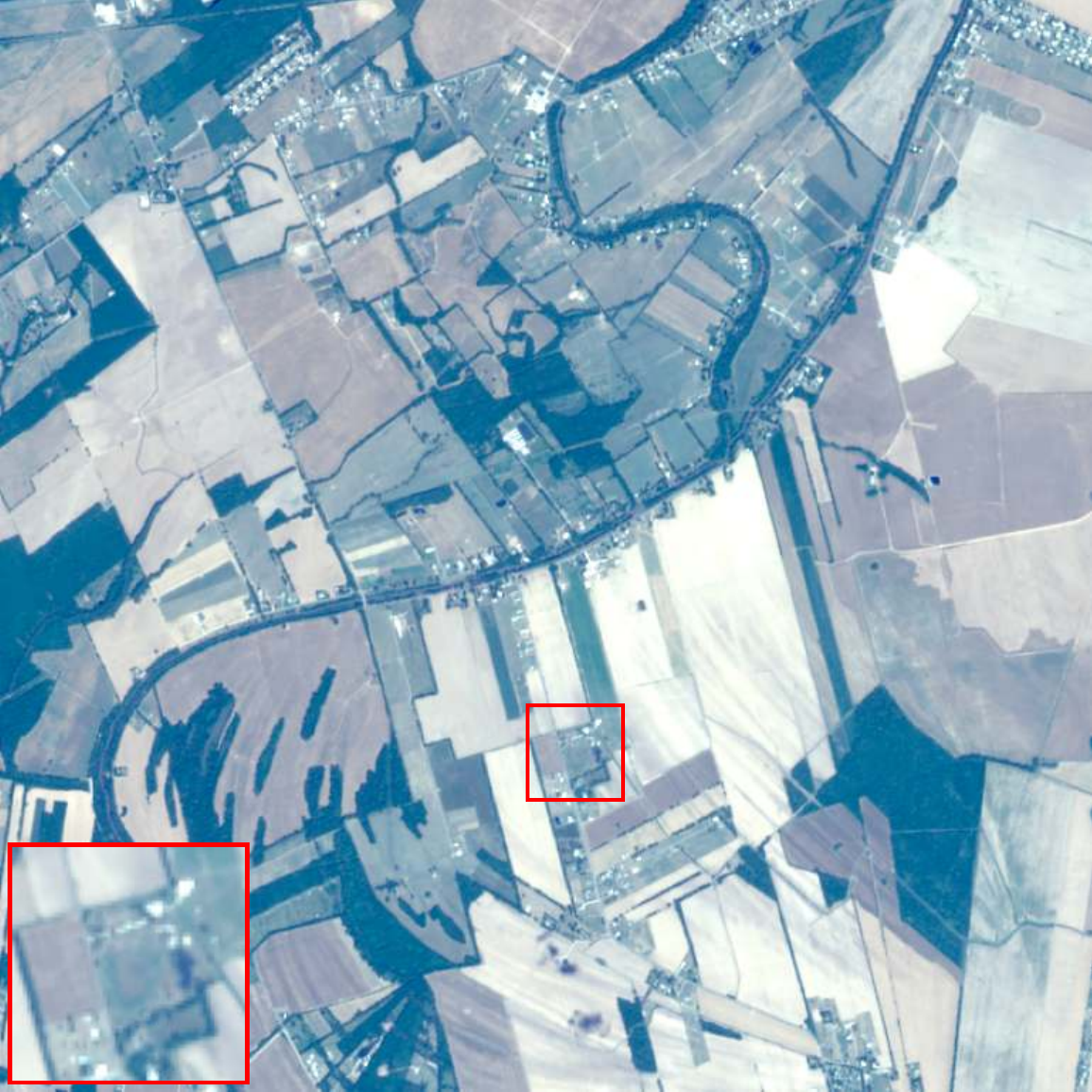}
        \captionsetup{labelformat=empty}
        \caption{\texttt{u2MDN} \cite{qu2021unsupervised}}
    \end{subfigure}
    \begin{subfigure}{0.15\linewidth}
        \centering
        \includegraphics[width=\linewidth]{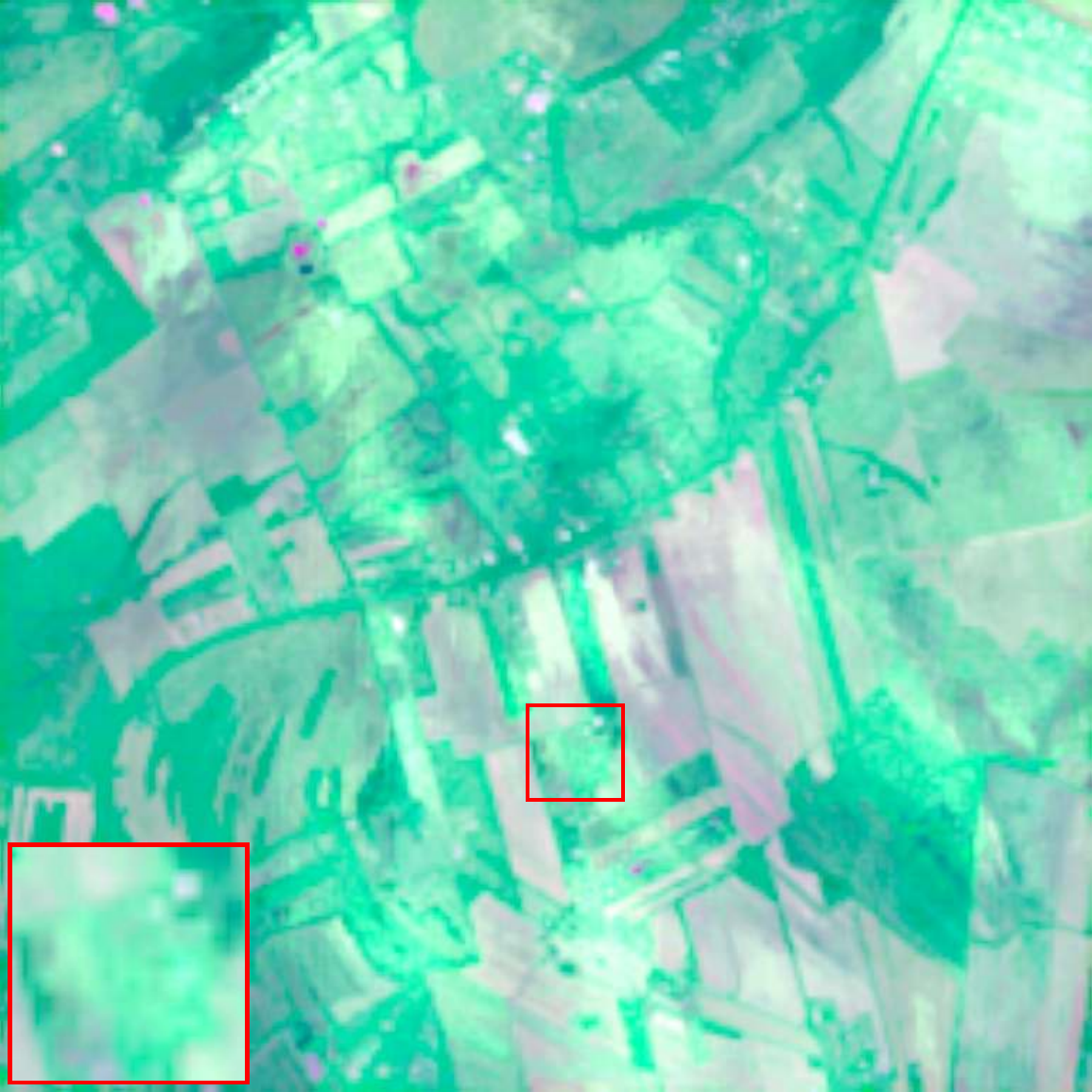}
        \captionsetup{labelformat=empty}
        \caption{\texttt{IARF} \cite{zhou2019integrated}}
    \end{subfigure}
    \begin{subfigure}{0.15\linewidth}
        \centering
        \includegraphics[width=\linewidth]{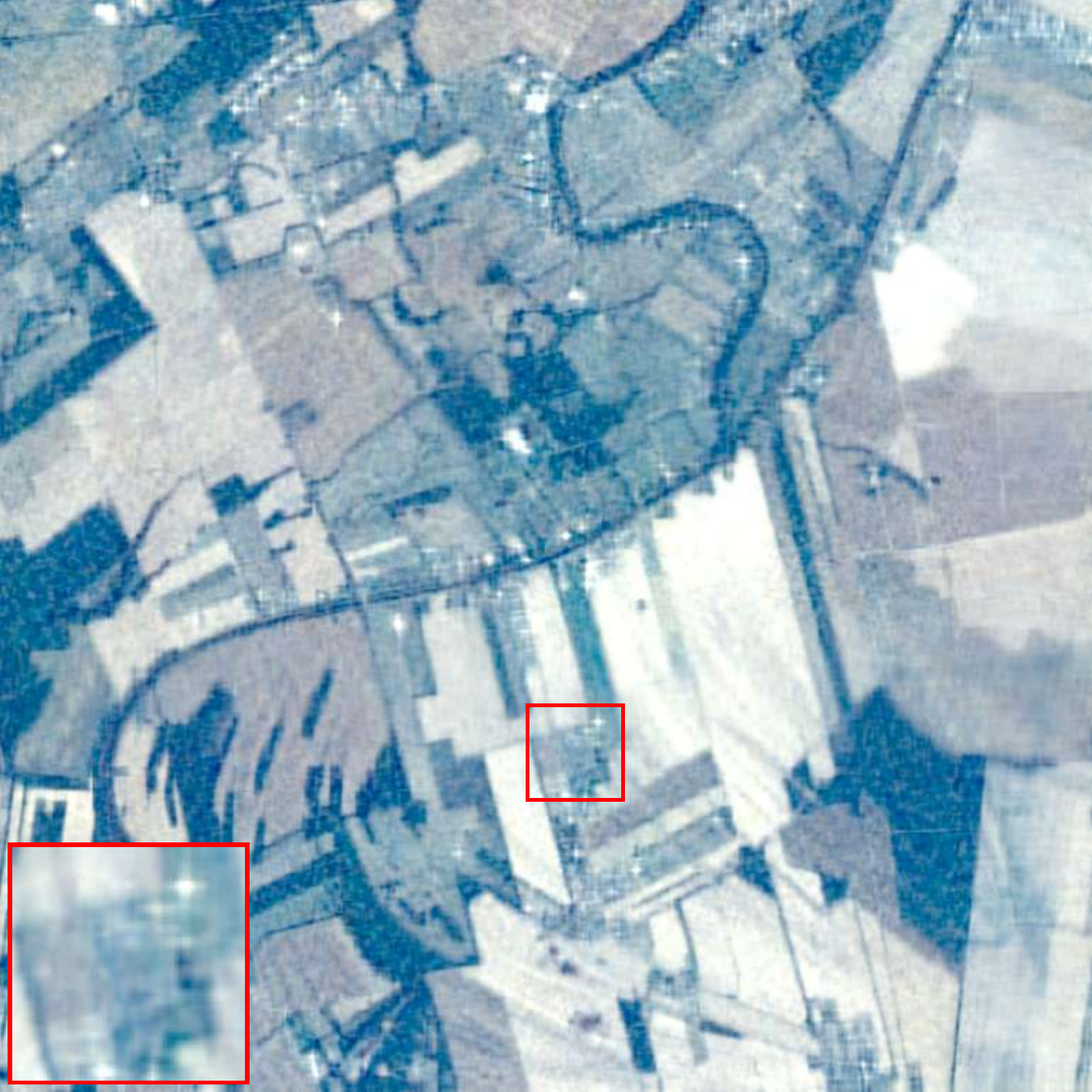}
        \captionsetup{labelformat=empty}
        \caption{\texttt{NBFusion} {\scriptsize \cite{chen2017normalized,kanatsoulis2018hyperspectral}}}
    \end{subfigure}
    \begin{subfigure}{0.15\linewidth}
        \centering
        \includegraphics[width=\linewidth]{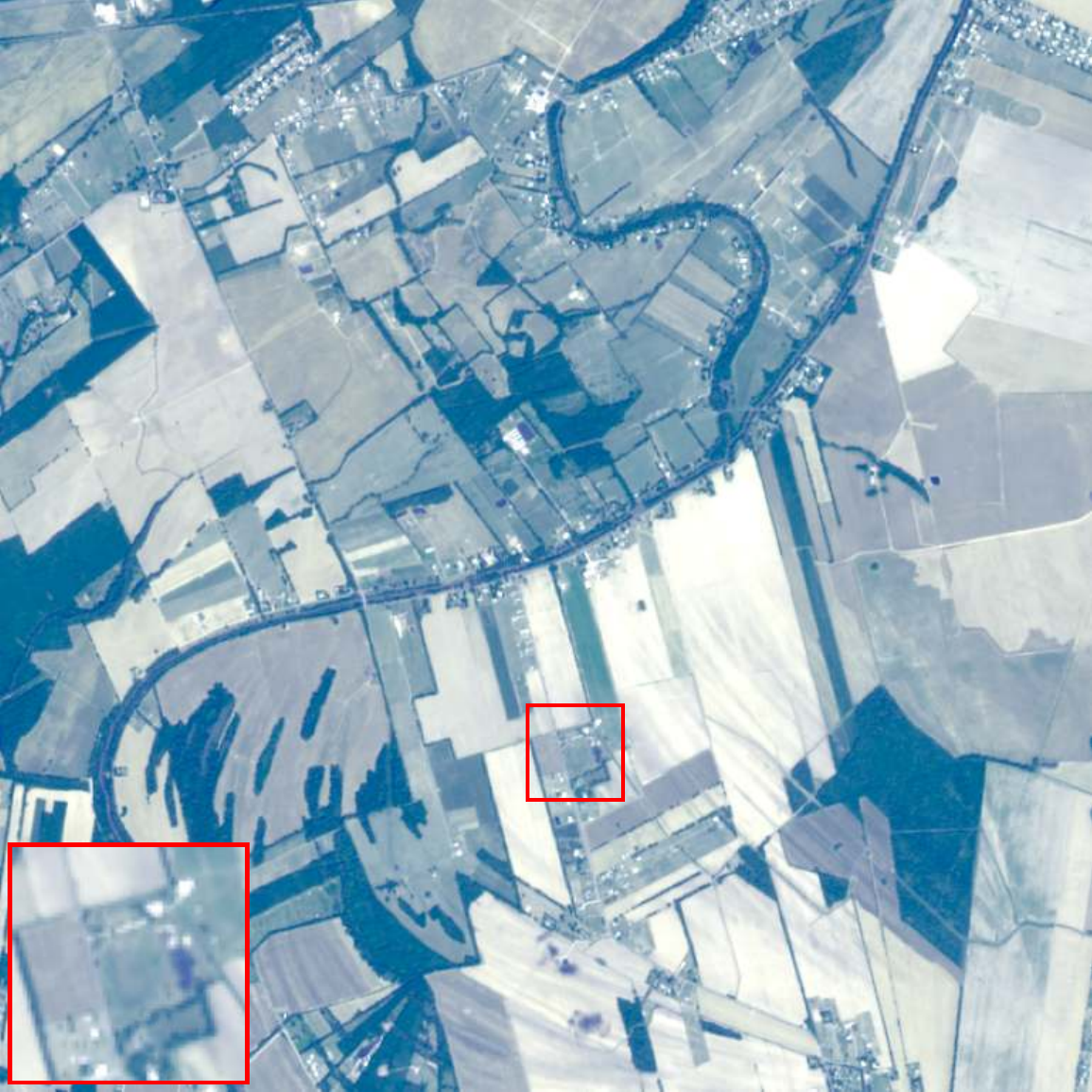}
        \captionsetup{labelformat=empty}
        \caption{\texttt{FRESCO} (\texttt{MSR})}
    \end{subfigure}
    \\
    \vspace{0.2cm}

    \begin{subfigure}{0.15\linewidth}
        \centering
        \includegraphics[width=\linewidth]{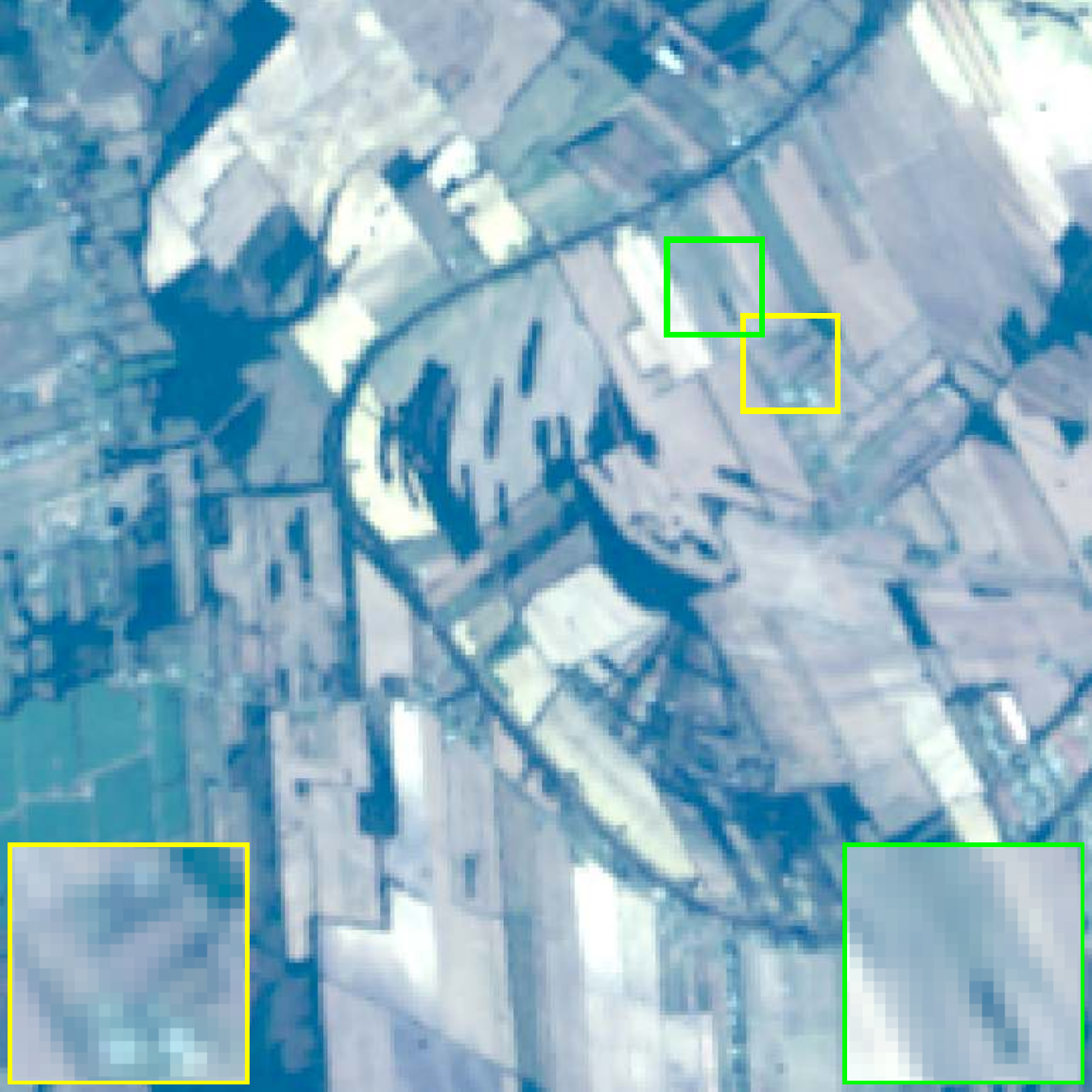}
        \captionsetup{labelformat=empty}
        \caption{HSI}
    \end{subfigure}
    \begin{subfigure}{0.15\linewidth}
        \centering
        \includegraphics[width=\linewidth]{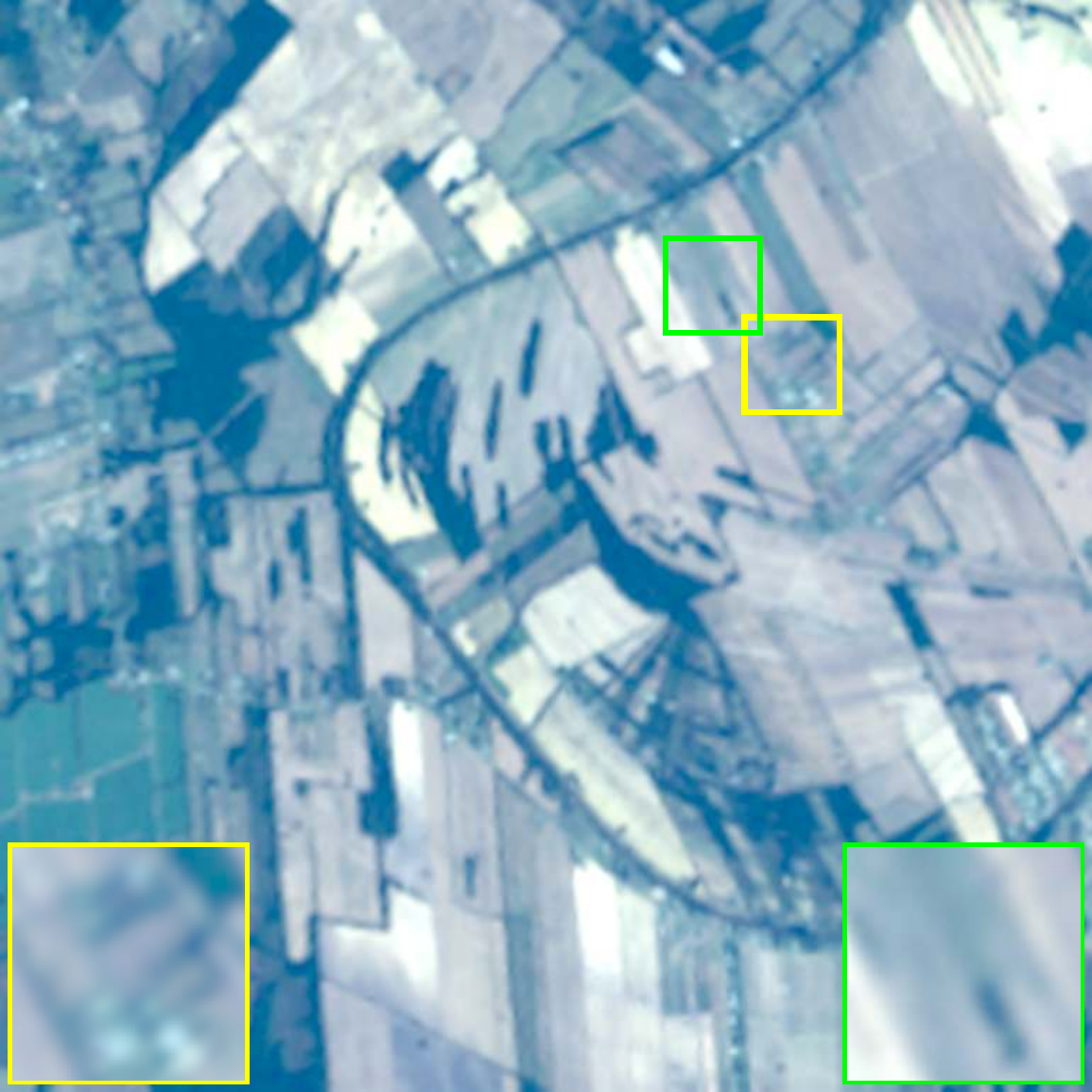}
        \captionsetup{labelformat=empty}
        \caption{\texttt{Lanczos} \cite{duchon1979lanczos}}
    \end{subfigure}
    \begin{subfigure}{0.15\linewidth}
        \centering
        \includegraphics[width=\linewidth]{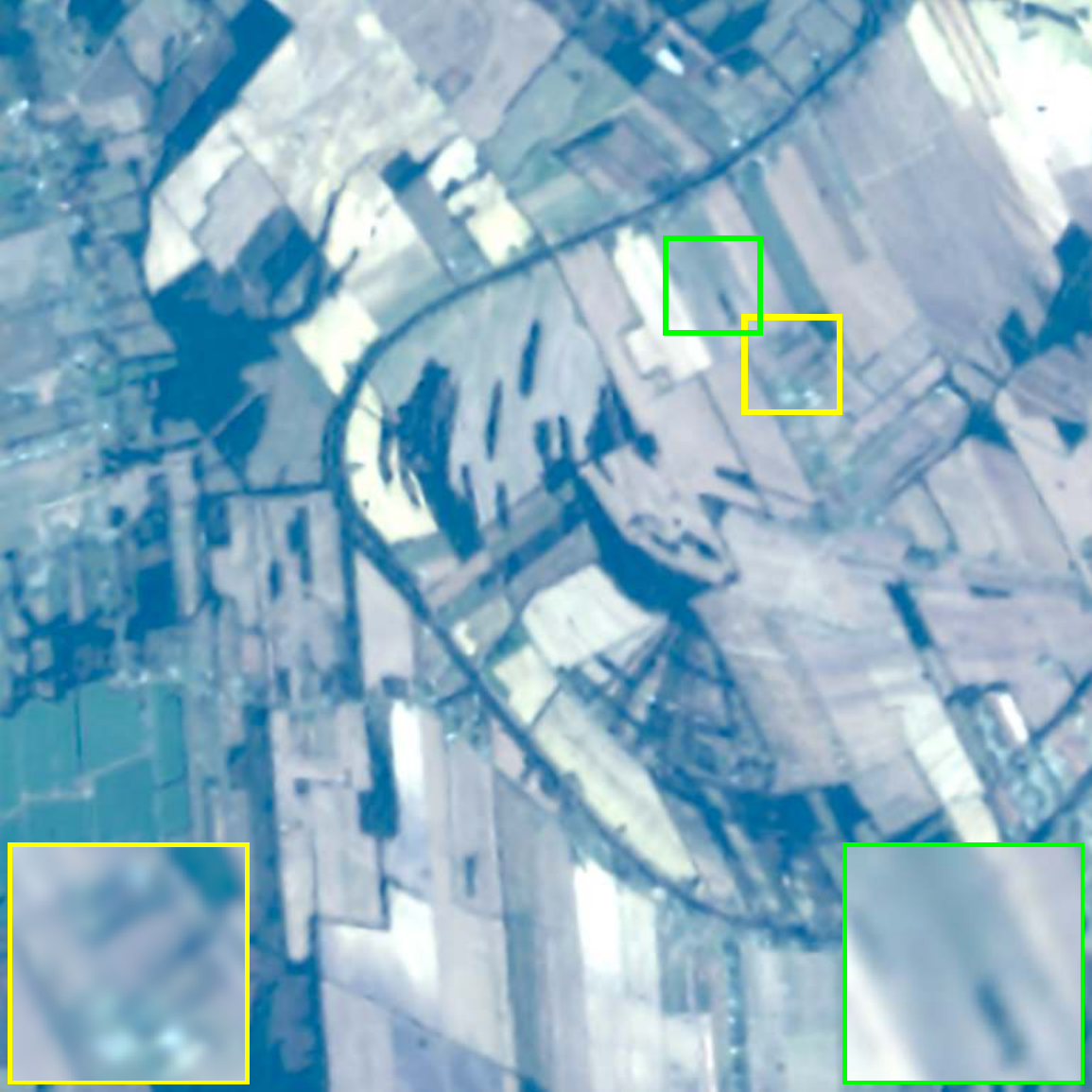}
        \captionsetup{labelformat=empty}
        \caption{\texttt{MC-Net} \cite{li2020mixed}}
    \end{subfigure}
    \begin{subfigure}{0.15\linewidth}
        \centering
        \includegraphics[width=\linewidth]{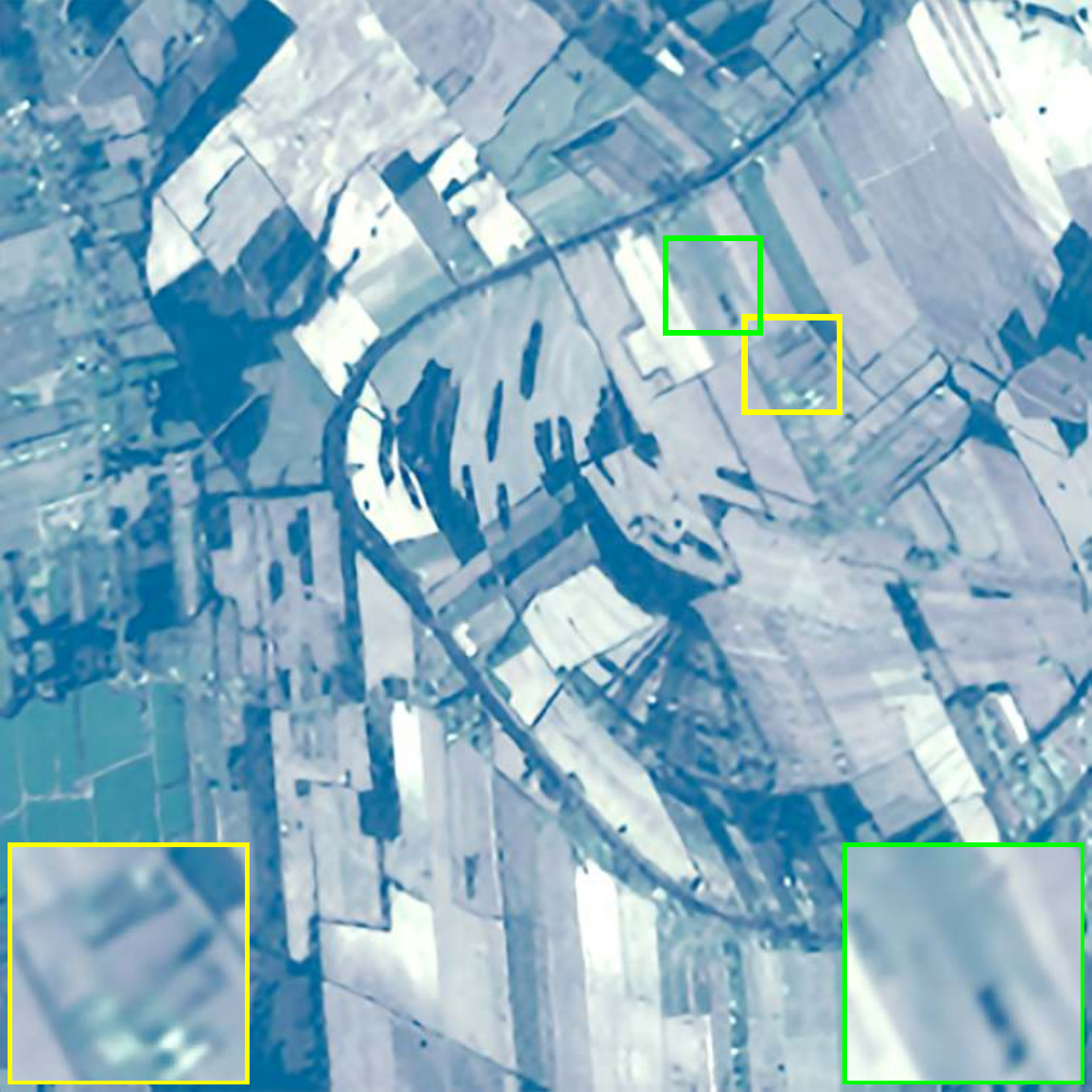}
        \captionsetup{labelformat=empty}
        \caption{\texttt{TSBSR} \cite{liang2023blind}}
    \end{subfigure}
    \begin{subfigure}{0.15\linewidth}
        \centering
        \includegraphics[width=\linewidth]{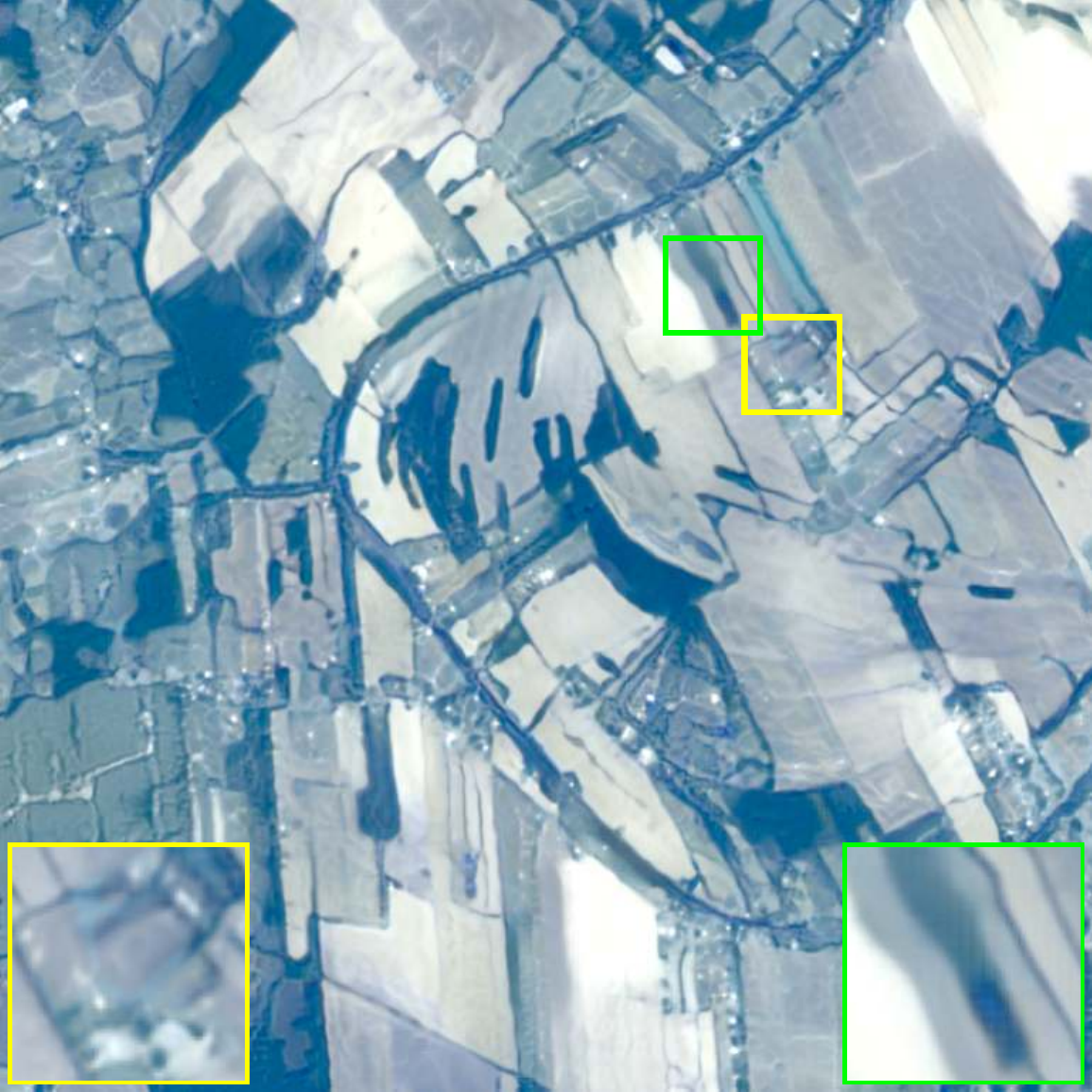}
        \captionsetup{labelformat=empty}
        \caption{\texttt{IAT} (cf. Eq.~\eqref{eq:iat})}
    \end{subfigure}
    \begin{subfigure}{0.15\linewidth}
        \centering
        \includegraphics[width=\linewidth]{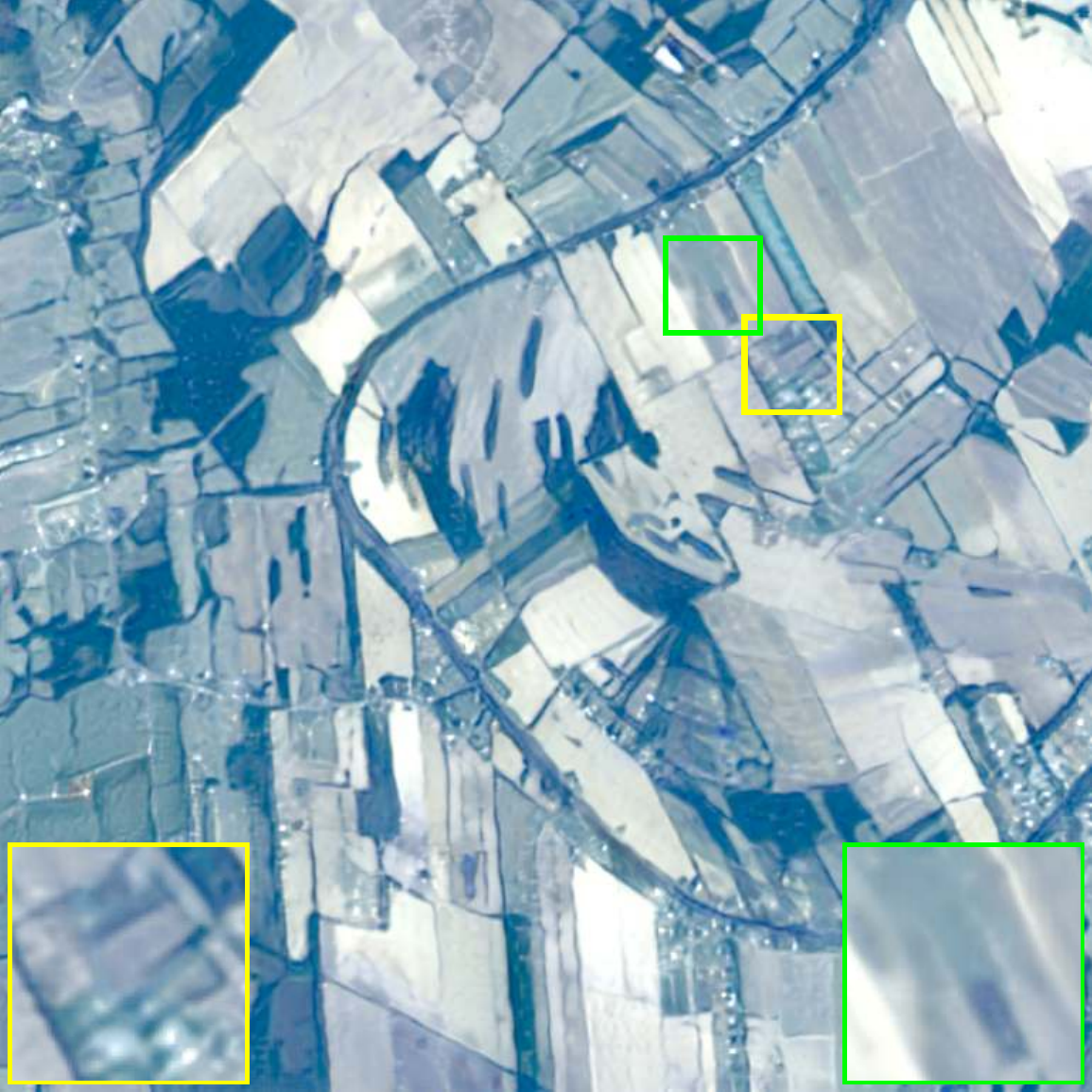}
        \captionsetup{labelformat=empty}
        \caption{\texttt{FRESCO} (\texttt{HSR})}
    \end{subfigure}
    
    \caption{RGB renderings of the recovered SRIs on real data. First row: The recovered SRIs for MSI region; Second row: The recovered SRIs for HSI region.}
    \label{fig:real_experiment_visual}
\end{figure*}

\subsubsection{Dataset}  
We use real HSI and MSI data over Jefferson, MO, USA.  
The HSI is from the Hyperion sensor on NASA’s EO-1 satellite~\cite{EarthExplore2025}, with 30\,m spatial resolution and 242 spectral bands (400–2500\,nm). After removing noisy and water-absorption bands, 71 bands remain.  
The MSI is from Sentinel-2A~\cite{EarthExplore2025}, which has 13 bands; we use the four bands with 10\,m resolution.
We manually crop roughly aligned regions: $\tY\pH \in \mathbb{R}^{230 \times 230 \times 71}$ and $\tY\pM \in \mathbb{R}^{690 \times 690 \times 4}$ (see Fig.~\ref{fig:real_experiment_visual}).  
Sensor spectral specifications are available in~\cite{EarthExplore2025}, allowing us to estimate $\bm{P}\pM \in \mathbb{R}^{4 \times 71}$ using Eq.~\eqref{eq:find_PM_unregistered_HMF}.  
This estimated $\bm P\pM$ is used across all methods. We set the number of materials to $R=3$.

Fig.~\ref{fig:real_experiment_visual} shows the spatial reconstruction results, visualized using RGB bands. These bands are not available in the original MSI, and thus the MSI is visualized by using the bands that are close to the RGB bands.
For MSR, all methods except for \texttt{IARF} produce visually reasonable reconstructions. However, \texttt{UHIF-RIM} and the \texttt{NBFusion} have noticeable noise artifacts. Both \texttt{u2MDN} and the proposed \texttt{FRESCO} method, achieves the best visual quality in this task. For HSR, the proposed framework demonstrates visually sharper edge preservation in the recovered $\tY\pH_{\mathrm{SRI}}$, as highlighted in the zoom-in box at the left side. In this experiment, \texttt{IAT} using \eqref{eq:iat} still generates ``hallucinations''. As highlighted in the zoom-in box on the right. Again, this is likely due to the lack of identifiability of $\bm f^\star$ using \eqref{eq:iat}.

\begin{table}[h]
\centering
\renewcommand{\arraystretch}{1.2}
\setlength{\tabcolsep}{18pt}
\caption{FID on real data.}
\begin{adjustbox}{max width=\linewidth}
\begin{tabular}{|c|c|c|}
\hline
Task & Methods & FID \\ \hline
\multirow{5}{*}{MSR} & \texttt{u2MDN} \cite{qu2021unsupervised}& 14.55 \\
 & \texttt{UHIF-RIM} \cite{ying2021unaligned}& 14.83 \\
 & \texttt{IARF} \cite{zhou2019integrated}& 51.63 \\
 & \texttt{NBFusion} {\cite{chen2017normalized,kanatsoulis2018hyperspectral}}& 32.19 \\
 & \texttt{FRESCO} (\texttt{MSR}) & \textbf{10.34} \\ \hline
\multirow{5}{*}{HSR} &\texttt{Lanczos} \cite{duchon1979lanczos}& 33.01 \\
 & \texttt{MC-Net} \cite{li2020mixed}& 29.40 \\
 & \texttt{TSBSR} \cite{liang2023blind}& 22.93 \\
 & \texttt{IAT} (cf. Eq.~\eqref{eq:iat})& 21.38 \\
 & \texttt{FRESCO} (\texttt{HSR}) & \textbf{20.73} \\ \hline
\end{tabular}
\end{adjustbox}
\label{tab:FID_real_experiment}
\end{table}

Table~\ref{tab:FID_real_experiment} presents the FID scores obtained by all the methods on this experiment. The FID is computed based on the three bands of the MSI and the corresponding bands of the recovered SRIs. For MSR, both \texttt{u2MDN} and the proposed method achieve comparably low FID values, suggesting effective reconstruction quality. For HSR, the proposed method yields a significantly lower FID compared to all baselines, indicating that the recovered $\tY\pH_{\mathrm{SRI}}$ better preserves high-resolution details.

\medskip

Code available at \url{https://github.com/XiaoFuLab/FRESCO_Unregistered-Spectral-Image-Fusion}. Benchmark datasets for illustration and experiments, including Pavia University, Terrain, and Indian Pines, were obtained from  \cite{pavia_university_dataset,hydice_dataset,baumgardner2015220}. The Hyperion and Sentinel-2A pair was obtained from \cite{EarthExplore2025}.

\section{Conclusion}
In this work, we proposed \texttt{FRESCO}, a new algorithmic framework to tackle the challenging problem of unregistered spectral image fusion for spatio-spectral super-resolution. \texttt{FRESCO} integrates coupled spectral unmixing (via coupled block-term tensor decomposition) and latent-space adversarial learning to jointly super-resolve both MSI and HSI in an unsupervised manner---without relying on training data or spatial co-registration. Importantly, we established the first theoretical guarantees for the recoverability of both super-resolved images under reasonable generative models in the unregistered spectral image fusion setting.
Extensive experiments on semi-real and real datasets demonstrate that \texttt{FRESCO} not only achieves state-of-the-art performance, but also offers a practical and theoretically grounded tool for unregistered hyperspectral-multispectral fusion. 

One limitation of this work is that it relies on a multimodal generative model in which the modality-specific mappings are assumed to be \textit{invertible} on the underlying patch manifolds. In practice, such invertibility assumptions may be challenged, since spatial resolution reduction can induce information loss. Extending the analytical framework to \textit{approximately} invertible settings could therefore be an important direction for future work. In addition, the current implementation is based on GAN-style architectures, which may suffer from mode collapse and optimization instability. Replacing the GAN components with diffusion- or flow-based generative models could potentially improve robustness and reconstruction quality.

\bibliographystyle{ieeetr}
\bibliography{refs}

\clearpage

\appendices
\begin{center}
    {\bf Supplementary Materials}
\end{center}

\section{Proof of Theorem~\ref{theorem:MSR}}\label{app:prof_theorem_1}
The proof is similar to that under co-registered HMF settings \cite{ding2020hyperspectral},
with some critical differences.
For example, the unregistered case allows the ranks of $\S_r\pM$ and $\S_r\pH$ to be different.
Nonetheless, the key steps in \cite{ding2020hyperspectral} can be used in our case, leading to the desired conclusion of Theorem~\ref{theorem:MSR}. 

\subsection{Preliminaries}
The proof will use properties of a tensor decomposition model, namely, the LL1 decomposition. The definition is as follows \cite{de2008decompositions}:
\begin{Def}[LL1 Decomposition]
A tensor $\tY\in\mathbb{R}^{I\times J\times K}$ admits a block-term decomposition (BTD) with multilinear rank-($L_r,L_r,1$) terms if it can be expressed as follows 
\begin{align}
    \tY = \sum_{r=1}^R \S_r \circ \c_r,~{\rm rank}(\S_r)= L_r,
\end{align}
for $r=1,\ldots,R$.
\end{Def}

We will need to use the following lemma:
\begin{Lemma}\label{lemma:LL1}
    \cite{de2008decompositions} For an LL1 tensor, denote $\S_r=\A_r\B_r^\T$ where $\A_r\in \mathbb{R}^{I\times L_r}$ and $\bm B_r\in \mathbb{R}^{J\times L_r}$. Assume that $L_r=L$ and all $r\in [R]$ are drawn from any joint absolutely continuous distribution. Then, the LL1 decomposition of $\underline{\bm Y}$ is essentially unique almost surely, if $IJ \geq L^2 R$ and 
    \begin{equation}
        \min\big( \lfloor\nicefrac{I}{L}\rfloor,R\big) + \min\big(\lfloor\nicefrac{J}{L}\rfloor,R\big) + \min\big(K,R\big) \geq 2R +2. 
    \end{equation}
\end{Lemma}
Here, ``essential uniqueness'' means that if $\tY = \sum_{r=1}^{R}\big(\widehat{\bm A}_r (\widehat{\bm B
}_r^\T\big)) \circ \widehat{\c}_r$, then we must have
\begin{align}\label{eq:permutation}
    \widehat{\bm S}_r  = \alpha_r\S_{\bm \pi(r)},~ \widehat{\bm c}_r = (\nicefrac{1}{\alpha_r}) \bm c_{\bm \pi(r)},
\end{align}
where $\widehat{\S}_r =\widehat{\A}_r\widehat{\B}_r^\T$, $\alpha_r\neq 0$ for all $r\in [R]$, and $\bm \pi$ is a permutation of $\{1,\ldots,R\}$.
We will also use the following:
\begin{Lemma}\label{lemma:joint_dis}
    \cite{kanatsoulis2018hyperspectral} Let $\widetilde{\bm A} = \bm P \bm A$, where $\bm A$ is drawn from any absolutely continuous joint distribution in $\mathbb{R}^{I \times L}$ and $\bm P \in \mathbb{R}^{I^{\prime} \times I}$ has full row rank. Then, $\widetilde{\bm A}$ follows an absolutely continuous joint distribution in $\mathbb{R}^{I^{\prime} \times L}$.
\end{Lemma}

\subsection{Proof of The Theorem}
Note that that $\{\widehat{\bm S}_r\pH, \widehat{\c}_r\pH, \widehat{\bm S}_r\pM\}_{r=1}^{R}$ is an optimal solution to Problem \eqref{eq:MSR}.  We will prove that the MSI and HSI decomposition parts share the same scaling and permutation ambiguities, and thus re-assembling $\widehat{\S}\pM_r$ and $\widehat{\bm c}_r\pH$ would correctly recover the desired super-resolution $\tY\pM$.

Lemma~\ref{lemma:LL1} indicates that the following equations hold
\begin{equation}\label{eq:prof_decompose_M}
    \widehat{\bm S}\pM_r = \alpha_r \bm S\pM_{\bm \pi_1 (r)},~ \bm P\pM \widehat{\bm c}\pH_r = \alpha_r^{-1} \bm P\pM \bm c\pH_{\bm \pi_1(r)}
\end{equation}
where $\alpha_r \neq 0$ for all $r\in[R]$ and $\bm \pi_1$ denotes a permutation of $\{1,\cdots,R\}$ resulted in the MSI decomposition (cf. Lemma~\ref{lemma:LL1}).

We first show that $\alpha_r = 1$ for all $r \in [R]$. Note that $\{\bm A_r\pM, \bm B_r\pM\}_{r=1}^R$ are drawn from continuous distribution, then the following matrix:
\begin{equation*}
    \begin{aligned}
        \bm S\pM &= \big[{\rm {\rm vec}}(\bm S\pM_1),\cdots,{\rm {\rm vec}}(\bm S\pM_R)\big] \\
        &= \big[{\rm vec}\big(\bm A\pM_1  (\bm B\pM_1)^\T\big),\cdots,{\rm vec}\big(\bm A\pM_R (\bm B\pM_R)^\T\big)\big],
    \end{aligned}
\end{equation*}
has full column rank almost surely under the condition $I\pM J\pM \geq (L\pM)^2 R \geq R$. Note that we have
\begin{equation}\label{eq:prof_scale_ambiguity}
    \begin{aligned}
        & \sum_{r=1}^{R} \bm S\pM_r = \sum_{r=1}^R \widehat{\bm S}\pM_r = \sum_{r=1}^R \alpha_r \bm S\pM_{\bm \pi_1(r)} = \bm 1_{I\pM \times J\pM}\\
        \Rightarrow &\sum_{r=1}^R (\alpha_r - 1)\bm S\pM_{\bm \pi_1(r)} = \bm 0_{I\pM \times J\pM}\\
        \Rightarrow & \sum_{r=1}^R (\alpha_r - 1) {\rm vec}\big(\bm S\pM_{\bm \pi_1(r)}\big) = \bm 0_{I\pM J\pM} 
    \end{aligned}
\end{equation}
As $\bm S\pM$ has full column rank indicates that ${\rm vec}\big(\bm S\pM_r\big)$ are linearly independent for $r\in [R]$. Thus, the only solution to \eqref{eq:prof_scale_ambiguity} is that $\alpha_r = 1$ for all $r \in [R]$.

Moreover, applying Lemma~\ref{lemma:LL1} to the $\tY\pH$, we have the following equations:
\begin{equation}\label{eq:prof_decompose_H}
\widehat{\bm S}\pH_r = \beta_r \bm S\pH_{\bm \pi_2(r)},~ \widehat{\bm c}\pH_r = \beta_r^{-1} \bm c\pH_{\bm \pi_2(r)} 
\end{equation}
With the arguments one can obtain that $\beta_1 = \cdots = \beta_R = 1$. 
Plugging \eqref{eq:prof_decompose_H} into \eqref{eq:prof_decompose_M}, we have
\begin{align}
    &\bm P\pM \bm c\pH_{\bm \pi_2 (r)} = \bm P\pM \bm c\pH_{\bm \pi_1(r)} ~, \forall r \in [R] \\
    \Rightarrow ~&\bm P\pM \bm C (\bm \Pi_1 - \bm \Pi_2) = \bm 0_{K\pM \times R}\label{eq:prof_Pi}
\end{align}
where $\bm C= \big[\bm c\pH_1,\cdots,\bm c\pH_R\big] \in \mathbb{R}^{K\pH \times R}$, $\bm \Pi_1 = [\bm e_{\bm \pi_1(1)},\cdots,\bm e_{\bm \pi_1(R)}]$ and $\bm \Pi_2 = [\bm e_{\bm \pi_2(1)},\cdots,\bm e_{\bm \pi_2(R)}]$, where $\bm e_i$ denotes the $i$-th column of identity matrix $\bm I \in \mathbb{R}^{R\times R}$. Note that when $R = 1$, then $\bm \pi_1(1) = \bm \pi_2(1) = 1$ holds trivially.

In what follows we claim that $\bm \Pi_1 = \bm \Pi_2$ still holds under the situation when $R \geq 2$. Since $\bm c\pH_r$ is drawn from any absolutely continuous joint distribution for all $r \in [R]$ and $\bm P\pM$ has full row rank, then Lemma~\ref{lemma:joint_dis} indicates that $\bm P\pM \bm C$ follows an absolutely continuous joint distribution in $\mathbb{R}^{K\pM \times R}$. It further indicates that Kruskal rank of $\bm P\pM \bm C$ is $\min (K\pM,R)$ with probability one. 

Therefore, with the assumption $K\pM \geq 2$, we have $\min (K\pM,R) \geq 2$, which means that any two columns of $\bm P\pM \bm C$ are linearly independent. Since $\bm \Pi_1$ and $\bm \Pi_2$ are both permutation matrices, every column of $\big(\bm \Pi_1 - \bm \Pi_2\big)$ always has at most two non-zero elements. This means that the only solution to \eqref{eq:prof_Pi} is 
\begin{equation*}
    \big[\bm \Pi_1 - \bm \Pi_2\big]_{:,r} = \bm 0, \ \forall r \in [R],
\end{equation*}
which leads to $\bm\pi_1(r) = \bm \pi_2(r) = \bm \pi(r)$ holds for all $r \in [R]$. Thus, we have
 \begin{align*}
        \widehat{\S}_r\pH = \S_{\bm \pi(r)}\pH,~ \widehat{\S}_r\pM= \S_{\bm \pi(r)}\pM,~ \widehat{\c}_r\pH = \c_{\bm \pi(r)}\pH,
\end{align*}
Consequently, we have
\begin{align*}
    \widehat{\tY}\pM_{\rm SRI} &=\sum_{r=1}^R \widehat{\S}_r\pM \circ \widehat{\c}_r\pH
    = \sum_{r=1}^R \bm S\pM_{\bm \pi(r)} \circ \bm c\pH_{\bm \pi(r)}\\
    &=\sum_{r=1}^R \bm S\pM_{r} \circ \bm c\pH_{r} =\tY\pM_{\rm SRI}.
\end{align*}
This completes the proof.

\section{Proof of Lemma~\ref{lemma:bijective_function_f}}
We show that there exists an $\bm f^\star$ that is a bijective and continuous function by construction.
Under the generative model in Assumption~\ref{assumption:rotate_generative_model}, both $\bm g\pH: \mathcal{Z} \to \mathrm{dom}(\bm S_r\pH[\mathcal{H}^{(\bm w,\theta)}])$ and $\bm g\pM: \mathcal{Z} \to \mathrm{dom}(\bm S_r\pM[\mathcal{M}^{(\bm w,\theta)}])$ are bijections, they are invertible between the domain and the range.

This means that that for any $\bm z_r$, we have $\bm z_r = (\bm g\pM)^{-1} \bullet \bm g\pM(\bm z_r)$ and $\bm z_r = (\bm g\pH)^{-1} \bullet \bm g\pH(\bm z_r)$. Then, one can construct $\bm f^\star = \bm g\pM \bullet (\bm g\pH)^{-1}$, which leads to
\begin{align}\label{eq: bij_f}
   &\underbrace{\bm g\pM \bullet (\bm g\pH)^{-1}}_{\bm f^\star} \bullet \bm g\pH(\z_r)  = \bm g\pM(\z_r),\\
\Longrightarrow& \bm f^\star(\S_r\pH[{\cal H}^{(\bm w,\theta)}]) = \S_r\pM[{\cal M}^{(\bm w,\theta)}].
\end{align}

As $\bm g\pM$ and $\bm g\pH$ are both continuous functions with continuous inverses, the composite function $\bm f^{\star} = \bm g\pM \bullet (\bm g\pH)^{-1}$ is guaranteed to be continuous as well. Similarly, one can also construct a continuous and bijective function $$\bm q^\star = \bm g\pH \bullet (\bm g\pM)^{-1}$$ that maps $\S_r\pM[{\cal M}^{(\bm w,\theta)}]$ to its corresponding $\S_r\pH[{\cal H}^{(\bm w,\theta)}]$. Therefore, $\bm f^{\star}$ is a continuous bijective function, 
which completes the proof.

\section{Proof of Theorem~\ref{theorem:HSR}}
\textit{Step 1)}: 
From Lemma~\ref{lemma:bijective_function_f}, we have established that
\begin{align}
    \bm f^\star\left(\S_r\pH[{\cal H}^{(\bm w,\theta)}]\right) = \S_r\pM[{\cal M}^{(\bm w,\theta)}],
\end{align}
for all the $\w,\theta$.
Consequently, we have
\begin{align}
&    [\bm f^{\star}]_{\#}p\pH_r = p\pM_r,~\forall r \in [R],\\
 &   \text{where~}\S_r\pM[{\cal M}^{(\bm w,\theta)}] \sim p_{r}\pM,~  \S_r\pH[{\cal H}^{(\bm w,\theta)}] \sim p_{r}\pH. \nonumber
\end{align}
That is, point-wise correspondence implies distribution matching (but not the other way around).

\textit{Step 2)}: Now we show that as long as the SDA condition is satisfied, the collection of PDFs $\{ p_{r}\pM\}_{r=1}^R$ also satisfies a similar condition.

Note that \( p(\bm z_r) \) is the marginal density obtained by averaging \( p(\bm z_r \mid \theta) \) over \( \theta \in [0,2\pi] \), i.e.,
\[
p(\bm z_r) = \frac{1}{2\pi} \int_0^{2\pi} p(\bm z_r \mid \theta) d\theta,
\]
where we have used the fact that $\theta$ follows the uniform distribution between $[0,2\pi]$.

Consider two disjoint, open, and connected sets $\mathcal{A}, \mathcal{B} \subseteq \mathcal{Z}$.
Assumption~\ref{assumption:SDA} indicates that their probability masses under the conditional density $p(\bm z_r \mid \theta)$ are unequal for all $\theta \in [0,2\pi]$, i.e.,
\begin{equation}\label{eq:inequality_conditional}
    \int_{\mathcal{A}} p(\bm z_r \mid \theta)\, d\bm z
    \neq
    \int_{\mathcal{B}} p(\bm z_r \mid \theta)\, d\bm z,
    \quad \forall\, \theta \in [0,2\pi].
\end{equation}
Now we define
\begin{equation}\label{eq:v_theta}
\bm v_r(\theta)
    \triangleq
    \int_{\mathcal{A}} p(\bm z_r \mid \theta)\, d\bm z
    -
    \int_{\mathcal{B}} p(\bm z_r \mid \theta)\, d\bm z .
\end{equation}
Since $p(\bm z_r \mid \theta)$ is continuous with respect to $\theta$, and the integrals are taken over fixed connected and open sets $\mathcal{A},\mathcal{B} \in \mathcal{Z}$, it follows that $\bm v_r(\theta)$ is also continuous on $[0,2\pi]$.
Moreover, \eqref{eq:inequality_conditional} implies that $\bm v_r(\theta) \neq 0$ for all $\theta \in [0,2\pi]$. By the Intermediate Value Theorem, $\bm v_r(\theta)$ is either strictly positive or strictly negative for all $\theta \in [0,2\pi]$.
Therefore, we have
\begin{equation}\label{eq:IVT_SDA}
\begin{aligned}
    &\int_{\mathcal{A}} p(\bm z_r)\, d\bm z -\int_{\mathcal{B}} p(\bm z_r)\, d\bm z\\
    &= \int_{0}^{2\pi}\left(\int_{\mathcal{A}} p(\bm z_r \mid \theta)\, d\bm z - \int_{\mathcal{B}} p(\bm z_r \mid \theta)\, d\bm z\right)d\theta \\
    &= \int_{0}^{2\pi} \bm v_r(\theta)\, d\theta \neq 0.
\end{aligned}
\end{equation}

Therefore, under the SDA, there is always an $({\cal A},{\cal B})$-dependent $r'=r({\cal A},{\cal B})\in\{1,\ldots,R\}$, such that for any $({\cal A},{\cal B})$ we have
\begin{align}
    \int_{\cal A} p(\bm z_{r'}) d\bm z \neq \int_{\cal B} p(\bm z_{r'}) d\bm z.
\end{align}

As \( \bm g\pM \) is bijective, we know that the following also holds:
\begin{align}\label{eq:munequal}
\int_{\bm g\pM({\cal A})} p_{r'}\pM(\bm{m})\, d\bm{m} \ \neq \ \int_{\bm g\pM({\cal B})} p_{r'}\pM(\bm{m})\, d\bm{m},    
\end{align}
where we have used $\bm m$ to represent the MSI patches for notation simplicity.
Here \( \bm g\pM({\cal A}) \) denotes the image of \( {\cal A} \) under \(\bm g\pM \), and \( p_{r'}\pM(\bm{m}) \) is the corresponding PDF of \( \bm{m} = \bm g\pM(\bm{z}) \).
Indeed, by change of variables, we can write:
\[
\int_{\bm g\pM({\cal A})} p_{r'}\pM(\bm{m})\, d\bm{m} = \int_{\cal A} p_{r'}(\bm{z})\, d\bm{z},
\]
and similarly for the integral over $\bm g\pM({\cal B})$,
leading to the conclusion in \eqref{eq:munequal}. That is, for any disjoint, connected, and open ${\cal A},{\cal B} \subseteq {\cal Z}$, there exists $r'=r({\cal A},{\cal B})\in \{1,\ldots,R\}$ such that
\begin{align}\label{eq:m_unequal}
    \int_{\bm g\pM({\cal A})} p_{r'}\pM(\bm{m})\, d\bm{m} \neq \int_{\bm g\pM({\cal B})} p_{r'}\pM(\bm{m})\, d\bm{m}
\end{align}

\textit{Step 3)}: 
We now show that solving \eqref{eq:diversified_gan_theta} in the limit as $T, L \rightarrow \infty$ recovers this ground-truth mapping.
To proceed, we first observe that any optimal solution $\widehat{\bm f}$ to \eqref{eq:diversified_gan_theta} satisfies
\begin{equation}\label{eq:proof_opt_solution}
[\widehat{\bm f}]_{\#}p\pH_r = p\pM_r, \quad \forall r \in [R],
\end{equation}
as established in \cite{goodfellow2014generative, shrestha2024towards}. 
Note that Lemma~\ref{lemma:bijective_function_f} guarantees the existence of a mapping $\bm f^{\star}$ that satisfies \eqref{eq:abundancemapping}. However, all the $\widehat{\bm f} = \bm h \bullet \bm f^{\star}$ functions with 
\begin{equation}\label{eq:mpa}
    [\bm h]_{\#} p\pM_r = p\pM_r,~\forall r \in [R]
\end{equation}
satisfy \eqref{eq:proof_opt_solution} as well. Our goal is to show that there does not exist a non-identity continuous $\bm h$ such that \eqref{eq:mpa} holds. To this end, we follow the steps in \cite[Theorem 1]{shrestha2024towards}, with slight modifications to accommodate our context. The proof is by contradiction. Suppose that there exists $\overline{\bm m} \in \bm g\pM (\mathcal{Z})$ such that
\begin{equation}
    \bm h(\overline{\bm m}) \neq \overline{\bm m}.
\end{equation}
Then, we can define an open set as follows:
\begin{equation}
    \mathcal{D} = \mathcal{N}_\epsilon (\overline{\bm m}) \cap \bm g\pM (\mathcal{Z}),
\end{equation}
where $\mathcal{N}_{\epsilon}(\overline{\bm m})$ defines a ball centered at $\overline{\bm m}$ with a radius of $\epsilon$. The above means that $\bm h(\mathcal{D}) \subseteq \bm g\pM (\mathcal{Z})$ under \eqref{eq:mpa}. The continuity of $\bm h$ also implies that
\begin{equation}
    \bm h(\overline{\bm m}) \in \bm h(\mathcal{D}).
\end{equation}
One can always choose a small enough $\epsilon$ such that
\begin{equation}
    \mathcal{D} \cap \bm h(\mathcal{D}) = \emptyset.
\end{equation}
Note that $\mathcal{D}$ is a connected set, which means that $\bm h(\mathcal{D})$ is also connected. Thus, one can construct two disjoint, open, and connected sets satisfying
\begin{equation}
    \int_{\mathcal{D}} p_r\pM(\bm m) d\bm m = \int_{\bm h(\mathcal{D})} p_r \pM(\bm m) d \bm m, ~ \forall r \in [R].
\end{equation}
Let $\mathcal{A^\prime} = (\bm g\pM)^{-1}(\mathcal{D})$ and $\mathcal{B^\prime} = (\bm g\pM)^{-1}\left( \bm h(\mathcal{D})\right)$. The continuity of $\bm g\pM$ and $ (\bm g\pM)^{-1}$ indicates that $\mathcal{A}^{\prime}$ and $\mathcal{B}^{\prime}$ are also disjoint, open, and connected, which is a contradiction to \eqref{eq:m_unequal}. Hence, the only $\bm h$ satisfying \eqref{eq:mpa} is an identity mapping. This means that the only solution $\widehat{\bm f}$ attaining \eqref{eq:proof_opt_solution} is $\widehat{\bm f} = \bm f^{\star}$. Subsequently, one can recover $\tY\pH_{\rm SRI}$ patch-by-patch, using \eqref{eq:patch_recover}. This completes the proof.

\section{Proof of Theorem~\ref{theorem:robust_HSR}}

The proof uses Lemma~A.3 in~\cite{shrestha2024towards}, which establishes a robustness guarantee for domain translation task. 

\textit{Step~1:)} We first show that, as long as the $\eta$-SDA condition is satisfied, the diameters of regions with similar densities in the collection of PDFs $\{p_r\pM\}_{r=1}^R$ are also bounded.

Under Assumption~\ref{assumption:relaxed_SDA}, for any $(\mathcal A,\mathcal B) \in \mathcal V$, there exists $\bar{\theta} \in [0,2\pi]$ such that
\begin{equation}
    \int_{\mathcal A} p(\bm z_r \mid \bar{\theta})\, d\bm z = \int_{\mathcal B} p(\bm z_r \mid \bar{\theta})\, d\bm z, ~~\forall r \in [R].
\end{equation}
By construction in \eqref{eq:v_theta}, one can obtain that $\bm v_{r}(\bar{\theta}) = 0, \forall r \in [R]$. Thus, the Intermediate Value Theorem based argument \eqref{eq:IVT_SDA} no longer holds. Consequently, after marginalizing over $\theta$, it is possible that
\begin{equation}\label{eq:latent_equality}
    \int_{\mathcal A} p(\bm z_r)\, d\bm z
    =
    \int_{\mathcal B} p(\bm z_r)\, d\bm z,~~~\forall r \in [R].
\end{equation}
Note that if the above did not hold for any $({\cal A}, {\cal B}) \in {\mathcal V}$, then SDA condition is satisfied, which trivially leads to the theorem's conclusion. Hence we only need to consider the worst case where \eqref{eq:latent_equality} holds.
Since $\bm g\pM$ is bijective, the above equality implies
\begin{equation}\label{eq:violation_pM}
    \int_{\bm g\pM(\mathcal A)} p_r\pM(\bm m)\, d\bm m = \int_{\bm g\pM(\mathcal B)} p_r\pM(\bm m)\, d\bm m , ~\forall r \in [R].
\end{equation}
where $\bm m = \bm g\pM(\bm z)$ denotes an MSI abundance patch. Moreover, continuity and bijectivity of $\bm g\pM$ ensure that
$\bm g\pM(\mathcal A)$ and $\bm g\pM(\mathcal B)$ remain non-empty, open, connected, and disjoint. 
Conversely, for any $(\mathcal A',\mathcal B') \notin \mathcal V$,
an argument analogous to~\eqref{eq:inequality_conditional}--%
\eqref{eq:IVT_SDA} shows that
\[
\int_{\mathcal{A}} p(\bm z_r) d\bm z \neq \int_{\mathcal{B}} p (\bm z_r) d\bm z, ~\forall r \in [R].
\]
It concludes that 
$$
\mathcal{V}\pM \triangleq \{\bm g\pM(\mathcal{A}), \bm g\pM(\mathcal{B}) \mid (\mathcal{A},\mathcal{B})\in \mathcal{V}\}
$$
already contains all the set pairs that induce equal densities under all distributions $\{p\pM_r\}_{r=1}^R$, i.e., those for which \eqref{eq:violation_pM} holds. Finally, by the Lipschitz condition~\eqref{eq:lipschitz_g_M}, one can obtain
\begin{equation}
\sup_{\bm w,\bm z \in \mathcal{A}} \| \bm g\pM(\bm w) - \bm g\pM(\bm z)\|_2 \leq \alpha\pM \sup_{\bm w,\bm z \in \mathcal{A}}\|\bm w - \bm z\|_2,
\end{equation}
which implies that 
$${\rm dia}(\bm g\pM(\mathcal{A})) \leq \alpha\pM {\rm dia}(\mathcal{A}),$$
and similarly,
$$
{\rm dia}(\bm g\pM(\mathcal{B})) \leq \alpha\pM {\rm dia}(\mathcal{B}).
$$
Hence,
\begin{equation}
    \max\!\left\{
    \mathrm{dia}(\bm g\pM(\mathcal A)),
    \mathrm{dia}(\bm g\pM(\mathcal B))
    \right\}
    \leq \alpha\pM \eta,
\end{equation}
holds for all $\left(\bm g\pM(\mathcal A), \bm g\pM(\mathcal B)\right) \in \mathcal V\pM$, which further implies that 
\begin{equation}\label{eq:relaxed_diversity_pM}
    \max_{(\mathcal{A}\pM,\mathcal{B}\pM)\in \mathcal{V}^{\pM}} \max\{{\rm dia}(\mathcal{A}\pM),{\rm dia}(\mathcal{B}\pM)\} \leq \alpha\pM \eta.
\end{equation}

\medskip
\textit{Step~2:)} Now we establish the Lipschitz continuity of $\bm f^{\star}$. Let $\bm h_1$ and $\bm h_2$ be any two HSI abundance patches.
By the bijectivity of $\bm g\pH$, there exist
$\bm z_1 = (\bm g\pH)^{-1}(\bm h_1)$ and
$\bm z_2 = (\bm g\pH)^{-1}(\bm h_2)$.
Using the definition $\bm f^{\star} = \bm g\pM\circ(\bm g\pH)^{-1}$,
we have
\begin{equation}
\begin{aligned}
    \|\bm f^{\star}(\bm h_1) - \bm f^{\star}(\bm h_2)\|_{\rm F}
    &= \|\bm g\pM(\bm z_1) - \bm g\pM(\bm z_2)\|_{\rm F} \\
    &\le \alpha\pM \|\bm z_1 - \bm z_2\|_2\\
    &\le \frac{\alpha\pM}{\alpha\pH}
    \|\bm g\pH(\bm z_1) - \bm g\pH(\bm z_2)\|_{\rm F} \\
    &= \frac{\alpha\pM}{\alpha\pH}
    \|\bm h_1 - \bm h_2\|_{\rm F},
\end{aligned}
\end{equation}
where the inequalities follow from the bi-Lipschitz conditions
\eqref{eq:lipschitz_g_M} and \eqref{eq:lipschitz_g_H}.
Hence, $\bm f^{\star}$ is $L_{\bm f^{\star}}$-Lipschitz continuous with
\begin{equation}\label{eq:lipschitz_f}
    L_{\bm f^{\star}} = \frac{\alpha\pM}{\alpha\pH}.
\end{equation}

\medskip
\textit{Step~3:)} Finally, we derive the reconstruction error bound. Since the proof follows exactly the same steps as those in Lemma A.3 of \cite{shrestha2024towards} under the Lipschitz continuity condition \eqref{eq:lipschitz_f} and the relaxed diversity condition \eqref{eq:relaxed_diversity_pM}, we omit the details and obtain
\begin{equation}
    \|\widehat{\bm f}(\bm h) - \bm f^{\star}(\bm h)\|_{\rm F}
    \le
    \frac{2(\alpha\pM)^2}{\alpha\pH}\,\eta,
    \quad \forall \bm h.
\end{equation}
Consequently, the reconstruction error of the super-resolved HSI patch satisfies
\begin{equation}
\begin{aligned}
    \|\widehat{\tY}_{\rm SRI}\pH[\bm m]
    - \tY_{\rm SRI}\pH[\bm m]\|_{\rm F}
    &=
    \left\|
    \sum_{r=1}^R
    \big(\widehat{\bm f}(\bm h_r) - \bm f^{\star}(\bm h_r)\big)
    \circ \bm c_r\pH
    \right\|_{\rm F} \\
    &\le
    \sum_{r=1}^R
    \|\widehat{\bm f}(\bm h_r) - \bm f^{\star}(\bm h_r)\|_{\rm F}
    \|\bm c_r\pH\|_2\\ 
    &\le
    \frac{2(\alpha\pM)^2}{\alpha\pH}\,\eta
    \sum_{r=1}^R \|\bm c_r\pH\|_2,~~\forall \bm m.
\end{aligned}
\end{equation}
This completes the proof.

\section{Neural Network Architectures}\label{app:networks}
\begin{table}[!t]
\centering
\caption{Notation for Network Components}
\label{tab:notation}
\renewcommand{\arraystretch}{1.2}
\begin{tabular}{c c}
\hline
\textbf{Symbol} & \textbf{Description} \\
\hline
Conv & 2D convolutional layer \\
BN & Batch Normalization \\
LReLU & LeakyReLU activation (slope = $0.2$) \\
MaxPool($u$) & Max pooling with kernel size of $u$ and stride of $u$.\\
Upsample($u$) & Bilinear upsampling by factor of $u$.\\
K-$k$ & convolution kernel of size $k$\\
S-$n$ & Stride of size $n$\\
ZP-$p$ & Zero padding of size $p$\\
RP-$p$ & Reflect padding of size $p$\\
\multirow{2}{*}{Skip(layer-$i$)} 
& Concatenate output feature maps of the $i$-th layer \\
& to the corresponding position in the decode path.\\
Average & Compute the mean value for each input feature map.\\
\hline
\end{tabular}
\end{table}

We describe the neural architectures used by $\bm f$, $\bm g$, and $\bm d_r$. Table~\ref{tab:notation} shows the symbols, following the definitions in \cite{shrestha2024towards}.

\noindent
{\bf Architecture of $\bm f$.} We apply a U-Net architecture to construct $\bm f$, whose architecture is detailed in Table~\ref{tab:gen_f}, where $s$ denotes the dowmsampling factor. The encoder blocks and decoder blocks are described as follows.

The encoder block Enc$(C_\text{in} \rightarrow C_\text{out})$ consists of a max pooling operation followed by two convolutional layers:

\begin{enumerate}
    \item MaxPool($2$)
    \item Conv($C_\text{in} \rightarrow C_\text{out}$, K-$3$, S-$1$, ZP-$1$, BN, LReLU)
    \item Conv($C_\text{out} \rightarrow C_\text{out}$, K-$3$, S-$1$, ZP-$1$, BN, LReLU)
\end{enumerate}

The decoder block Dec($C_\text{in} \rightarrow C_\text{out}$, Skip-layer-$i$) performs bilinear upsampling, followed by convolution and concatenation with the corresponding encoder feature map, and two additional convolutional layers:

\begin{enumerate}
    \item Upsample($2$)
    \item Conv($C_\text{in} \rightarrow C_\text{out}$, K-$1$, S-$1$, ZP-$0$, BN, LReLU)
    \item Skip(layer-$i$)
    \item Conv($2 \times C_\text{out} \rightarrow C_\text{out}$, K-$3$, S-$1$, ZP-$1$, BN, LReLU)
    \item Conv($C_\text{out} \rightarrow C_\text{out}$, K-$3$, S-$1$, ZP-$1$, BN, LReLU)
\end{enumerate}

\begin{table}[!t]
\centering
\caption{Spatial super-resolution neural network $\bm f$; if no normalization or activation function is applied, the corresponding position is denoted by $\#$.}
\label{tab:gen_f}
\renewcommand{\arraystretch}{1.2}
\begin{tabular}{cc}
\hline
\textbf{Layer Number} & \textbf{Layer Details} \\
\hline
1 & Upsample($s$)\\
2 & Conv($1 \rightarrow 64$, K-$1$, S-$1$, ZP-$0$, $\#$, LReLU)\\
3 & Conv($64 \rightarrow 64$, K-$3$, S-$1$, ZP-$1$, BN, LReLU)\\
4 & Conv($64 \rightarrow 64$, K-$3$, S-$1$, ZP-$1$, BN, LReLU)\\
5 & Enc($64 \rightarrow 128$)\\
6 & Enc($128 \rightarrow 256$)\\
7 & Enc($256 \rightarrow 512$)\\
8 & Enc($512 \rightarrow 1024$)\\
9 & Dec($1024 \rightarrow 512$, Skip-layer-$7$)\\
10 & Dec($512 \rightarrow 256$, Skip-layer-$6$)\\
11 & Dec($256 \rightarrow 128$, Skip-layer-$5$)\\
12 & Dec($128 \rightarrow 64$, Skip-layer-$4$)\\
13 & Conv($64 \rightarrow 1$, K-$1$, S-$1$, ZP-$0$, $\#$, LReLU)\\
\hline
\end{tabular}
\end{table}

\noindent
{\bf Inverse Mapping Neural Network $\bm g$.}
we adopt a simpler architecture for inverse mapping $\bm g$, consisting of a sequence of residual blocks and a bilinear downsampling in the middle.
The residual block \text{ResBlock}($C_\text{in}\rightarrow C_\text{out}$) operates two convolution layers with a residual connection:
\begin{enumerate}
    \item Conv($C_\text{in}\rightarrow C_\text{out}$, K-$3$, S-$1$, RP-$1$, BN, LReLU)
    \item Conv($C_\text{out}\rightarrow C_\text{out}$, K-$3$, S-$1$, RP-$1$, BN, $\#$)
\end{enumerate}
The architecture of the inverse mapping neural network $\bm g$ is presented in Table~\ref{tab:gen_g}.

\begin{table}[!t]
\centering
\caption{Inverse mapping neural network $\bm g$.}
\label{tab:gen_g}
\renewcommand{\arraystretch}{1.2}
\begin{tabular}{cc}
\hline
\textbf{Layer Number} & \textbf{Layer Details} \\
\hline
1 & Conv($1 \rightarrow 64$, K-$1$, S-$1$, ZP-$0$, $\#$, LReLU)\\
2 & ResBlock($64 \rightarrow 64$) \\
3 & ResBlock($64 \rightarrow 64$) \\
4 & Conv($64 \rightarrow 64$, K-$3$, S-$s$, RP-$1$, $\#$, LReLU) \\
5 & Conv($64 \rightarrow 1$, K-$3$, S-$1$, RP-$1$, $\#$, LReLU)\\
\hline
\end{tabular}
\end{table}

\begin{table}[!t]
\centering
\caption{Discriminator $\bm d$}
\label{tab:discriminator_m}
\renewcommand{\arraystretch}{1.2}
\begin{tabular}{cc}
\hline
\textbf{Layer Number} & \textbf{Layer Details} \\
\hline
1 & Conv($1 \rightarrow 64$, K-$3$, S-$1$, ZP-$1$, $\#$, LReLU).\\
2 & Conv($64 \rightarrow 64$, K-$3$, S-$2$, ZP-$1$, BN, LReLU)\\
3 & Conv($64 \rightarrow 128$, K-$3$, S-$1$, ZP-$1$, BN, LReLU)\\
4 & Conv($128 \rightarrow 128$, K-$3$, S-$2$, ZP-$1$, BN, LReLU)\\
5 & Conv($128 \rightarrow 256$, K-$3$, S-$1$, ZP-$1$, BN, LReLU)\\
6 & Conv($256 \rightarrow 256$, K-$3$, S-$2$, ZP-$1$, BN, LReLU)\\
7 & Conv($256 \rightarrow 512$, K-$3$, S-$1$, ZP-$1$, $\#$, LReLU)\\
8 & Conv($512 \rightarrow R$, K-$1$, S-$1$, ZP-$0$)\\
9 & Average\\
\hline
\end{tabular}
\end{table}

\noindent
{\bf Discriminator $\bm d_r$.}
Finally, to handle the multiple abundance maps translation task, we employ a multi-task discriminator described in Table.~\ref{tab:discriminator_m}, where $R$ denotes the number of materials, and the $r$th outputs of $\bm d$ is served as $\bm d_r$ for all $r\in [R]$.

\end{document}